\documentclass[longauth]{aaV2}

\usepackage{euclid}
\usepackage{lastpage}
\usepackage{graphicx}
\usepackage{natbib}
\usepackage{scalerel}
\usepackage{multirow}

\usepackage[version=4]{mhchem}

\usepackage[nolist,nohyperlinks]{acronym}

\usepackage[table]{xcolor}

\usepackage{txfonts}
\usepackage[pdfencoding=auto,psdextra]{hyperref}
\hypersetup{
    colorlinks=true,
    linkcolor=blue,
    filecolor=magenta,      
    urlcolor=blue,
    citecolor=blue
}
\makeatletter
\renewcommand*\aa@pageof{, page \thepage{} of \pageref*{LastPage}}
\makeatother

\usepackage[utf8]{inputenc}

\usepackage[switch, modulo]{lineno}

\DeclareSIUnit \kelvin{K}
\DeclareSIUnit \erg{erg}
\DeclareSIUnit \grv{grooves}

\newcommand{\Ha}{\ensuremath{\text{H}\alpha}\,}

\usepackage{orcidlink} 
\newcommand{\orcid}[1]{\orcidlink{#1}}

\begin{document}

\acrodef{ASIC}{application specific integrated circuit}
\acrodef{BFE}{brighter-fatter effect}
\acrodef{CaLA}{camera-lens assembly}
\acrodef{CCD}{charge-coupled device}
\acrodef{CoLA}{corrector-lens assembly}
\acrodef{CDS}{Correlated Double Sampling}
\acrodef{CFC}{cryo-flex cable}
\acrodef{CGH}{computer-generated hologram}
\acrodef{CNES}{Centre National d'Etude Spacial}
\acrodef{CPPM}{Centre de Physique des Particules de Marseille}
\acrodef{CPU}{central processing unit}
\acrodef{CTE}{coefficient of thermal expansion}
\acrodef{DCU}{Detector Control Unit}
\acrodef{DPU}{Data Processing Unit}
\acrodef{DS}{Detector System}
\acrodef{EDS}{\Euclid Deep Survey}
\acrodef{EE}{encircled energy}
\acrodef{ESA}{European Space Agency}
\acrodef{EWS}{\Euclid Wide Survey}
\acrodef{FDIR}{Fault Detection, Isolation and Recovery}
\acrodef{FOM}{figure of merit}
\acrodef{FOV}{field of view}
\acrodef{FPA}{focal plane array}
\acrodef{FWA}{filter-wheel assembly}
\acrodef{FWC}{full-well capacity}
\acrodef{FWHM}{full width at half maximum}
\acrodef{GWA}{grism-wheel assembly}
\acrodef{H2RG}{HAWAII-2RG}
\acrodef{IP2I}{Institut de Physique des 2 Infinis de Lyon}
\acrodef{JWST}{{\em James Webb} Space Telescope}
\acrodef{IAD}{ion-assisted deposition}
\acrodef{ICU}{Instrument Control Unit}
\acrodef{IPC}{inter-pixel capacitance}
\acrodef{LAM}{Laboratoire d'Astrophysique de Marseille}
\acrodef{LED}{light-emitting diode}
\acrodef{MACC}{Multiple Accumulated}
\acrodef{MCT}{Mercury Cadmium Telluride}
\acrodef{MLI}{multi-layer insulation}
\acrodef{MMU}{Mass Memory Unit}
\acrodef{MPE}{Max-Planck-Institut für extraterrestrische Physik}
\acrodef{MPIA}{Max-Planck-Institut für Astronomie}
\acrodef{NA}{numerical aperture}
\acrodef{NASA}{National Aeronautic and Space Administration}
\acrodef{JPL}{NASA Jet Propulsion Laboratory}
\acrodef{MZ-CGH}{multi-zonal computer-generated hologram}
\acrodef{NI-CU}{NISP calibration unit}
\acrodef{NI-OA}{near-infrared optical assembly}
\acrodef{NI-GWA}{NISP Grism Wheel Assembly}
\acrodef{NISP}{Near-Infrared Spectrometer and Photometer}
\acrodef{PARMS}{plasma-assisted reactive magnetron sputtering}
\acrodef{PLM}{payload module}
\acrodef{PTFE}{polytetrafluoroethylene}
\acrodef{PV}{performance verification}
\acrodef{PWM}{pulse-width modulation}
\acrodef{PSF}{point spread function}
\acrodef{QE}{quantum efficiency}
\acrodef{ROIC}{readout-integrated chip}
\acrodef{ROS}{reference observing sequence}
\acrodef{SCA}{sensor chip array}
\acrodef{SCE}{sensor chip electronic}
\acrodef{SCS}{sensor chip system}
\acrodef{SGS}{science ground segment}
\acrodef{SHS}{Shack-Hartmann sensor}
\acrodef{SNR}[SNR]{signal-to-noise ratio}
\acrodef{SED}{spectral energy distribution}
\acrodef{SiC}{silicon carbide}
\acrodef{SVM}{service module}
\acrodef{VIS}{visible imager}
\acrodef{WD}{white dwarf}
\acrodef{WFE}{wavefront error}
\acrodef{ZP}{zero point}


\newcommand{\BGlow}{\ensuremath{926}}
\newcommand{\BGhigh}{\ensuremath{1366}}
\newcommand{\RGlow}{\ensuremath{1206}}
\newcommand{\RGhigh}{\ensuremath{1892}}
\newcommand{\BGrange}{\ensuremath{[\BGlow\,;\,\BGhigh]}}
\newcommand{\RGrange}{\ensuremath{[\RGlow\,;\,\RGhigh]}}
\newcommand{\BGrangenew}{\BGlow--\BGhigh}
\newcommand{\RGrangenew}{\RGlow--\RGhigh}

\ifdefined\BGE
    \renewcommand{\BGE}{\ensuremath{BG_\sfont{E}}}
\else
    \newcommand{\BGE}{\ensuremath{BG_\sfont{E}}}
\fi

\ifdefined\RGE
    \renewcommand{\RGE}{\ensuremath{RG_\sfont{E}}}
\else
    \newcommand{\RGE}{\ensuremath{RG_\sfont{E}}}
\fi

\newcommand{\TBfirst}{\ensuremath{84.2}} 
\newcommand{\TBzero}{\ensuremath{2.0}} 
\newcommand{\TRfirst}{\ensuremath{84.7}} 
\newcommand{\TRzero}{\ensuremath{2.4}} 

\newcommand{\TRRfirst}{\ensuremath{84.3}} 
\newcommand{\TRRzero}{\ensuremath{2.3}} 

%
%
   \title{\Euclid. III. The NISP Instrument\thanks{Dedicated to our friend and colleague Favio Bortoletto (1951--2019) and his central contributions to NISP.}}


\author{Euclid Collaboration: K.~Jahnke\orcid{0000-0003-3804-2137}\thanks{\email{jahnke@mpia.de}}\inst{\ref{aff1}}
\and W.~Gillard\orcid{0000-0003-4744-9748}\inst{\ref{aff2}}
\and M.~Schirmer\orcid{0000-0003-2568-9994}\inst{\ref{aff1}}
\and A.~Ealet\orcid{0000-0003-3070-014X}\inst{\ref{aff3}}
\and T.~Maciaszek\inst{\ref{aff4}}
\and E.~Prieto\inst{\ref{aff5}}
\and R.~Barbier\inst{\ref{aff3}}
\and C.~Bonoli\inst{\ref{aff6}}
\and L.~Corcione\orcid{0000-0002-6497-5881}\inst{\ref{aff7}}
\and S.~Dusini\orcid{0000-0002-1128-0664}\inst{\ref{aff8}}
\and F.~Grupp\inst{\ref{aff9},\ref{aff10}}
\and F.~Hormuth\inst{\ref{aff11},\ref{aff1}}
\and S.~Ligori\orcid{0000-0003-4172-4606}\inst{\ref{aff7}}
\and L.~Martin\inst{\ref{aff5}}
\and G.~Morgante\inst{\ref{aff12}}
\and C.~Padilla\orcid{0000-0001-7951-0166}\inst{\ref{aff13}}
\and R.~Toledo-Moreo\orcid{0000-0002-2997-4859}\inst{\ref{aff14}}
\and M.~Trifoglio\orcid{0000-0002-2505-3630}\inst{\ref{aff12}}
\and L.~Valenziano\orcid{0000-0002-1170-0104}\inst{\ref{aff12},\ref{aff15}}
\and R.~Bender\orcid{0000-0001-7179-0626}\inst{\ref{aff9},\ref{aff10}}
\and F.~J.~Castander\orcid{0000-0001-7316-4573}\inst{\ref{aff16},\ref{aff17}}
\and B.~Garilli\orcid{0000-0001-7455-8750}\inst{\ref{aff18}}
\and P.~B.~Lilje\orcid{0000-0003-4324-7794}\inst{\ref{aff19}}
\and H.-W.~Rix\orcid{0000-0003-4996-9069}\inst{\ref{aff1}}
\and N.~Auricchio\orcid{0000-0003-4444-8651}\inst{\ref{aff12}}
\and A.~Balestra\orcid{0000-0002-6967-261X}\inst{\ref{aff6}}
\and J.-C.~Barriere\inst{\ref{aff20},\ref{aff21}}
\and P.~Battaglia\orcid{0000-0002-7337-5909}\inst{\ref{aff12}}
\and M.~Berthe\inst{\ref{aff22}}
\and C.~Bodendorf\inst{\ref{aff9}}
\and T.~Boenke\inst{\ref{aff23}}
\and W.~Bon\inst{\ref{aff5}}
\and A.~Bonnefoi\inst{\ref{aff5}}
\and A.~Caillat\inst{\ref{aff5}}
\and V.~Capobianco\orcid{0000-0002-3309-7692}\inst{\ref{aff7}}
\and M.~Carle\inst{\ref{aff5}}
\and R.~Casas\orcid{0000-0002-8165-5601}\inst{\ref{aff17},\ref{aff16}}
\and H.~Cho\inst{\ref{aff24}}
\and A.~Costille\inst{\ref{aff5}}
\and F.~Ducret\inst{\ref{aff5}}
\and S.~Ferriol\inst{\ref{aff3}}
\and E.~Franceschi\orcid{0000-0002-0585-6591}\inst{\ref{aff12}}
\and J.-L.~Gimenez\inst{\ref{aff5}}
\and W.~Holmes\inst{\ref{aff24}}
\and A.~Hornstrup\orcid{0000-0002-3363-0936}\inst{\ref{aff25},\ref{aff26}}
\and M.~Jhabvala\inst{\ref{aff27}}
\and R.~Kohley\inst{\ref{aff28}}
\and B.~Kubik\orcid{0009-0006-5823-4880}\inst{\ref{aff3}}
\and R.~Laureijs\inst{\ref{aff23}}
\and D.~Le~Mignant\orcid{0000-0002-5339-5515}\inst{\ref{aff5}}
\and I.~Lloro\inst{\ref{aff29}}
\and E.~Medinaceli\orcid{0000-0002-4040-7783}\inst{\ref{aff12}}
\and Y.~Mellier\inst{\ref{aff30},\ref{aff31}}
\and G.~Polenta\orcid{0000-0003-4067-9196}\inst{\ref{aff32}}
\and G.~D.~Racca\inst{\ref{aff23}}
\and A.~Renzi\orcid{0000-0001-9856-1970}\inst{\ref{aff33},\ref{aff8}}
\and J.-C.~Salvignol\inst{\ref{aff23}}
\and A.~Secroun\orcid{0000-0003-0505-3710}\inst{\ref{aff2}}
\and G.~Seidel\orcid{0000-0003-2907-353X}\inst{\ref{aff1}}
\and M.~Seiffert\orcid{0000-0002-7536-9393}\inst{\ref{aff24}}
\and C.~Sirignano\orcid{0000-0002-0995-7146}\inst{\ref{aff33},\ref{aff8}}
\and G.~Sirri\orcid{0000-0003-2626-2853}\inst{\ref{aff34}}
\and P.~Strada\inst{\ref{aff23}}
\and G.~Smadja\inst{\ref{aff3}}
\and L.~Stanco\orcid{0000-0002-9706-5104}\inst{\ref{aff8}}
\and S.~Wachter\inst{\ref{aff35}}
\and S.~Anselmi\orcid{0000-0002-3579-9583}\inst{\ref{aff8},\ref{aff33},\ref{aff36}}
\and E.~Borsato\inst{\ref{aff33},\ref{aff8}}
\and L.~Caillat\inst{\ref{aff2}}
\and F.~Cogato\orcid{0000-0003-4632-6113}\inst{\ref{aff37},\ref{aff12}}
\and C.~Colodro-Conde\inst{\ref{aff38}}
\and P.-E.~Crouzet\inst{\ref{aff23}}
\and V.~Conforti\orcid{0000-0002-0007-3520}\inst{\ref{aff12}}
\and M.~D'Alessandro\inst{\ref{aff6}}
\and Y.~Copin\orcid{0000-0002-5317-7518}\inst{\ref{aff3}}
\and J.-C.~Cuillandre\orcid{0000-0002-3263-8645}\inst{\ref{aff22}}
\and J.~E.~Davies\orcid{0000-0002-5079-9098}\inst{\ref{aff1}}
\and S.~Davini\orcid{0000-0003-3269-1718}\inst{\ref{aff39}}
\and A.~Derosa\inst{\ref{aff12}}
\and J.~J.~Diaz\inst{\ref{aff40}}
\and S.~Di~Domizio\orcid{0000-0003-2863-5895}\inst{\ref{aff41},\ref{aff39}}
\and D.~Di~Ferdinando\inst{\ref{aff34}}
\and R.~Farinelli\inst{\ref{aff12}}
\and A.~G.~Ferrari\orcid{0009-0005-5266-4110}\inst{\ref{aff42},\ref{aff34}}
\and F.~Fornari\orcid{0000-0003-2979-6738}\inst{\ref{aff15}}
\and L.~Gabarra\orcid{0000-0002-8486-8856}\inst{\ref{aff43}}
\and C.~M.~Gutierrez\orcid{0000-0001-7854-783X}\inst{\ref{aff44}}
\and F.~Giacomini\orcid{0000-0002-3129-2814}\inst{\ref{aff34}}
\and P.~Lagier\inst{\ref{aff2}}
\and F.~Gianotti\orcid{0000-0003-4666-119X}\inst{\ref{aff12}}
\and O.~Krause\inst{\ref{aff1}}
\and F.~Madrid\orcid{0000-0002-5380-7970}\inst{\ref{aff16}}
\and F.~Laudisio\inst{\ref{aff8}}
\and J.~Macias-Perez\orcid{0000-0002-5385-2763}\inst{\ref{aff45}}
\and G.~Naletto\orcid{0000-0003-2007-3138}\inst{\ref{aff33}}
\and M.~Niclas\orcid{0009-0001-1882-5313}\inst{\ref{aff2}}
\and J.~Marpaud\inst{\ref{aff45}}
\and N.~Mauri\orcid{0000-0001-8196-1548}\inst{\ref{aff42},\ref{aff34}}
\and R.~da~Silva\orcid{0000-0003-4788-677X}\inst{\ref{aff46},\ref{aff32}}
\and F.~Passalacqua\inst{\ref{aff33},\ref{aff8}}
\and K.~Paterson\orcid{0000-0001-8340-3486}\inst{\ref{aff1}}
\and L.~Patrizii\inst{\ref{aff34}}
\and I.~Risso\orcid{0000-0003-2525-7761}\inst{\ref{aff47}}
\and B.~G.~B.~Solheim\orcid{0009-0008-2307-2978}\inst{\ref{aff48}}
\and M.~Scodeggio\inst{\ref{aff18}}
\and P.~Stassi\orcid{0000-0001-5584-8410}\inst{\ref{aff45}}
\and J.~Steinwagner\inst{\ref{aff9}}
\and M.~Tenti\orcid{0000-0002-4254-5901}\inst{\ref{aff34}}
\and G.~Testera\inst{\ref{aff39}}
\and R.~Travaglini\orcid{0000-0002-5288-1407}\inst{\ref{aff34}}
\and S.~Tosi\orcid{0000-0002-7275-9193}\inst{\ref{aff41},\ref{aff39}}
\and A.~Troja\orcid{0000-0003-0239-4595}\inst{\ref{aff33},\ref{aff8}}
\and O.~Tubio\inst{\ref{aff38}}
\and C.~Valieri\inst{\ref{aff34}}
\and C.~Vescovi\inst{\ref{aff45}}
\and S.~Ventura\inst{\ref{aff8}}
\and N.~Aghanim\orcid{0000-0002-6688-8992}\inst{\ref{aff49}}
\and B.~Altieri\orcid{0000-0003-3936-0284}\inst{\ref{aff28}}
\and A.~Amara\inst{\ref{aff50}}
\and J.~Amiaux\inst{\ref{aff22}}
\and S.~Andreon\orcid{0000-0002-2041-8784}\inst{\ref{aff51}}
\and H.~Aussel\orcid{0000-0002-1371-5705}\inst{\ref{aff22}}
\and M.~Baldi\orcid{0000-0003-4145-1943}\inst{\ref{aff52},\ref{aff12},\ref{aff34}}
\and S.~Bardelli\orcid{0000-0002-8900-0298}\inst{\ref{aff12}}
\and A.~Basset\inst{\ref{aff4}}
\and A.~Bonchi\orcid{0000-0002-2667-5482}\inst{\ref{aff32}}
\and D.~Bonino\orcid{0000-0002-3336-9977}\inst{\ref{aff7}}
\and E.~Branchini\orcid{0000-0002-0808-6908}\inst{\ref{aff41},\ref{aff39},\ref{aff51}}
\and M.~Brescia\orcid{0000-0001-9506-5680}\inst{\ref{aff53},\ref{aff54},\ref{aff55}}
\and J.~Brinchmann\orcid{0000-0003-4359-8797}\inst{\ref{aff56}}
\and S.~Camera\orcid{0000-0003-3399-3574}\inst{\ref{aff57},\ref{aff58},\ref{aff7}}
\and C.~Carbone\orcid{0000-0003-0125-3563}\inst{\ref{aff18}}
\and V.~F.~Cardone\inst{\ref{aff46},\ref{aff59}}
\and J.~Carretero\orcid{0000-0002-3130-0204}\inst{\ref{aff60},\ref{aff61}}
\and S.~Casas\orcid{0000-0002-4751-5138}\inst{\ref{aff62}}
\and M.~Castellano\orcid{0000-0001-9875-8263}\inst{\ref{aff46}}
\and S.~Cavuoti\orcid{0000-0002-3787-4196}\inst{\ref{aff54},\ref{aff55}}
\and P.-Y.~Chabaud\inst{\ref{aff5}}
\and A.~Cimatti\inst{\ref{aff42}}
\and G.~Congedo\orcid{0000-0003-2508-0046}\inst{\ref{aff63}}
\and C.~J.~Conselice\orcid{0000-0003-1949-7638}\inst{\ref{aff64}}
\and L.~Conversi\orcid{0000-0002-6710-8476}\inst{\ref{aff65},\ref{aff28}}
\and F.~Courbin\orcid{0000-0003-0758-6510}\inst{\ref{aff66}}
\and H.~M.~Courtois\orcid{0000-0003-0509-1776}\inst{\ref{aff67}}
\and M.~Cropper\orcid{0000-0003-4571-9468}\inst{\ref{aff68}}
\and J.-G.~Cuby\orcid{0000-0002-8767-1442}\inst{\ref{aff69},\ref{aff5}}
\and A.~Da~Silva\orcid{0000-0002-6385-1609}\inst{\ref{aff70},\ref{aff71}}
\and H.~Degaudenzi\orcid{0000-0002-5887-6799}\inst{\ref{aff72}}
\and A.~M.~Di~Giorgio\orcid{0000-0002-4767-2360}\inst{\ref{aff73}}
\and J.~Dinis\orcid{0000-0001-5075-1601}\inst{\ref{aff70},\ref{aff71}}
\and M.~Douspis\orcid{0000-0003-4203-3954}\inst{\ref{aff49}}
\and F.~Dubath\orcid{0000-0002-6533-2810}\inst{\ref{aff72}}
\and C.~A.~J.~Duncan\inst{\ref{aff64}}
\and X.~Dupac\inst{\ref{aff28}}
\and M.~Fabricius\orcid{0000-0002-7025-6058}\inst{\ref{aff9},\ref{aff10}}
\and M.~Farina\orcid{0000-0002-3089-7846}\inst{\ref{aff73}}
\and S.~Farrens\orcid{0000-0002-9594-9387}\inst{\ref{aff22}}
\and F.~Faustini\orcid{0000-0001-6274-5145}\inst{\ref{aff32},\ref{aff46}}
\and P.~Fosalba\orcid{0000-0002-1510-5214}\inst{\ref{aff17},\ref{aff74}}
\and S.~Fotopoulou\orcid{0000-0002-9686-254X}\inst{\ref{aff75}}
\and N.~Fourmanoit\inst{\ref{aff2}}
\and M.~Frailis\orcid{0000-0002-7400-2135}\inst{\ref{aff76}}
\and P.~Franzetti\inst{\ref{aff18}}
\and S.~Galeotta\orcid{0000-0002-3748-5115}\inst{\ref{aff76}}
\and B.~Gillis\orcid{0000-0002-4478-1270}\inst{\ref{aff63}}
\and C.~Giocoli\orcid{0000-0002-9590-7961}\inst{\ref{aff12},\ref{aff77}}
\and P.~G\'omez-Alvarez\orcid{0000-0002-8594-5358}\inst{\ref{aff78},\ref{aff28}}
\and B.~R.~Granett\orcid{0000-0003-2694-9284}\inst{\ref{aff51}}
\and A.~Grazian\orcid{0000-0002-5688-0663}\inst{\ref{aff6}}
\and L.~Guzzo\orcid{0000-0001-8264-5192}\inst{\ref{aff79},\ref{aff51}}
\and M.~Hailey\inst{\ref{aff68}}
\and S.~V.~H.~Haugan\orcid{0000-0001-9648-7260}\inst{\ref{aff19}}
\and J.~Hoar\inst{\ref{aff28}}
\and H.~Hoekstra\orcid{0000-0002-0641-3231}\inst{\ref{aff80}}
\and I.~Hook\orcid{0000-0002-2960-978X}\inst{\ref{aff81}}
\and P.~Hudelot\inst{\ref{aff31}}
\and B.~Joachimi\orcid{0000-0001-7494-1303}\inst{\ref{aff82}}
\and E.~Keih\"anen\orcid{0000-0003-1804-7715}\inst{\ref{aff83}}
\and S.~Kermiche\orcid{0000-0002-0302-5735}\inst{\ref{aff2}}
\and A.~Kiessling\orcid{0000-0002-2590-1273}\inst{\ref{aff24}}
\and M.~Kilbinger\orcid{0000-0001-9513-7138}\inst{\ref{aff22}}
\and T.~Kitching\orcid{0000-0002-4061-4598}\inst{\ref{aff68}}
\and M.~K\"ummel\orcid{0000-0003-2791-2117}\inst{\ref{aff10}}
\and M.~Kunz\orcid{0000-0002-3052-7394}\inst{\ref{aff84}}
\and H.~Kurki-Suonio\orcid{0000-0002-4618-3063}\inst{\ref{aff85},\ref{aff86}}
\and O.~Lahav\orcid{0000-0002-1134-9035}\inst{\ref{aff82}}
\and V.~Lindholm\orcid{0000-0003-2317-5471}\inst{\ref{aff85},\ref{aff86}}
\and J.~Lorenzo~Alvarez\orcid{0000-0002-6845-993X}\inst{\ref{aff23}}
\and D.~Maino\inst{\ref{aff79},\ref{aff18},\ref{aff87}}
\and E.~Maiorano\orcid{0000-0003-2593-4355}\inst{\ref{aff12}}
\and O.~Mansutti\orcid{0000-0001-5758-4658}\inst{\ref{aff76}}
\and O.~Marggraf\orcid{0000-0001-7242-3852}\inst{\ref{aff88}}
\and K.~Markovic\orcid{0000-0001-6764-073X}\inst{\ref{aff24}}
\and J.~Martignac\inst{\ref{aff22}}
\and N.~Martinet\orcid{0000-0003-2786-7790}\inst{\ref{aff5}}
\and F.~Marulli\orcid{0000-0002-8850-0303}\inst{\ref{aff37},\ref{aff12},\ref{aff34}}
\and R.~Massey\orcid{0000-0002-6085-3780}\inst{\ref{aff89},\ref{aff90}}
\and D.~C.~Masters\orcid{0000-0001-5382-6138}\inst{\ref{aff91}}
\and S.~Maurogordato\inst{\ref{aff92}}
\and H.~J.~McCracken\orcid{0000-0002-9489-7765}\inst{\ref{aff31}}
\and S.~Mei\orcid{0000-0002-2849-559X}\inst{\ref{aff93}}
\and M.~Melchior\inst{\ref{aff94}}
\and M.~Meneghetti\orcid{0000-0003-1225-7084}\inst{\ref{aff12},\ref{aff34}}
\and E.~Merlin\orcid{0000-0001-6870-8900}\inst{\ref{aff46}}
\and G.~Meylan\inst{\ref{aff66}}
\and J.~J.~Mohr\orcid{0000-0002-6875-2087}\inst{\ref{aff10},\ref{aff9}}
\and M.~Moresco\orcid{0000-0002-7616-7136}\inst{\ref{aff37},\ref{aff12}}
\and L.~Moscardini\orcid{0000-0002-3473-6716}\inst{\ref{aff37},\ref{aff12},\ref{aff34}}
\and R.~Nakajima\inst{\ref{aff88}}
\and R.~C.~Nichol\orcid{0000-0003-0939-6518}\inst{\ref{aff50}}
\and S.-M.~Niemi\inst{\ref{aff23}}
\and T.~Nutma\inst{\ref{aff95},\ref{aff80}}
\and K.~Paech\orcid{0000-0003-0625-2367}\inst{\ref{aff9}}
\and S.~Paltani\orcid{0000-0002-8108-9179}\inst{\ref{aff72}}
\and F.~Pasian\orcid{0000-0002-4869-3227}\inst{\ref{aff76}}
\and J.~A.~Peacock\orcid{0000-0002-1168-8299}\inst{\ref{aff63}}
\and K.~Pedersen\inst{\ref{aff96}}
\and W.~J.~Percival\orcid{0000-0002-0644-5727}\inst{\ref{aff97},\ref{aff98},\ref{aff99}}
\and V.~Pettorino\inst{\ref{aff23}}
\and S.~Pires\orcid{0000-0002-0249-2104}\inst{\ref{aff22}}
\and M.~Poncet\inst{\ref{aff4}}
\and L.~A.~Popa\inst{\ref{aff100}}
\and L.~Pozzetti\orcid{0000-0001-7085-0412}\inst{\ref{aff12}}
\and F.~Raison\orcid{0000-0002-7819-6918}\inst{\ref{aff9}}
\and R.~Rebolo\inst{\ref{aff38},\ref{aff101}}
\and A.~Refregier\inst{\ref{aff102}}
\and J.~Rhodes\orcid{0000-0002-4485-8549}\inst{\ref{aff24}}
\and G.~Riccio\inst{\ref{aff54}}
\and E.~Romelli\orcid{0000-0003-3069-9222}\inst{\ref{aff76}}
\and M.~Roncarelli\orcid{0000-0001-9587-7822}\inst{\ref{aff12}}
\and C.~Rosset\orcid{0000-0003-0286-2192}\inst{\ref{aff93}}
\and E.~Rossetti\orcid{0000-0003-0238-4047}\inst{\ref{aff52}}
\and H.~J.~A.~Rottgering\orcid{0000-0001-8887-2257}\inst{\ref{aff80}}
\and R.~Saglia\orcid{0000-0003-0378-7032}\inst{\ref{aff10},\ref{aff9}}
\and D.~Sapone\orcid{0000-0001-7089-4503}\inst{\ref{aff103}}
\and M.~Sauvage\orcid{0000-0002-0809-2574}\inst{\ref{aff22}}
\and R.~Scaramella\orcid{0000-0003-2229-193X}\inst{\ref{aff46},\ref{aff59}}
\and P.~Schneider\orcid{0000-0001-8561-2679}\inst{\ref{aff88}}
\and T.~Schrabback\orcid{0000-0002-6987-7834}\inst{\ref{aff104}}
\and S.~Serrano\orcid{0000-0002-0211-2861}\inst{\ref{aff17},\ref{aff105},\ref{aff16}}
\and P.~Tallada-Cresp\'{i}\orcid{0000-0002-1336-8328}\inst{\ref{aff60},\ref{aff61}}
\and D.~Tavagnacco\orcid{0000-0001-7475-9894}\inst{\ref{aff76}}
\and A.~N.~Taylor\inst{\ref{aff63}}
\and H.~I.~Teplitz\orcid{0000-0002-7064-5424}\inst{\ref{aff91}}
\and I.~Tereno\inst{\ref{aff70},\ref{aff106}}
\and F.~Torradeflot\orcid{0000-0003-1160-1517}\inst{\ref{aff61},\ref{aff60}}
\and I.~Tutusaus\orcid{0000-0002-3199-0399}\inst{\ref{aff107}}
\and T.~Vassallo\orcid{0000-0001-6512-6358}\inst{\ref{aff10},\ref{aff76}}
\and G.~Verdoes~Kleijn\orcid{0000-0001-5803-2580}\inst{\ref{aff95}}
\and A.~Veropalumbo\orcid{0000-0003-2387-1194}\inst{\ref{aff51},\ref{aff39},\ref{aff47}}
\and D.~Vibert\orcid{0009-0008-0607-631X}\inst{\ref{aff5}}
\and Y.~Wang\orcid{0000-0002-4749-2984}\inst{\ref{aff91}}
\and J.~Weller\orcid{0000-0002-8282-2010}\inst{\ref{aff10},\ref{aff9}}
\and A.~Zacchei\orcid{0000-0003-0396-1192}\inst{\ref{aff76},\ref{aff108}}
\and G.~Zamorani\orcid{0000-0002-2318-301X}\inst{\ref{aff12}}
\and F.~M.~Zerbi\inst{\ref{aff51}}
\and J.~Zoubian\inst{\ref{aff2}}
\and E.~Zucca\orcid{0000-0002-5845-8132}\inst{\ref{aff12}}
\and P.~N.~Appleton\orcid{0000-0002-7607-8766}\inst{\ref{aff109},\ref{aff91}}
\and C.~Baccigalupi\orcid{0000-0002-8211-1630}\inst{\ref{aff108},\ref{aff76},\ref{aff110},\ref{aff111}}
\and A.~Biviano\orcid{0000-0002-0857-0732}\inst{\ref{aff76},\ref{aff108}}
\and M.~Bolzonella\orcid{0000-0003-3278-4607}\inst{\ref{aff12}}
\and A.~Boucaud\orcid{0000-0001-7387-2633}\inst{\ref{aff93}}
\and E.~Bozzo\orcid{0000-0002-8201-1525}\inst{\ref{aff72}}
\and C.~Burigana\orcid{0000-0002-3005-5796}\inst{\ref{aff112},\ref{aff15}}
\and M.~Calabrese\orcid{0000-0002-2637-2422}\inst{\ref{aff113},\ref{aff18}}
\and P.~Casenove\inst{\ref{aff4}}
\and M.~Crocce\orcid{0000-0002-9745-6228}\inst{\ref{aff16},\ref{aff74}}
\and G.~De~Lucia\orcid{0000-0002-6220-9104}\inst{\ref{aff76}}
\and J.~A.~Escartin~Vigo\inst{\ref{aff9}}
\and G.~Fabbian\orcid{0000-0002-3255-4695}\inst{\ref{aff114},\ref{aff115}}
\and F.~Finelli\orcid{0000-0002-6694-3269}\inst{\ref{aff12},\ref{aff15}}
\and K.~George\orcid{0000-0002-1734-8455}\inst{\ref{aff10}}
\and J.~Gracia-Carpio\inst{\ref{aff9}}
\and S.~Ili\'c\orcid{0000-0003-4285-9086}\inst{\ref{aff116},\ref{aff107}}
\and P.~Liebing\inst{\ref{aff68}}
\and C.~Liu\inst{\ref{aff117}}
\and G.~Mainetti\orcid{0000-0003-2384-2377}\inst{\ref{aff118}}
\and S.~Marcin\inst{\ref{aff94}}
\and M.~Martinelli\orcid{0000-0002-6943-7732}\inst{\ref{aff46},\ref{aff59}}
\and P.~W.~Morris\orcid{0000-0002-5186-4381}\inst{\ref{aff117}}
\and C.~Neissner\orcid{0000-0001-8524-4968}\inst{\ref{aff13},\ref{aff61}}
\and A.~Pezzotta\orcid{0000-0003-0726-2268}\inst{\ref{aff9}}
\and M.~P\"ontinen\orcid{0000-0001-5442-2530}\inst{\ref{aff85}}
\and C.~Porciani\orcid{0000-0002-7797-2508}\inst{\ref{aff88}}
\and Z.~Sakr\orcid{0000-0002-4823-3757}\inst{\ref{aff119},\ref{aff107},\ref{aff120}}
\and V.~Scottez\inst{\ref{aff30},\ref{aff121}}
\and E.~Sefusatti\orcid{0000-0003-0473-1567}\inst{\ref{aff76},\ref{aff108},\ref{aff110}}
\and M.~Viel\orcid{0000-0002-2642-5707}\inst{\ref{aff108},\ref{aff76},\ref{aff111},\ref{aff110},\ref{aff122}}
\and M.~Wiesmann\orcid{0009-0000-8199-5860}\inst{\ref{aff19}}
\and Y.~Akrami\orcid{0000-0002-2407-7956}\inst{\ref{aff123},\ref{aff124}}
\and V.~Allevato\orcid{0000-0001-7232-5152}\inst{\ref{aff54}}
\and E.~Aubourg\orcid{0000-0002-5592-023X}\inst{\ref{aff93},\ref{aff20}}
\and M.~Ballardini\orcid{0000-0003-4481-3559}\inst{\ref{aff125},\ref{aff12},\ref{aff126}}
\and D.~Bertacca\orcid{0000-0002-2490-7139}\inst{\ref{aff33},\ref{aff6},\ref{aff8}}
\and M.~Bethermin\orcid{0000-0002-3915-2015}\inst{\ref{aff127},\ref{aff5}}
\and A.~Blanchard\orcid{0000-0001-8555-9003}\inst{\ref{aff107}}
\and L.~Blot\orcid{0000-0002-9622-7167}\inst{\ref{aff128},\ref{aff36}}
\and S.~Borgani\orcid{0000-0001-6151-6439}\inst{\ref{aff129},\ref{aff108},\ref{aff76},\ref{aff110}}
\and A.~S.~Borlaff\orcid{0000-0003-3249-4431}\inst{\ref{aff130},\ref{aff131}}
\and S.~Bruton\orcid{0000-0002-6503-5218}\inst{\ref{aff132}}
\and R.~Cabanac\orcid{0000-0001-6679-2600}\inst{\ref{aff107}}
\and A.~Calabro\orcid{0000-0003-2536-1614}\inst{\ref{aff46}}
\and G.~Calderone\orcid{0000-0002-7738-5389}\inst{\ref{aff76}}
\and G.~Canas-Herrera\orcid{0000-0003-2796-2149}\inst{\ref{aff23},\ref{aff133}}
\and A.~Cappi\inst{\ref{aff12},\ref{aff92}}
\and C.~S.~Carvalho\inst{\ref{aff106}}
\and G.~Castignani\orcid{0000-0001-6831-0687}\inst{\ref{aff12}}
\and T.~Castro\orcid{0000-0002-6292-3228}\inst{\ref{aff76},\ref{aff110},\ref{aff108},\ref{aff122}}
\and K.~C.~Chambers\orcid{0000-0001-6965-7789}\inst{\ref{aff134}}
\and Y.~Charles\inst{\ref{aff5}}
\and R.~Chary\orcid{0000-0001-7583-0621}\inst{\ref{aff91}}
\and J.~Colbert\orcid{0000-0001-6482-3020}\inst{\ref{aff91}}
\and S.~Contarini\orcid{0000-0002-9843-723X}\inst{\ref{aff9}}
\and T.~Contini\orcid{0000-0003-0275-938X}\inst{\ref{aff107}}
\and A.~R.~Cooray\orcid{0000-0002-3892-0190}\inst{\ref{aff135}}
\and M.~Costanzi\orcid{0000-0001-8158-1449}\inst{\ref{aff129},\ref{aff76},\ref{aff108}}
\and O.~Cucciati\orcid{0000-0002-9336-7551}\inst{\ref{aff12}}
\and B.~De~Caro\inst{\ref{aff18}}
\and S.~de~la~Torre\inst{\ref{aff5}}
\and G.~Desprez\inst{\ref{aff136}}
\and A.~D\'iaz-S\'anchez\orcid{0000-0003-0748-4768}\inst{\ref{aff137}}
\and H.~Dole\orcid{0000-0002-9767-3839}\inst{\ref{aff49}}
\and S.~Escoffier\orcid{0000-0002-2847-7498}\inst{\ref{aff2}}
\and P.~G.~Ferreira\orcid{0000-0002-3021-2851}\inst{\ref{aff43}}
\and I.~Ferrero\orcid{0000-0002-1295-1132}\inst{\ref{aff19}}
\and A.~Finoguenov\orcid{0000-0002-4606-5403}\inst{\ref{aff85}}
\and A.~Fontana\orcid{0000-0003-3820-2823}\inst{\ref{aff46}}
\and K.~Ganga\orcid{0000-0001-8159-8208}\inst{\ref{aff93}}
\and J.~Garc\'ia-Bellido\orcid{0000-0002-9370-8360}\inst{\ref{aff123}}
\and V.~Gautard\inst{\ref{aff138}}
\and E.~Gaztanaga\orcid{0000-0001-9632-0815}\inst{\ref{aff16},\ref{aff17},\ref{aff139}}
\and G.~Gozaliasl\orcid{0000-0002-0236-919X}\inst{\ref{aff140},\ref{aff85}}
\and A.~Gregorio\orcid{0000-0003-4028-8785}\inst{\ref{aff129},\ref{aff76},\ref{aff110}}
\and A.~Hall\orcid{0000-0002-3139-8651}\inst{\ref{aff63}}
\and W.~G.~Hartley\inst{\ref{aff72}}
\and S.~Hemmati\orcid{0000-0003-2226-5395}\inst{\ref{aff109}}
\and H.~Hildebrandt\orcid{0000-0002-9814-3338}\inst{\ref{aff141}}
\and J.~Hjorth\orcid{0000-0002-4571-2306}\inst{\ref{aff142}}
\and S.~Hosseini\inst{\ref{aff107}}
\and M.~Huertas-Company\orcid{0000-0002-1416-8483}\inst{\ref{aff38},\ref{aff40},\ref{aff143},\ref{aff144}}
\and O.~Ilbert\orcid{0000-0002-7303-4397}\inst{\ref{aff5}}
\and J.~Jacobson\inst{\ref{aff109}}
\and S.~Joudaki\orcid{0000-0001-8820-673X}\inst{\ref{aff139}}
\and J.~J.~E.~Kajava\orcid{0000-0002-3010-8333}\inst{\ref{aff145},\ref{aff146}}
\and V.~Kansal\orcid{0000-0002-4008-6078}\inst{\ref{aff147},\ref{aff148}}
\and D.~Karagiannis\orcid{0000-0002-4927-0816}\inst{\ref{aff149},\ref{aff150}}
\and C.~C.~Kirkpatrick\inst{\ref{aff83}}
\and F.~Lacasa\orcid{0000-0002-7268-3440}\inst{\ref{aff151},\ref{aff49},\ref{aff84}}
\and V.~Le~Brun\orcid{0000-0002-5027-1939}\inst{\ref{aff5}}
\and J.~Le~Graet\orcid{0000-0001-6523-7971}\inst{\ref{aff2}}
\and L.~Legrand\orcid{0000-0003-0610-5252}\inst{\ref{aff152}}
\and G.~Libet\inst{\ref{aff4}}
\and S.~J.~Liu\orcid{0000-0001-7680-2139}\inst{\ref{aff73}}
\and A.~Loureiro\orcid{0000-0002-4371-0876}\inst{\ref{aff153},\ref{aff154}}
\and M.~Magliocchetti\orcid{0000-0001-9158-4838}\inst{\ref{aff73}}
\and C.~Mancini\orcid{0000-0002-4297-0561}\inst{\ref{aff18}}
\and F.~Mannucci\orcid{0000-0002-4803-2381}\inst{\ref{aff155}}
\and R.~Maoli\orcid{0000-0002-6065-3025}\inst{\ref{aff156},\ref{aff46}}
\and C.~J.~A.~P.~Martins\orcid{0000-0002-4886-9261}\inst{\ref{aff157},\ref{aff56}}
\and S.~Matthew\orcid{0000-0001-8448-1697}\inst{\ref{aff63}}
\and L.~Maurin\orcid{0000-0002-8406-0857}\inst{\ref{aff49}}
\and C.~J.~R.~McPartland\orcid{0000-0003-0639-025X}\inst{\ref{aff26},\ref{aff158}}
\and R.~B.~Metcalf\orcid{0000-0003-3167-2574}\inst{\ref{aff37},\ref{aff12}}
\and M.~Migliaccio\inst{\ref{aff159},\ref{aff160}}
\and M.~Miluzio\inst{\ref{aff28}}
\and P.~Monaco\orcid{0000-0003-2083-7564}\inst{\ref{aff129},\ref{aff76},\ref{aff110},\ref{aff108}}
\and C.~Moretti\orcid{0000-0003-3314-8936}\inst{\ref{aff111},\ref{aff122},\ref{aff76},\ref{aff108},\ref{aff110}}
\and S.~Nadathur\orcid{0000-0001-9070-3102}\inst{\ref{aff139}}
\and L.~Nicastro\orcid{0000-0001-8534-6788}\inst{\ref{aff12}}
\and Nicholas~A.~Walton\orcid{0000-0003-3983-8778}\inst{\ref{aff161}}
\and J.~Odier\orcid{0000-0002-1650-2246}\inst{\ref{aff45}}
\and M.~Oguri\orcid{0000-0003-3484-399X}\inst{\ref{aff162},\ref{aff163}}
\and V.~Popa\inst{\ref{aff100}}
\and D.~Potter\orcid{0000-0002-0757-5195}\inst{\ref{aff164}}
\and A.~Pourtsidou\orcid{0000-0001-9110-5550}\inst{\ref{aff63},\ref{aff165}}
\and P.-F.~Rocci\inst{\ref{aff49}}
\and R.~P.~Rollins\orcid{0000-0003-1291-1023}\inst{\ref{aff63}}
\and B.~Rusholme\orcid{0000-0001-7648-4142}\inst{\ref{aff109}}
\and M.~Sahl\'en\orcid{0000-0003-0973-4804}\inst{\ref{aff166}}
\and A.~G.~S\'anchez\orcid{0000-0003-1198-831X}\inst{\ref{aff9}}
\and C.~Scarlata\orcid{0000-0002-9136-8876}\inst{\ref{aff132}}
\and J.~Schaye\orcid{0000-0002-0668-5560}\inst{\ref{aff80}}
\and J.~A.~Schewtschenko\inst{\ref{aff63}}
\and A.~Schneider\orcid{0000-0001-7055-8104}\inst{\ref{aff164}}
\and M.~Schultheis\inst{\ref{aff92}}
\and M.~Sereno\orcid{0000-0003-0302-0325}\inst{\ref{aff12},\ref{aff34}}
\and F.~Shankar\orcid{0000-0001-8973-5051}\inst{\ref{aff167}}
\and A.~Shulevski\orcid{0000-0002-1827-0469}\inst{\ref{aff168},\ref{aff95},\ref{aff169}}
\and G.~Sikkema\inst{\ref{aff95}}
\and A.~Silvestri\orcid{0000-0001-6904-5061}\inst{\ref{aff133}}
\and P.~Simon\inst{\ref{aff88}}
\and A.~Spurio~Mancini\orcid{0000-0001-5698-0990}\inst{\ref{aff170},\ref{aff68}}
\and J.~Stadel\orcid{0000-0001-7565-8622}\inst{\ref{aff164}}
\and S.~A.~Stanford\orcid{0000-0003-0122-0841}\inst{\ref{aff171}}
\and K.~Tanidis\inst{\ref{aff43}}
\and C.~Tao\orcid{0000-0001-7961-8177}\inst{\ref{aff2}}
\and N.~Tessore\orcid{0000-0002-9696-7931}\inst{\ref{aff82}}
\and R.~Teyssier\orcid{0000-0001-7689-0933}\inst{\ref{aff172}}
\and S.~Toft\orcid{0000-0003-3631-7176}\inst{\ref{aff26},\ref{aff173},\ref{aff158}}
\and M.~Tucci\inst{\ref{aff72}}
\and J.~Valiviita\orcid{0000-0001-6225-3693}\inst{\ref{aff85},\ref{aff86}}
\and D.~Vergani\orcid{0000-0003-0898-2216}\inst{\ref{aff12}}
\and F.~Vernizzi\orcid{0000-0003-3426-2802}\inst{\ref{aff174}}
\and G.~Verza\orcid{0000-0002-1886-8348}\inst{\ref{aff175},\ref{aff115}}
\and P.~Vielzeuf\orcid{0000-0003-2035-9339}\inst{\ref{aff2}}
\and J.~R.~Weaver\orcid{0000-0003-1614-196X}\inst{\ref{aff176}}
\and L.~Zalesky\orcid{0000-0001-5680-2326}\inst{\ref{aff134}}
\and I.~A.~Zinchenko\inst{\ref{aff10}}
\and M.~Archidiacono\orcid{0000-0003-4952-9012}\inst{\ref{aff79},\ref{aff87}}
\and F.~Atrio-Barandela\orcid{0000-0002-2130-2513}\inst{\ref{aff177}}
\and C.~L.~Bennett\orcid{0000-0001-8839-7206}\inst{\ref{aff178}}
\and T.~Bouvard\orcid{0009-0002-7959-312X}\inst{\ref{aff179}}
\and F.~Caro\inst{\ref{aff46}}
\and S.~Conseil\orcid{0000-0002-3657-4191}\inst{\ref{aff3}}
\and P.~Dimauro\orcid{0000-0001-7399-2854}\inst{\ref{aff46},\ref{aff180}}
\and P.-A.~Duc\orcid{0000-0003-3343-6284}\inst{\ref{aff127}}
\and Y.~Fang\inst{\ref{aff10}}
\and A.~M.~N.~Ferguson\inst{\ref{aff63}}
\and T.~Gasparetto\orcid{0000-0002-7913-4866}\inst{\ref{aff76}}
\and I.~Kova{\v{c}}i{\'{c}}\orcid{0000-0001-6751-3263}\inst{\ref{aff181}}
\and S.~Kruk\orcid{0000-0001-8010-8879}\inst{\ref{aff28}}
\and A.~M.~C.~Le~Brun\orcid{0000-0002-0936-4594}\inst{\ref{aff36}}
\and T.~I.~Liaudat\orcid{0000-0002-9104-314X}\inst{\ref{aff20}}
\and A.~Montoro\orcid{0000-0003-4730-8590}\inst{\ref{aff16},\ref{aff17}}
\and A.~Mora\orcid{0000-0002-1922-8529}\inst{\ref{aff182}}
\and C.~Murray\inst{\ref{aff93}}
\and L.~Pagano\orcid{0000-0003-1820-5998}\inst{\ref{aff125},\ref{aff126}}
\and D.~Paoletti\orcid{0000-0003-4761-6147}\inst{\ref{aff12},\ref{aff15}}
\and M.~Radovich\orcid{0000-0002-3585-866X}\inst{\ref{aff6}}
\and E.~Sarpa\orcid{0000-0002-1256-655X}\inst{\ref{aff111},\ref{aff122},\ref{aff110}}
\and E.~Tommasi\inst{\ref{aff183}}
\and A.~Viitanen\orcid{0000-0001-9383-786X}\inst{\ref{aff83},\ref{aff46}}
\and J.~Lesgourgues\orcid{0000-0001-7627-353X}\inst{\ref{aff62}}
\and M.~E.~Levi\orcid{0000-0003-1887-1018}\inst{\ref{aff184}}
\and J.~Mart\'{i}n-Fleitas\orcid{0000-0002-8594-569X}\inst{\ref{aff182}}}
										   
\institute{Max-Planck-Institut f\"ur Astronomie, K\"onigstuhl 17, 69117 Heidelberg, Germany\label{aff1}
\and
Aix-Marseille Universit\'e, CNRS/IN2P3, CPPM, Marseille, France\label{aff2}
\and
Universit\'e Claude Bernard Lyon 1, CNRS/IN2P3, IP2I Lyon, UMR 5822, Villeurbanne, F-69100, France\label{aff3}
\and
Centre National d'Etudes Spatiales -- Centre spatial de Toulouse, 18 avenue Edouard Belin, 31401 Toulouse Cedex 9, France\label{aff4}
\and
Aix-Marseille Universit\'e, CNRS, CNES, LAM, Marseille, France\label{aff5}
\and
INAF-Osservatorio Astronomico di Padova, Via dell'Osservatorio 5, 35122 Padova, Italy\label{aff6}
\and
INAF-Osservatorio Astrofisico di Torino, Via Osservatorio 20, 10025 Pino Torinese (TO), Italy\label{aff7}
\and
INFN-Padova, Via Marzolo 8, 35131 Padova, Italy\label{aff8}
\and
Max Planck Institute for Extraterrestrial Physics, Giessenbachstr. 1, 85748 Garching, Germany\label{aff9}
\and
Universit\"ats-Sternwarte M\"unchen, Fakult\"at f\"ur Physik, Ludwig-Maximilians-Universit\"at M\"unchen, Scheinerstrasse 1, 81679 M\"unchen, Germany\label{aff10}
\and
Felix Hormuth Engineering, Goethestr. 17, 69181 Leimen, Germany\label{aff11}
\and
INAF-Osservatorio di Astrofisica e Scienza dello Spazio di Bologna, Via Piero Gobetti 93/3, 40129 Bologna, Italy\label{aff12}
\and
Institut de F\'{i}sica d'Altes Energies (IFAE), The Barcelona Institute of Science and Technology, Campus UAB, 08193 Bellaterra (Barcelona), Spain\label{aff13}
\and
Universidad Polit\'ecnica de Cartagena, Departamento de Electr\'onica y Tecnolog\'ia de Computadoras,  Plaza del Hospital 1, 30202 Cartagena, Spain\label{aff14}
\and
INFN-Bologna, Via Irnerio 46, 40126 Bologna, Italy\label{aff15}
\and
Institute of Space Sciences (ICE, CSIC), Campus UAB, Carrer de Can Magrans, s/n, 08193 Barcelona, Spain\label{aff16}
\and
Institut d'Estudis Espacials de Catalunya (IEEC),  Edifici RDIT, Campus UPC, 08860 Castelldefels, Barcelona, Spain\label{aff17}
\and
INAF-IASF Milano, Via Alfonso Corti 12, 20133 Milano, Italy\label{aff18}
\and
Institute of Theoretical Astrophysics, University of Oslo, P.O. Box 1029 Blindern, 0315 Oslo, Norway\label{aff19}
\and
IRFU, CEA, Universit\'e Paris-Saclay 91191 Gif-sur-Yvette Cedex, France\label{aff20}
\and
CEA-Saclay, DRF/IRFU, departement d'ingenierie des systemes, bat472, 91191 Gif sur Yvette cedex, France\label{aff21}
\and
Universit\'e Paris-Saclay, Universit\'e Paris Cit\'e, CEA, CNRS, AIM, 91191, Gif-sur-Yvette, France\label{aff22}
\and
European Space Agency/ESTEC, Keplerlaan 1, 2201 AZ Noordwijk, The Netherlands\label{aff23}
\and
Jet Propulsion Laboratory, California Institute of Technology, 4800 Oak Grove Drive, Pasadena, CA, 91109, USA\label{aff24}
\and
Technical University of Denmark, Elektrovej 327, 2800 Kgs. Lyngby, Denmark\label{aff25}
\and
Cosmic Dawn Center (DAWN), Denmark\label{aff26}
\and
NASA Goddard Space Flight Center, Greenbelt, MD 20771, USA\label{aff27}
\and
ESAC/ESA, Camino Bajo del Castillo, s/n., Urb. Villafranca del Castillo, 28692 Villanueva de la Ca\~nada, Madrid, Spain\label{aff28}
\and
NOVA optical infrared instrumentation group at ASTRON, Oude Hoogeveensedijk 4, 7991PD, Dwingeloo, The Netherlands\label{aff29}
\and
Institut d'Astrophysique de Paris, 98bis Boulevard Arago, 75014, Paris, France\label{aff30}
\and
Institut d'Astrophysique de Paris, UMR 7095, CNRS, and Sorbonne Universit\'e, 98 bis boulevard Arago, 75014 Paris, France\label{aff31}
\and
Space Science Data Center, Italian Space Agency, via del Politecnico snc, 00133 Roma, Italy\label{aff32}
\and
Dipartimento di Fisica e Astronomia "G. Galilei", Universit\`a di Padova, Via Marzolo 8, 35131 Padova, Italy\label{aff33}
\and
INFN-Sezione di Bologna, Viale Berti Pichat 6/2, 40127 Bologna, Italy\label{aff34}
\and
Carnegie Observatories, Pasadena, CA 91101, USA\label{aff35}
\and
Laboratoire Univers et Th\'eorie, Observatoire de Paris, Universit\'e PSL, Universit\'e Paris Cit\'e, CNRS, 92190 Meudon, France\label{aff36}
\and
Dipartimento di Fisica e Astronomia "Augusto Righi" - Alma Mater Studiorum Universit\`a di Bologna, via Piero Gobetti 93/2, 40129 Bologna, Italy\label{aff37}
\and
Instituto de Astrof\'isica de Canarias, Calle V\'ia L\'actea s/n, 38204, San Crist\'obal de La Laguna, Tenerife, Spain\label{aff38}
\and
INFN-Sezione di Genova, Via Dodecaneso 33, 16146, Genova, Italy\label{aff39}
\and
Instituto de Astrof\'isica de Canarias (IAC); Departamento de Astrof\'isica, Universidad de La Laguna (ULL), 38200, La Laguna, Tenerife, Spain\label{aff40}
\and
Dipartimento di Fisica, Universit\`a di Genova, Via Dodecaneso 33, 16146, Genova, Italy\label{aff41}
\and
Dipartimento di Fisica e Astronomia "Augusto Righi" - Alma Mater Studiorum Universit\`a di Bologna, Viale Berti Pichat 6/2, 40127 Bologna, Italy\label{aff42}
\and
Department of Physics, Oxford University, Keble Road, Oxford OX1 3RH, UK\label{aff43}
\and
Instituto de Astrof\'\i sica de Canarias, c/ Via Lactea s/n, La Laguna E-38200, Spain. Departamento de Astrof\'\i sica de la Universidad de La Laguna, Avda. Francisco Sanchez, La Laguna, E-38200, Spain\label{aff44}
\and
Univ. Grenoble Alpes, CNRS, Grenoble INP, LPSC-IN2P3, 53, Avenue des Martyrs, 38000, Grenoble, France\label{aff45}
\and
INAF-Osservatorio Astronomico di Roma, Via Frascati 33, 00078 Monteporzio Catone, Italy\label{aff46}
\and
Dipartimento di Fisica, Universit\`a degli studi di Genova, and INFN-Sezione di Genova, via Dodecaneso 33, 16146, Genova, Italy\label{aff47}
\and
Clara Venture Labs AS, Fantoftvegen 38, 5072 Bergen, Norway\label{aff48}
\and
Universit\'e Paris-Saclay, CNRS, Institut d'astrophysique spatiale, 91405, Orsay, France\label{aff49}
\and
School of Mathematics and Physics, University of Surrey, Guildford, Surrey, GU2 7XH, UK\label{aff50}
\and
INAF-Osservatorio Astronomico di Brera, Via Brera 28, 20122 Milano, Italy\label{aff51}
\and
Dipartimento di Fisica e Astronomia, Universit\`a di Bologna, Via Gobetti 93/2, 40129 Bologna, Italy\label{aff52}
\and
Department of Physics "E. Pancini", University Federico II, Via Cinthia 6, 80126, Napoli, Italy\label{aff53}
\and
INAF-Osservatorio Astronomico di Capodimonte, Via Moiariello 16, 80131 Napoli, Italy\label{aff54}
\and
INFN section of Naples, Via Cinthia 6, 80126, Napoli, Italy\label{aff55}
\and
Instituto de Astrof\'isica e Ci\^encias do Espa\c{c}o, Universidade do Porto, CAUP, Rua das Estrelas, PT4150-762 Porto, Portugal\label{aff56}
\and
Dipartimento di Fisica, Universit\`a degli Studi di Torino, Via P. Giuria 1, 10125 Torino, Italy\label{aff57}
\and
INFN-Sezione di Torino, Via P. Giuria 1, 10125 Torino, Italy\label{aff58}
\and
INFN-Sezione di Roma, Piazzale Aldo Moro, 2 - c/o Dipartimento di Fisica, Edificio G. Marconi, 00185 Roma, Italy\label{aff59}
\and
Centro de Investigaciones Energ\'eticas, Medioambientales y Tecnol\'ogicas (CIEMAT), Avenida Complutense 40, 28040 Madrid, Spain\label{aff60}
\and
Port d'Informaci\'{o} Cient\'{i}fica, Campus UAB, C. Albareda s/n, 08193 Bellaterra (Barcelona), Spain\label{aff61}
\and
Institute for Theoretical Particle Physics and Cosmology (TTK), RWTH Aachen University, 52056 Aachen, Germany\label{aff62}
\and
Institute for Astronomy, University of Edinburgh, Royal Observatory, Blackford Hill, Edinburgh EH9 3HJ, UK\label{aff63}
\and
Jodrell Bank Centre for Astrophysics, Department of Physics and Astronomy, University of Manchester, Oxford Road, Manchester M13 9PL, UK\label{aff64}
\and
European Space Agency/ESRIN, Largo Galileo Galilei 1, 00044 Frascati, Roma, Italy\label{aff65}
\and
Institute of Physics, Laboratory of Astrophysics, Ecole Polytechnique F\'ed\'erale de Lausanne (EPFL), Observatoire de Sauverny, 1290 Versoix, Switzerland\label{aff66}
\and
UCB Lyon 1, CNRS/IN2P3, IUF, IP2I Lyon, 4 rue Enrico Fermi, 69622 Villeurbanne, France\label{aff67}
\and
Mullard Space Science Laboratory, University College London, Holmbury St Mary, Dorking, Surrey RH5 6NT, UK\label{aff68}
\and
Canada-France-Hawaii Telescope, 65-1238 Mamalahoa Hwy, Kamuela, HI 96743, USA\label{aff69}
\and
Departamento de F\'isica, Faculdade de Ci\^encias, Universidade de Lisboa, Edif\'icio C8, Campo Grande, PT1749-016 Lisboa, Portugal\label{aff70}
\and
Instituto de Astrof\'isica e Ci\^encias do Espa\c{c}o, Faculdade de Ci\^encias, Universidade de Lisboa, Campo Grande, 1749-016 Lisboa, Portugal\label{aff71}
\and
Department of Astronomy, University of Geneva, ch. d'Ecogia 16, 1290 Versoix, Switzerland\label{aff72}
\and
INAF-Istituto di Astrofisica e Planetologia Spaziali, via del Fosso del Cavaliere, 100, 00100 Roma, Italy\label{aff73}
\and
Institut de Ciencies de l'Espai (IEEC-CSIC), Campus UAB, Carrer de Can Magrans, s/n Cerdanyola del Vall\'es, 08193 Barcelona, Spain\label{aff74}
\and
School of Physics, HH Wills Physics Laboratory, University of Bristol, Tyndall Avenue, Bristol, BS8 1TL, UK\label{aff75}
\and
INAF-Osservatorio Astronomico di Trieste, Via G. B. Tiepolo 11, 34143 Trieste, Italy\label{aff76}
\and
Istituto Nazionale di Fisica Nucleare, Sezione di Bologna, Via Irnerio 46, 40126 Bologna, Italy\label{aff77}
\and
FRACTAL S.L.N.E., calle Tulip\'an 2, Portal 13 1A, 28231, Las Rozas de Madrid, Spain\label{aff78}
\and
Dipartimento di Fisica "Aldo Pontremoli", Universit\`a degli Studi di Milano, Via Celoria 16, 20133 Milano, Italy\label{aff79}
\and
Leiden Observatory, Leiden University, Einsteinweg 55, 2333 CC Leiden, The Netherlands\label{aff80}
\and
Department of Physics, Lancaster University, Lancaster, LA1 4YB, UK\label{aff81}
\and
Department of Physics and Astronomy, University College London, Gower Street, London WC1E 6BT, UK\label{aff82}
\and
Department of Physics and Helsinki Institute of Physics, Gustaf H\"allstr\"omin katu 2, 00014 University of Helsinki, Finland\label{aff83}
\and
Universit\'e de Gen\`eve, D\'epartement de Physique Th\'eorique and Centre for Astroparticle Physics, 24 quai Ernest-Ansermet, CH-1211 Gen\`eve 4, Switzerland\label{aff84}
\and
Department of Physics, P.O. Box 64, 00014 University of Helsinki, Finland\label{aff85}
\and
Helsinki Institute of Physics, Gustaf H{\"a}llstr{\"o}min katu 2, University of Helsinki, Helsinki, Finland\label{aff86}
\and
INFN-Sezione di Milano, Via Celoria 16, 20133 Milano, Italy\label{aff87}
\and
Universit\"at Bonn, Argelander-Institut f\"ur Astronomie, Auf dem H\"ugel 71, 53121 Bonn, Germany\label{aff88}
\and
Department of Physics, Centre for Extragalactic Astronomy, Durham University, South Road, DH1 3LE, UK\label{aff89}
\and
Department of Physics, Institute for Computational Cosmology, Durham University, South Road, DH1 3LE, UK\label{aff90}
\and
Infrared Processing and Analysis Center, California Institute of Technology, Pasadena, CA 91125, USA\label{aff91}
\and
Universit\'e C\^{o}te d'Azur, Observatoire de la C\^{o}te d'Azur, CNRS, Laboratoire Lagrange, Bd de l'Observatoire, CS 34229, 06304 Nice cedex 4, France\label{aff92}
\and
Universit\'e Paris Cit\'e, CNRS, Astroparticule et Cosmologie, 75013 Paris, France\label{aff93}
\and
University of Applied Sciences and Arts of Northwestern Switzerland, School of Engineering, 5210 Windisch, Switzerland\label{aff94}
\and
Kapteyn Astronomical Institute, University of Groningen, PO Box 800, 9700 AV Groningen, The Netherlands\label{aff95}
\and
Department of Physics and Astronomy, University of Aarhus, Ny Munkegade 120, DK-8000 Aarhus C, Denmark\label{aff96}
\and
Waterloo Centre for Astrophysics, University of Waterloo, Waterloo, Ontario N2L 3G1, Canada\label{aff97}
\and
Department of Physics and Astronomy, University of Waterloo, Waterloo, Ontario N2L 3G1, Canada\label{aff98}
\and
Perimeter Institute for Theoretical Physics, Waterloo, Ontario N2L 2Y5, Canada\label{aff99}
\and
Institute of Space Science, Str. Atomistilor, nr. 409 M\u{a}gurele, Ilfov, 077125, Romania\label{aff100}
\and
Departamento de Astrof\'isica, Universidad de La Laguna, 38206, La Laguna, Tenerife, Spain\label{aff101}
\and
Institute for Particle Physics and Astrophysics, Dept. of Physics, ETH Zurich, Wolfgang-Pauli-Strasse 27, 8093 Zurich, Switzerland\label{aff102}
\and
Departamento de F\'isica, FCFM, Universidad de Chile, Blanco Encalada 2008, Santiago, Chile\label{aff103}
\and
Universit\"at Innsbruck, Institut f\"ur Astro- und Teilchenphysik, Technikerstr. 25/8, 6020 Innsbruck, Austria\label{aff104}
\and
Satlantis, University Science Park, Sede Bld 48940, Leioa-Bilbao, Spain\label{aff105}
\and
Instituto de Astrof\'isica e Ci\^encias do Espa\c{c}o, Faculdade de Ci\^encias, Universidade de Lisboa, Tapada da Ajuda, 1349-018 Lisboa, Portugal\label{aff106}
\and
Institut de Recherche en Astrophysique et Plan\'etologie (IRAP), Universit\'e de Toulouse, CNRS, UPS, CNES, 14 Av. Edouard Belin, 31400 Toulouse, France\label{aff107}
\and
IFPU, Institute for Fundamental Physics of the Universe, via Beirut 2, 34151 Trieste, Italy\label{aff108}
\and
Caltech/IPAC, 1200 E. California Blvd., Pasadena, CA 91125, USA\label{aff109}
\and
INFN, Sezione di Trieste, Via Valerio 2, 34127 Trieste TS, Italy\label{aff110}
\and
SISSA, International School for Advanced Studies, Via Bonomea 265, 34136 Trieste TS, Italy\label{aff111}
\and
INAF, Istituto di Radioastronomia, Via Piero Gobetti 101, 40129 Bologna, Italy\label{aff112}
\and
Astronomical Observatory of the Autonomous Region of the Aosta Valley (OAVdA), Loc. Lignan 39, I-11020, Nus (Aosta Valley), Italy\label{aff113}
\and
School of Physics and Astronomy, Cardiff University, The Parade, Cardiff, CF24 3AA, UK\label{aff114}
\and
Center for Computational Astrophysics, Flatiron Institute, 162 5th Avenue, 10010, New York, NY, USA\label{aff115}
\and
Universit\'e Paris-Saclay, CNRS/IN2P3, IJCLab, 91405 Orsay, France\label{aff116}
\and
California institute of Technology, 1200 E California Blvd, Pasadena, CA 91125, USA\label{aff117}
\and
Centre de Calcul de l'IN2P3/CNRS, 21 avenue Pierre de Coubertin 69627 Villeurbanne Cedex, France\label{aff118}
\and
Institut f\"ur Theoretische Physik, University of Heidelberg, Philosophenweg 16, 69120 Heidelberg, Germany\label{aff119}
\and
Universit\'e St Joseph; Faculty of Sciences, Beirut, Lebanon\label{aff120}
\and
Junia, EPA department, 41 Bd Vauban, 59800 Lille, France\label{aff121}
\and
ICSC - Centro Nazionale di Ricerca in High Performance Computing, Big Data e Quantum Computing, Via Magnanelli 2, Bologna, Italy\label{aff122}
\and
Instituto de F\'isica Te\'orica UAM-CSIC, Campus de Cantoblanco, 28049 Madrid, Spain\label{aff123}
\and
CERCA/ISO, Department of Physics, Case Western Reserve University, 10900 Euclid Avenue, Cleveland, OH 44106, USA\label{aff124}
\and
Dipartimento di Fisica e Scienze della Terra, Universit\`a degli Studi di Ferrara, Via Giuseppe Saragat 1, 44122 Ferrara, Italy\label{aff125}
\and
Istituto Nazionale di Fisica Nucleare, Sezione di Ferrara, Via Giuseppe Saragat 1, 44122 Ferrara, Italy\label{aff126}
\and
Universit\'e de Strasbourg, CNRS, Observatoire astronomique de Strasbourg, UMR 7550, 67000 Strasbourg, France\label{aff127}
\and
Kavli Institute for the Physics and Mathematics of the Universe (WPI), University of Tokyo, Kashiwa, Chiba 277-8583, Japan\label{aff128}
\and
Dipartimento di Fisica - Sezione di Astronomia, Universit\`a di Trieste, Via Tiepolo 11, 34131 Trieste, Italy\label{aff129}
\and
NASA Ames Research Center, Moffett Field, CA 94035, USA\label{aff130}
\and
Bay Area Environmental Research Institute, Moffett Field, California 94035, USA\label{aff131}
\and
Minnesota Institute for Astrophysics, University of Minnesota, 116 Church St SE, Minneapolis, MN 55455, USA\label{aff132}
\and
Institute Lorentz, Leiden University, Niels Bohrweg 2, 2333 CA Leiden, The Netherlands\label{aff133}
\and
Institute for Astronomy, University of Hawaii, 2680 Woodlawn Drive, Honolulu, HI 96822, USA\label{aff134}
\and
Department of Physics \& Astronomy, University of California Irvine, Irvine CA 92697, USA\label{aff135}
\and
Department of Astronomy \& Physics and Institute for Computational Astrophysics, Saint Mary's University, 923 Robie Street, Halifax, Nova Scotia, B3H 3C3, Canada\label{aff136}
\and
Departamento F\'isica Aplicada, Universidad Polit\'ecnica de Cartagena, Campus Muralla del Mar, 30202 Cartagena, Murcia, Spain\label{aff137}
\and
CEA Saclay, DFR/IRFU, Service d'Astrophysique, Bat. 709, 91191 Gif-sur-Yvette, France\label{aff138}
\and
Institute of Cosmology and Gravitation, University of Portsmouth, Portsmouth PO1 3FX, UK\label{aff139}
\and
Department of Computer Science, Aalto University, PO Box 15400, Espoo, FI-00 076, Finland\label{aff140}
\and
Ruhr University Bochum, Faculty of Physics and Astronomy, Astronomical Institute (AIRUB), German Centre for Cosmological Lensing (GCCL), 44780 Bochum, Germany\label{aff141}
\and
DARK, Niels Bohr Institute, University of Copenhagen, Jagtvej 155, 2200 Copenhagen, Denmark\label{aff142}
\and
Universit\'e PSL, Observatoire de Paris, Sorbonne Universit\'e, CNRS, LERMA, 75014, Paris, France\label{aff143}
\and
Universit\'e Paris-Cit\'e, 5 Rue Thomas Mann, 75013, Paris, France\label{aff144}
\and
Department of Physics and Astronomy, Vesilinnantie 5, 20014 University of Turku, Finland\label{aff145}
\and
Serco for European Space Agency (ESA), Camino bajo del Castillo, s/n, Urbanizacion Villafranca del Castillo, Villanueva de la Ca\~nada, 28692 Madrid, Spain\label{aff146}
\and
ARC Centre of Excellence for Dark Matter Particle Physics, Melbourne, Australia\label{aff147}
\and
Centre for Astrophysics \& Supercomputing, Swinburne University of Technology, Victoria 3122, Australia\label{aff148}
\and
School of Physics and Astronomy, Queen Mary University of London, Mile End Road, London E1 4NS, UK\label{aff149}
\and
Department of Physics and Astronomy, University of the Western Cape, Bellville, Cape Town, 7535, South Africa\label{aff150}
\and
Universit\'e Libre de Bruxelles (ULB), Service de Physique Th\'eorique CP225, Boulevard du Triophe, 1050 Bruxelles, Belgium\label{aff151}
\and
ICTP South American Institute for Fundamental Research, Instituto de F\'{\i}sica Te\'orica, Universidade Estadual Paulista, S\~ao Paulo, Brazil\label{aff152}
\and
Oskar Klein Centre for Cosmoparticle Physics, Department of Physics, Stockholm University, Stockholm, SE-106 91, Sweden\label{aff153}
\and
Astrophysics Group, Blackett Laboratory, Imperial College London, London SW7 2AZ, UK\label{aff154}
\and
INAF-Osservatorio Astrofisico di Arcetri, Largo E. Fermi 5, 50125, Firenze, Italy\label{aff155}
\and
Dipartimento di Fisica, Sapienza Universit\`a di Roma, Piazzale Aldo Moro 2, 00185 Roma, Italy\label{aff156}
\and
Centro de Astrof\'{\i}sica da Universidade do Porto, Rua das Estrelas, 4150-762 Porto, Portugal\label{aff157}
\and
Niels Bohr Institute, University of Copenhagen, Jagtvej 128, 2200 Copenhagen, Denmark\label{aff158}
\and
Dipartimento di Fisica, Universit\`a di Roma Tor Vergata, Via della Ricerca Scientifica 1, Roma, Italy\label{aff159}
\and
INFN, Sezione di Roma 2, Via della Ricerca Scientifica 1, Roma, Italy\label{aff160}
\and
Institute of Astronomy, University of Cambridge, Madingley Road, Cambridge CB3 0HA, UK\label{aff161}
\and
Center for Frontier Science, Chiba University, 1-33 Yayoi-cho, Inage-ku, Chiba 263-8522, Japan\label{aff162}
\and
Department of Physics, Graduate School of Science, Chiba University, 1-33 Yayoi-Cho, Inage-Ku, Chiba 263-8522, Japan\label{aff163}
\and
Department of Astrophysics, University of Zurich, Winterthurerstrasse 190, 8057 Zurich, Switzerland\label{aff164}
\and
Higgs Centre for Theoretical Physics, School of Physics and Astronomy, The University of Edinburgh, Edinburgh EH9 3FD, UK\label{aff165}
\and
Theoretical astrophysics, Department of Physics and Astronomy, Uppsala University, Box 515, 751 20 Uppsala, Sweden\label{aff166}
\and
School of Physics \& Astronomy, University of Southampton, Highfield Campus, Southampton SO17 1BJ, UK\label{aff167}
\and
ASTRON, the Netherlands Institute for Radio Astronomy, Postbus 2, 7990 AA, Dwingeloo, The Netherlands\label{aff168}
\and
Anton Pannekoek Institute for Astronomy, University of Amsterdam, Postbus 94249, 1090 GE Amsterdam, The Netherlands\label{aff169}
\and
Department of Physics, Royal Holloway, University of London, TW20 0EX, UK\label{aff170}
\and
Department of Physics and Astronomy, University of California, Davis, CA 95616, USA\label{aff171}
\and
Department of Astrophysical Sciences, Peyton Hall, Princeton University, Princeton, NJ 08544, USA\label{aff172}
\and
Cosmic Dawn Center (DAWN)\label{aff173}
\and
Institut de Physique Th\'eorique, CEA, CNRS, Universit\'e Paris-Saclay 91191 Gif-sur-Yvette Cedex, France\label{aff174}
\and
Center for Cosmology and Particle Physics, Department of Physics, New York University, New York, NY 10003, USA\label{aff175}
\and
Department of Astronomy, University of Massachusetts, Amherst, MA 01003, USA\label{aff176}
\and
Departamento de F{\'\i}sica Fundamental. Universidad de Salamanca. Plaza de la Merced s/n. 37008 Salamanca, Spain\label{aff177}
\and
Johns Hopkins University 3400 North Charles Street Baltimore, MD 21218, USA\label{aff178}
\and
Thales~Services~S.A.S., 290 All\'ee du Lac, 31670 Lab\`ege, France\label{aff179}
\and
Observatorio Nacional, Rua General Jose Cristino, 77-Bairro Imperial de Sao Cristovao, Rio de Janeiro, 20921-400, Brazil\label{aff180}
\and
Sterrenkundig Observatorium, Universiteit Gent, Krijgslaan 281 S9, 9000 Gent, Belgium\label{aff181}
\and
Aurora Technology for European Space Agency (ESA), Camino bajo del Castillo, s/n, Urbanizacion Villafranca del Castillo, Villanueva de la Ca\~nada, 28692 Madrid, Spain\label{aff182}
\and
Italian Space Agency, via del Politecnico snc, 00133 Roma, Italy\label{aff183}
\and
Lawrence Berkeley National Laboratory, One Cyclotron Road, Berkeley, CA 94720, USA\label{aff184}}    


%
%
\abstract{
The Near-Infrared Spectrometer and Photometer (NISP) on board the \Euclid satellite provides multiband photometry and $R\gtrsim450$ slitless grism spectroscopy in the 950--2020\,nm wavelength range. 
In this reference article we illuminate the background of NISP's functional and calibration requirements, describe the instrument's integral components, and provide all its key properties. We also sketch the processes needed to understand how NISP operates and is calibrated, and its technical potentials and limitations. Links to articles providing more details and technical background are included.
%
NISP's 16 \ac{H2RG} detectors with a plate scale of \ang{;;0.3}\,pixel$^{-1}$ deliver a field-of-view of 0.57\,deg$^2$.
In photometric mode, NISP reaches a limiting magnitude of $\sim$\,24.5\,AB\,mag in three photometric exposures of about 100\,s exposure time, for point sources and with a \ac{SNR} of 5. For spectroscopy, NISP's point-source sensitivity is a ${\rm \ac{SNR}}=3.5$ detection of an emission line with flux $\sim$\,$2\times10^{-16}$\,erg\,s$^{-1}$\,cm$^{-2}$ integrated over two resolution elements of 13.4\,\AA, in 3$\times$ 560\,s grism exposures at 1.6\,\micron\ (redshifted H$\alpha$). Our calibration includes on-ground and in-flight characterisation and monitoring of pixel-based detector baseline, dark current, non-linearity, and sensitivity, to guarantee a relative photometric accuracy of better than 1.5\%, and relative spectrophotometry to better than 0.7\%. The wavelength calibration must be accurate to 5\,\AA\ or better.
%
NISP is the state-of-the-art instrument in the near-infrared for all science beyond small areas available from HST and JWST -- and an enormous advance from any existing instrumentation due to its combination of field size and high throughput of telescope and instrument. 
During \Euclid's 6-year survey covering 14\,000\,deg$^2$ of extragalactic sky, NISP will be the backbone for determining distances of more than a billion galaxies. Its near-infrared data will become a rich reference imaging and spectroscopy data set for the coming decades.
    }
%
%

%
%
   \titlerunning{The \Euclid NISP instrument }
   \authorrunning{Euclid Collaboration: K.\ Jahnke et al.}
   
   \maketitle


\section{Introduction}
\label{sc:Intro}

Physics currently has a good understanding of what astronomers call `baryons', meaning the fundamental composition of atoms and molecules that make up stars, planets, gas, and dust. However, observations of the dynamics of galaxies and galaxy clusters demonstrated the need for an extra component of mass, or in the accepted gravitational laws, in order to reconcile observed motions with inferred gravitational forces.

Such an extra `dark matter' mass component would very well fit into the structure formation theory of the Universe, but to this point no first principles predictions of the particle or field making up dark matter exist from the particle physics side.

At the same time, an extra `dark energy' is needed. The clear signal found of an accelerated expansion of the Universe at the current epoch and near zero curvature require an additional energy component that will influence the Universe's expansion history.

With together $\sim$\,95\% of the total mass-energy content of the Universe at present time, dark matter and dark energy pose one if not the largest open questions in physical sciences today -- at the heart of our understanding of gravitation and the composition and history of our Universe itself.

The \Euclid mission is designed to bring light into this dark sector of the Universe \citep{laureijs2011,euclid2024}, within the `Cosmic Vision 2015--2025' programme of the \ac{ESA}. Its primary cosmological probes of (i) weak lensing and (ii) baryon acoustic oscillations will for the first time be using data from a large-volume, space-based survey to measure the expansion history of the Universe, as well as characterise the details of structure formation to a level of accuracy not possible from the ground. \Euclid, its instruments and its data analysis were designed to differentiate between a cosmological constant and a time-variable dark energy, and at the same time probe the very nature of gravity \citep{amendola2018}.

\Euclid has two predecessor concepts within ESA's Cosmic Vision programme: SPACE \citep{cimatti2009} and DUNE \citep{refregier2006}.  
While they were proposed to carry out each one of these two main probes, \Euclid's payload was designed to cover both.

The overall design for \Euclid was solely driven by the so-called `dark energy \acl{FOM}' \citep[][]{albrecht2006} that represents a combined precision measure for dark energy properties. This very downstream number was subsequently broken down into required precision goals of the individual cosmological probes, from there to number statistics of galaxies, leading to requirements on survey volume, as well as on the abilities of the \Euclid telescope and instruments plus cosmology data analysis of the planned data. It was clear that a dramatic improvement over the existing or planned ground-based projects would come from (i) a very large survey volume, requiring a wide-field telescope and high sensitivity to cover a large area of the sky to a sufficient depth quickly, (ii) very high-fidelity, high precision measurement of galaxy shapes, and (iii) both near-infrared spectroscopy as well as multi-band photometry. This defined \Euclid's capabilities from the cosmology side.

At the same time, it was made clear from the beginning that \Euclid data will be usable for a very large variety of non-cosmology (`legacy') science projects, from solar system to objects in the early Universe. While none of these projects would be allowed to drive \Euclid's design, at all times data processing and data releases were being planned to enable as much broad astrophysical science as possible.

All these considerations led to the definition of two instruments onboard \Euclid: \acsu{NISP} and VIS. \ac{NISP} is the \acl{NISP}. On one side its near-infrared spectroscopy channel (NISP-S) will provide three-dimensional clustering information for $\sim$\,35 million galaxies to infer the growth of the Universe over cosmic time, using baryon acoustic oscillation measurements as a tracer of scales. 

On the other side its photometry channel (NISP-P) will provide photometry for more than a billion galaxies to derive photometric redshift estimates (photo-$z$s) when joined with ground-based data at shorter wavelengths. In combination with the high-resolution, high-fidelity, 530--920\,nm images of the \ac{VIS} -- the other instrument onboard \Euclid \citep{cropper2018,cropper2024} -- 
one can build three-dimensional weak lensing maps of matter across space and time.

\Euclid will test the large-scale structure and its constituents also beyond these two diagnostics and beyond the standard cosmological model, constraining alternatives to standard gravity as well as the neutrino mass hierarchy \citep{amendola2018}. Aside from cosmology, \Euclid's extensive and unprecedented survey data, be it high-resolution imaging over $\sim$\,14\,000\,deg$^2$ from VIS or slitless spectroscopy and multiband-imaging in the near-infrared from NISP, will enable a large variety of astrophysics studies from the Solar System to the earliest times after the Big Bang.
\medskip

In the following, first an overview is given of NISP and the requirements on which its design and functionality are based (Sect.~\ref{sc:overview}) -- then NISP's various components and partially design decisions are described to a level necessary for understanding how NISP functions and which capabilities NISP provides (Sect.~\ref{sec:components}). This is followed by a description of how NISP survey data will be calibrated (Sect.~\ref{sec:cal}), and an initial assessment of NISP's performance in flight (Sect.~\ref{sc:performance}). The paper closes with a description of NISP's operational options and limits (Sect.~\ref{sec:operations}), and an outlook (Sect.~\ref{sec:outlook}) into NISP's operations in the coming decade.


\section{\label{sc:overview}NISP overview and requirements}

The need to use both weak lensing and galaxy clustering over a large part of the sky via \Euclid as a single mission created a clear design outline for spacecraft, instruments, survey, and data analysis \citep{racca2016}. As a result the mission required for \Euclid a 1.20\,m main mirror with a flat and low-distortion field of view using a three-mirror anastigmat design.
Furthermore, three instrument channels were defined, distributed over two instruments: On one side the VIS instrument provides a wide-field,  high-fidelity imaging capability at visible wavelengths \citep{cropper2018,cropper2024}. 
VIS has one single wide passband between $\sim$530 and 920\,nm, with the main aim of imaging galaxies with a very stable and very well characterized point spread function, in order to extract ultra-precise galaxy shapes or weak lensing measurements.
The other two channels, near-infrared multi-passband photometry and slitless spectrometry, were combined into NISP -- which is the subject of this overview. 
For cosmology, NISP provides both spectroscopic redshifts directly, and for photometric redshifts contributes the near-infrared passbands to be combined with external, ground-based data in the spectral range $\sim$\,400--900\,nm.

These instrumental channels are meant to be used in the \acl{EWS} over $\sim$14\,000\,deg$^2$ -- and an associated 2\,mag deeper \acl{EDS} over $\sim$\,53\,deg$^2$ -- in the extra-galactic and extra-ecliptic area of the sky, to avoid both the dusty galactic plane and high zodiacal light background, and carried out over six years \citep{scaramella2022} at the thermally stable and low-background Sun-Earth Lagrange point L2.

Within this framework, the fundamental instrumental requirements that drove NISP's design and capabilities -- in conjunction with the capabilities of the telescope and subsequent data processing -- are as follows. For actual in-flight performance numbers please see Table~\ref{tab:zp}:

\bi
\item Wavelength range: near-infrared, $\sim$\,900--2000\,nm
\item \ac{FOV}: $\sim$\,0.5\,deg$^2$
\item Sampling: 0\farcs3\,pix$^{-1}$, with very compact, near diffraction-limited optics
\item Spectroscopy: 2 grism passbands, $\mathcal{R}\ge380$, with $3.5\sigma$ flux limit $\le$\,2$\times10^{-16}$\,erg\,cm$^{-2}$\,s$^{-1}$ for the $H_\alpha$ line at 1600\,nm from a \ang{;;0.5} diameter source
\item Photometry: 3 passbands, depth 24.0\,mag at 5$\sigma$ for point sources
\item Calibration: photometric calibration $\le$\,1.5\% over the whole survey, spectrophotometric fluctuation of the zero-point in flux limit $\le$0\,.7\% with wavelength calibration precision $\le$38\% of one resolution element
\item Structural thermal stability\footnote{In telescope design practices the material ratio between the thermal conductivity and the coefficient of thermal expansion measured in Wm$^{-1}$ is an important parameter and is called steady state stability. Materials with high steady state stability, like Silicon Carbide ($\sim$\,85\,W\micron$^{-1}$ at 300\,K) are very well suited to realising stable optical structures with easy thermal control. For NISP this permits to build a lightweight structure with very low thermal deformation which is able to conduct heat quickly.} better than 180\,W\,m$^{-1}$\,K$^{-1}$/ 2.2$\times$10$^{-6}$\,K$^{-1}$  
\item Thermal stability: detector temperature variation $\le$\,4\,mK over $\sim$\,1200\,s.

\ei

These fundamental requirements led to the NISP design that is now operating in space at L2. Together with other constraints such as limits on mass, volume, power consumption, thermal stability, and downlink data-rate, as well as boundary conditions to materials and technological readiness-levels NISP was designed, manufactured, and then tested, integrated into \Euclid and finally launched in mid 2023.

For \Euclid, the responsibility for the telescope, spacecraft including mission operations, as well as data distribution lies with \ac{ESA}, while design and manufacturing of the instruments including their onboard application software, as well as for the \ac{SGS}, responsible for data analysis, lies with consortia of diverse institutes, funded by their respective national agencies. The NISP instrument is led by the French national space agency (CNES), with the \ac{LAM} as the central institution for coordinating, integrating, and testing NISP. Major NISP subsystems, components, and contributions were provided through a number of institutions within a framework contract between the national agencies of France, Italy, Germany, the USA, Spain, Denmark, and Norway. 
Instrument operations, data processing, and scientific evaluation are taking place within the \Euclid Consortium,\footnote{\url{https://www.euclid-ec.org}} its \ac{SGS} and science working groups with more than 2000 scientists and engineers, jointly working with \ac{ESA} to reach the mission goals and provide \Euclid's data to the world.


The above instrumental requirements resulted in a NISP design with the following basic properties -- the resulting flight model instrument is shown in Fig.~\ref{fig:NISP_FM}:

\begin{figure*}[!thb]
    \centering
    \includegraphics[width=0.8\textwidth]{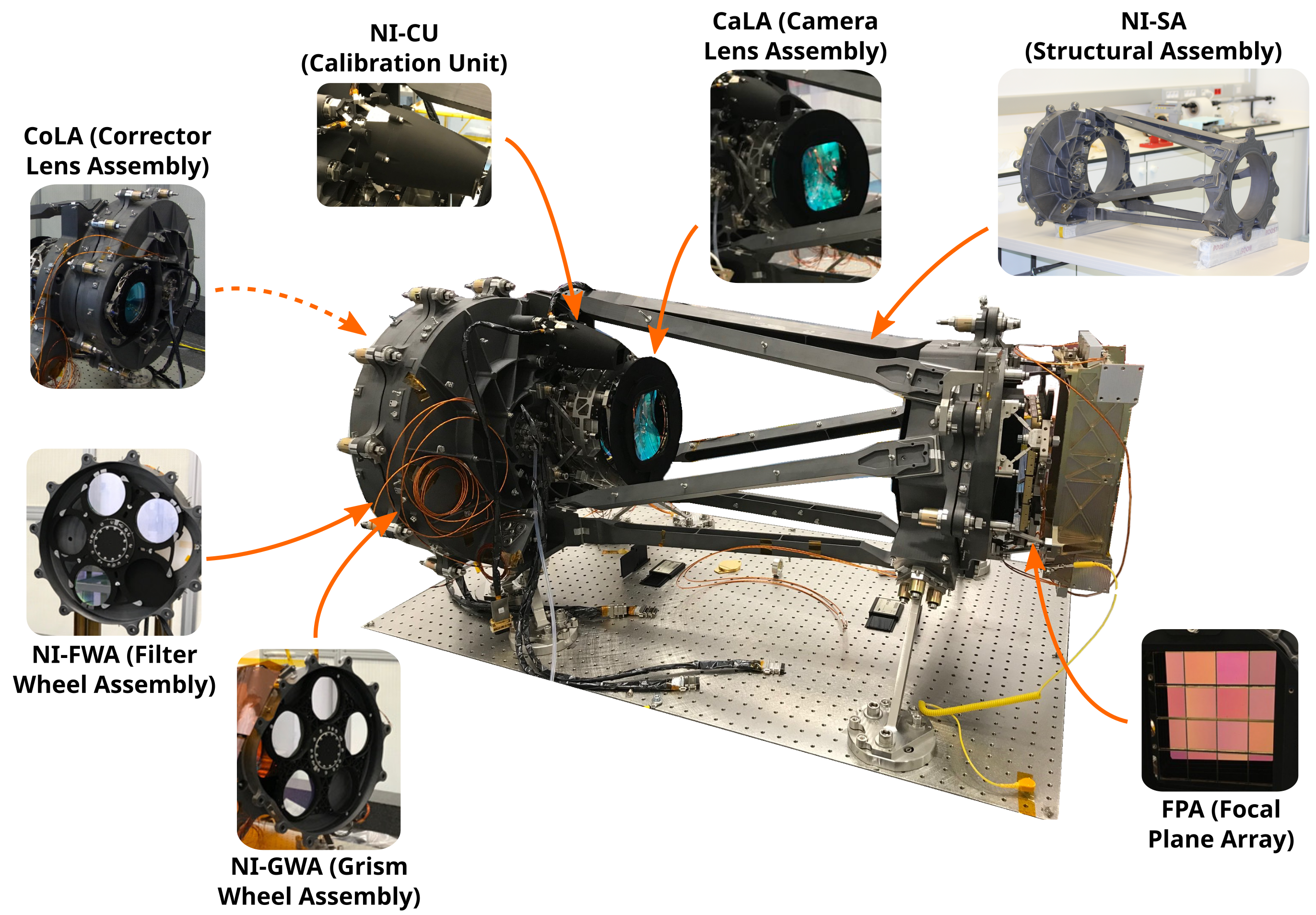}
    \caption{
The NISP flight model after completion but without its light-tight multi-layer insulation. As a scale reference: NISP fits into a volume of 100\,cm\,$\times$\,50\,cm\,$\times$\,50\,cm. Not pictured are the warm electronics (Sect.~\ref{sec:warmelectronics}) that are located in the \Euclid \ac{SVM}.
    }
    \label{fig:NISP_FM}
\end{figure*}

\begin{figure}[t]
\centering
\includegraphics[angle=0,width=1.0\hsize]{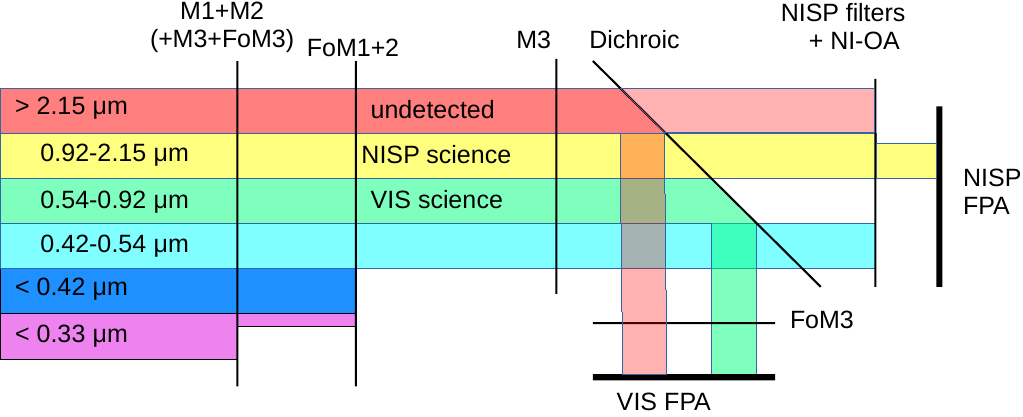}
\caption{The chromatic selection function of \Euclid's optical elements. M1--3 are the primary, secondary, and tertiary mirror, FoM1--3 are planar folding mirrors. The behaviour of the dichroic element above $2.2$\,\micron\ is not controlled by specifications; longer wavelengths could enter NISP and would be blocked by the filters.}
\label{fig:chromatic_selection}
\end{figure}

\begin{enumerate}

\item General layout: NISP has a F/10.4 camera, transforming the initial F/20.4 from the telescope to a field-size matching the available instrumental volume, and providing a \ac{FOV} of 0.57\,deg$^2$ with \ang{;;0.3}\,pix$^{-1}$ over a set of 4$\times$4 detectors. The optics are matched to the telescope's Korsch off-axis design \citep{korsch1977}. NISP has to function inside a rather limited volume, with a maximum mass of 155\,kg (actual mass 85.3\,kg for the main instrument, 40.7\,kg for the warm electronics) and maximum power consumption of 200\,W.

\item Shared \ac{FOV}: NISP receives light transmitted by a dichroic element outside of the instrument (Fig.~\ref{fig:chromatic_selection}); the reflected component is used by VIS. 
The \ac{FOV} is almost, but not perfectly concurrent \citep[see][]{euclid2024}. VIS and NISP can hence take data simultaneously, requiring a well coordinated sequence of observations and mechanism movements.

\item \ac{FPA}: NISP uses 4$\times$4 H2RGs with each 2040$\times$2040 science pixels as well as a 4-pixel wide border of light-insensitive reference pixels. H2RG detectors are operated at a temperature of about 95\,K to reduce thermal noise.   
NISP's plate scale is 0\farcs30 per pixel in both directions with a 18\,\micron\ pixel pitch. This makes the NISP \ac{PSF} undersampled for a 1.20\,m primary mirror. 
The \ac{FPA} is read out by 16 \acp{SCE} at 135\,K, each interfaced with \ac{DCU} electronics using a \ac{MACC} readout scheme, thereby transferring its data to the Data Processing Unit (DPU) -- warm electronics (293\,K) located outside of the payload cavity in the \Euclid \acf{SVM}.

\item Spectroscopy (NISP-S mode): The grism wheel assembly contains one blue grism (\BGE, 926--1366\,nm, $\mathcal{R}>400$), and three red grisms (\RGE, 1206--1892\,nm, $\mathcal{R}>480$) at different orientations of dispersion direction. 
A 5th slot is an open aperture to let the full beam pass for photometry observations.

\item Photometry (NISP-P mode): The filter wheel assembly contains three passband filters \YE, \JE, \HE, splitting the wavelength range $\sim$\,950--2020\,nm almost evenly in log-space of wavelength, one open slot to pass light for spectroscopic observations, and a closed position blocking any light towards the \ac{FPA}. 
The latter is used for any dark calibration images.

\item Detector calibration: The \ac{LED}-based calibration unit NI-CU provides light to repeatedly calibrate the detector's pixel-to-pixel sensitivity, as well as linearity. 
It can directly illuminate the \ac{FPA} area in each one of five wavelengths from $\sim$\,930 to 1880\,nm.

\item Control and processing electronics: NISP's warm electronics command all instrument functions, and process the received science data to reduce their volume consistent with downlink data-rate limits. It also adds housekeeping data to be used as diagnostics on ground with every science frame.

\item NISP has a data-rate limit for downlink to Earth of 290\,Gbit/day.

\end{enumerate}

The resulting design is described in Sect.~\ref{sec:components} below and as shown in Fig.~\ref{fig:NISP_FM}: a thermo-mechanically ultra-stable \ac{SiC} structure that allows mounting NISP to the payload module and that connects optics, filter and grism wheel, and a calibration source to the left in the image with the detector system and cold readout electronics on the right. In a separate location in the \ac{SVM}, NISP's warm electronics provides commanding and data processing capabilities.

In its design and manufacturing NISP has gone through a standard space hardware development cycle \citep{prieto2012,maciaszek2014,maciaszek2016,maciaszek2022}: Breadboard and engineering models were used on component level to assess materials and functionality of designed parts. A Structural and Thermal Model (delivered to ESA in 2017) was used to validate the mechanical design and thermal control. 

The NISP Engineering Qualification Model was used to qualify all subsystems individually, including structure and simplified optics. All electronics and their connections were functional in this model, and tests were conducted in thermal vacuum and thermal balance. This model will continue to be used during \Euclid operations as a testbed for NISP-internal software maintenance.

The Avionic Model has been designed to accurately represent the electric and electronic functionalities of NISP.  It contains all movable mechanisms, a representation of the calibration light source, as well as part of a detector array, but neither mechanical structure nor optics. This model has been used in the development of the \Euclid service module and in the testing of satellite operation procedures. The NISP Avionic Model has been delivered to ESA and will remain there to be used by mission operations for command testing.

Finally, the NISP Flight Model (delivered to ESA in 2020 as in Fig.~\ref{fig:NISP_FM}) was ultimately integrated into \Euclid and is now operating in orbit.

Sub-system integration, assembly and testing took place under the contributor's responsibility in collaboration with various industries. At instrument level assembly and integration of all models, as well as most of the testing activities, took place at LAM and its large cryo-vacuum laboratory.

\section{NISP components}
\label{sec:components}

NISP has a number of almost self-contained subsystems, which we describe in the following. A partial background on development decisions is included to motivate the designs that in the end were used in the flight model.

\subsection{\label{sc:structure}Mechanical structure}

NISP's main mechanical structure (central part in Fig.~\ref{fig:NISP_FM}) supports the optics, filter and grism wheel, calibration source, and detector system, with the goal to keep all these sub-systems aligned for the duration of the mission -- while providing extreme thermo-mechanical stability, high stiffness, and a thermally controlled environment, separated from the rest of the telescope.
The design and choice of material for this structure were directly driven by technical and performance specifications and boundary conditions:
the proximity of the telescope's optical beam folding mirror \#2 constrained the available volume, and the mass allocation budget of 37\,kg for the main mechanical structure, drove the structure design and choice of materials. 
The position of the collimator lens -- located at the entry of the NISP optical beam before the filter and grism wheels -- and the large distance between optical axis and the mounting interface to the payload module baseplate -- that is NISP's `legs' -- were challenges for the structural concept.

The driving criteria for the material selection for the main structure were (i) a high intrinsic material stiffness -- ratio of the elasticity modulus $E$ and the density $\rho$ -- in order to obtain a good stiffness-to-mass ratio. The goal was that the first eigenfrequency of the structure should be above 80\,Hz to survive vibrations during launch; and (ii) a good structural thermal stability, as explained above.  
The latter is central to achieving less than 0.3\,K variation in the optics over the 6 years of flight operation, enabling the continued alignment of the NISP's optical system with its focal plane. This is particularly challenging due to the different temperature levels present in NISP: the optics and mechanical structure are operated at a cryogenic temperature of $\sim$135\,K, the detector array at around 95\,K, its read-out electronics again at 135\,K. 
The choice of material and design were therefore also important to minimise the thermal gradients and simplify the associated thermal control required to achieve the targeted thermal stability of the mechanical structure -- and to avoid having to add a dedicated NISP focusing mechanism to retain co-focality with the VIS instrument. 

All these reasons led to the choice of sintered \ac{SiC} with a stiffness of 420\,GPa/3.15\,g\,cm$^{-3}$ and a thermal stability of 180\,W\,m$^{-1}$\,K$^{-1}$/$2.2\times10^{-6}$\,K$^{-1}$ for the instrument's main structure \citep{2017SPIE10562E..4JB}. 
A second, smaller structure also made of \ac{SiC} forms the protective cavity for the grism and filter wheels.

The interface between the NISP instrument and the \Euclid spacecraft was designed to ensure a thermo-mechanical decoupling from the payload module to minimise heat transfers from the satellite to the NISP instrument, and vice-versa.
The final concept is based on a hexapod system made of Invar that ensures a quasi-isostatic mechanical link. The hexapod also reduces the total conductance from the NISP main structure to the payload module baseplate to $\simeq$\,0.035\,W\,K$^{-1}$ while providing NISP with the required stiffness.

\subsection{Optics}
\label{sec:optics}

\begin{figure*}[ht]
    \centering
    \includegraphics[width=\textwidth]{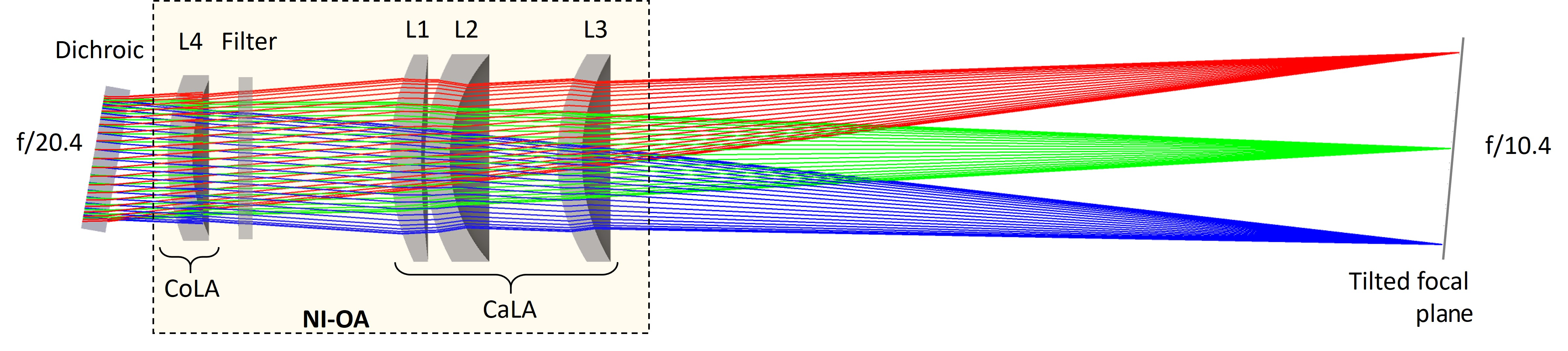}
    \caption{
    Beam path through the NISP optical assembly (NI-OA) in the photometer mode, consisting of CoLA, CaLA, and a bandpass filter. The dichroic element is located in the exit pupil of the telescope in front of NISP.
    The three cones of rays, shown in different colours, correspond to the centre and two opposing edges of the NISP field of view.
    }
    \label{fig:NI-OA}
\end{figure*}
%
We refer to the optics of NISP as the \ac{NI-OA}. It reduces the telescope's F/20.4 focal ratio to F/10.4, halving the magnification by a factor of two. This matches the illuminated optical field to the physical dimensions of the \ac{FPA}, and enables a more compact instrument architecture.

\subsubsection{Optical design}
The optical design of NI-OA is shown in Fig.~\ref{fig:NI-OA}. 
The dichroic beam splitter in the exit pupil of the telescope in front of NISP reflects the 530--920\,nm range into the VIS channel \citep[not shown here, see][]{euclid2024,cropper2024} and transmits the range $>$\,920\,nm into NISP -- for more details see Fig.~\ref{fig:chromatic_selection}.  
The NI-OA consists of two opto-mechanical groups \citep{2017SPIE10563E..4GG, 2014SPIE.9143E..2XG}, namely the \ac{CoLA} and the \ac{CaLA}. In between are the filter and the grism wheels (see Fig.~\ref{fig:NISP_FM}). Figure~\ref{fig:NI-OA} schematically shows the photometer mode with a filter in the beam path. The grism mode is set up similarly. 

In total NI-OA consists of four meniscus lenses. Since all materials must be resistant to cosmic radiation \citep{2013SPIE.8860E..0NG}, the choice of optical glasses is very limited. While the CoLA lens L4 is made of fused silica, the three CaLA lenses consist of \ce{CaF2} (L1) and S-FTM16 (L2 and L3). \ce{CaF2} has unique optical properties, such as low dispersion and high thermal stability. At the same time it is not a glass, but a brittle crystal with a cubic lattice structure that is challenging to work with. Other substrate materials suitable for the space environment would have resulted in a considerable loss in optical performance.

The final design emerged from a more complex precursor in which CoLA formed a multi-lens collimator so that the filter was in a parallel-beam path. To save mass and volume, the number of lenses was gradually reduced, dispensing with the parallel path. This increases somewhat the angle-of-incidence variations across the filter surface, and thus the passband variations across the \ac{FOV}. These are present -- and dominating -- even if the filter was in a parallel beam, see also Sect.~\ref{sc:nifs} and \cite{schirmer2022}.

The four lenses have a spherical surface on the convex front side and an aspherical surface on the concave back side. We found this system to have the minimum number of degrees of freedom required to meet the high demands on optical imaging quality. Additional aspherical surfaces on the convex sides could improve the design even further or perhaps further reduce the required number of lenses. However, aspherical convex surfaces of these large lenses with diameters between 130 and 170\,mm can in practice not be measured interferometrically and thus are not manufacturable because  \acp{CGH} and Fizeau interferometers of the required size are not available in industry. 

\Euclid's wide-angle telescope is of a Korsch off-axis design: The rays hitting the centre of NISP's \ac{FPA} do not enter the telescope perpendicularly, but with an angle of \ang{0.8357} to the optical axis of the primary mirror. This design is a result of studies trying to combine both a \ac{FOV} large enough for \Euclid's survey with flatness of the \ac{FOV} in order to have excellent image quality using planar detectors. The need to also have a geometrically sufficiently large and accessible focal plane for placement of instruments led to an off-axis telescope design. One consequence of this off-axis configuration is the tilted focal plane of NISP, clearly visible in Fig.~\ref{fig:NI-OA}.

\subsubsection{Imaging quality requirements} 
The requirements for NI-OA's optical imaging quality are predefined by the telescope, which is designed to feed the instrument with a diffraction-limited signal over the full off-centred \ac{FOV} \citep{2013SPIE.8860E..0GG, 2014SPIE.9143E..2XG, 2016SPIE.9904E..2MG}.
Here we refer to the Maréchal criterion \citep[Eq.~30-53 in][]{gross2006}, which requires a wavefront-error root mean square (WFE RMS) value smaller than $\lambda/14$ to be considered as diffraction limited. This requirement emphasises the need for excellent mechanical thermal stability.

Accordingly, the NI-OA design must also be almost diffraction limited over the entire \ac{FOV} to preserve the imaging quality. In addition, it must maintain the flatness of the \ac{FOV} and it has to guarantee low ghost-image intensities (see Sect.~\ref{sec:ghosting}). 

\subsubsection{Filters and grisms with refractive power\label{sec:filtergrismsubstrates}} 
In contrast to the telescope's exclusively reflective mirror design, NI-OA's refractive lens design is not completely free from chromatic aberrations.
With the wide wavelength range of NISP -- covering more than one octave -- such errors could not be completely avoided. This problem is however overcome by dividing NISP-P  into three (Tab.~\ref{tab:photo_passbands}) and NISP-S into two (Tab.~\ref{tab:nisp_grism}) wavelength bands.

The three bandpass filters (Sect.~\ref{sc:nifs}) are made from `Suprasil 3001' fused silica. They have a diameter of 130\,mm with a clear aperture of 126\,mm, a central thickness of 11.20\,mm to 11.96\,mm, and are the largest NIR filters flown on an astronomy space mission to date. The filters' exit surface is planar, while their sky-facing entrance surface is spherically convex, with curvature radii between 9968\,mm and 10\,027\,mm. This slight optical power compensates for the residual chromatic aberrations of the refractive \ac{NI-OA}. The filter substrates were glued into holders on metal-blade springs that allow for compensation of differential thermal contraction, which may occur between the Suprasil substrate and the metals of the filter wheel.

The four grisms (Sect.~\ref{sc:nigs}) are also made of fused silica and -- for the same reason as the filters -- have a mildly powered spherical entrance surface. The filter and grism blanks were jointly manufactured by Winlight (France), all compared to a common curvature reference surface to achieve a high accuracy in the slightly different target curvature radii for the different elements. The latter facilitates co-focality of photometric and spectroscopic channels, that is they share the same focal plane. For this reason it is not possible to use both a filter and a grisms simultaneously, as this would result in strongly defocused images.

\subsubsection{Lens manufacturing} 
With a size of over 170\,mm and a weight of more than 2.5\,kg for the most massive lens, NI-OA has the largest civilian lens system ever launched into space.
The manufacturing of this large and high-precision assembly required extraordinarily sophisticated technical methods. This included an accurate measurement of the refractive indices that affect the focal length and -- across the large \ac{FOV} -- the field curvature and thus the imaging quality.

The measurements were performed with NASA's CHARMS cryo-refractometer \citep{2016SPIE.9904E..2MG} for all lens materials used, over a wide temperature (100\,K--300\,K) and a broad wavelength range  (420\,nm--3000\,nm). The wide temperature range is required because of \ac{NI-OA}'s operating temperature of about 135\,K, whereas it is manufactured and aligned at room temperature. This also called for a precise conversion of the lens dimensions from cold to warm, using precision measurements of the temperature-dependent \ac{CTE}.  

The eight optical surfaces -- four spherical and four aspherical -- were produced using magnetorheological finishing in the last manufacturing step, which was required to fulfil our specifications. 

\subsubsection{Assembly and Integration} 
Monte-Carlo tolerance analysis showed that the reduction of the optical design to five refractive elements -- four lenses and one filter/grism -- in combination with the large \ac{FOV} and the aspherical lens surfaces leads to particularly critical centring tolerances of each lens. These range from 10\,$\micron$ to 20\,$\micron$ from the assembly of the components in the warm laboratory to operation under cold conditions in a vacuum.

To overcome this challenge, we developed and tested an interferometric alignment method based on \acp{MZ-CGH}. These holograms generate multiple wavefronts with different focal lengths that build a series of spots along a straight line. The thus-defined optical axis has an extraordinary straightness with only tiny deviations in the sub-micron range, that makes it possible to simultaneously align several optical elements on and along this axis, using one wavefront-zone per element. This is the key feature of our interferometric alignment method \citep[for more details see][]{Bodendorf:19}. 

\subsubsection{Common focus and test in warm environment} 

As stated, neither VIS nor NISP possess a refocusing mechanism. In-flight focusing is solely achieved by adjusting the tip-tilt and piston of the secondary mirror -- a fine-tuning of the focus is done in zero gravity during commissioning in space. Hence the focus offset between both instruments must be almost zero in space, and within NISP the three imaging (Tab.\,\ref{tab:photo_passbands}) and four spectroscopy modes (Tab.\,\ref{tab:nisp_grism}) must have identical focal lengths. To avoid mechanically stressful and time-consuming iterative cool-down circles to find the common focus, a unique combination of \acp{MZ-CGH} and a coordinate-measuring machine was used to determine NISP's back-focal distance -- and thus the position of the focal plane -- with an accuracy of single microns. With the precise knowledge of the refractive indices and the \acp{CTE} we performed these measurements in a warm environment, these agreed with the corresponding simulations of the instrument-model at these temperatures. This confirmation allowed us to correctly predict the properties also at cold. The procedure is described in \cite{2019SPIE11116E..18G}.

\subsubsection{Cryogenic optical performance test} \label{sec:opt_perf_test}
\begin{figure} 
    \centering
    \includegraphics[width=0.8\hsize]{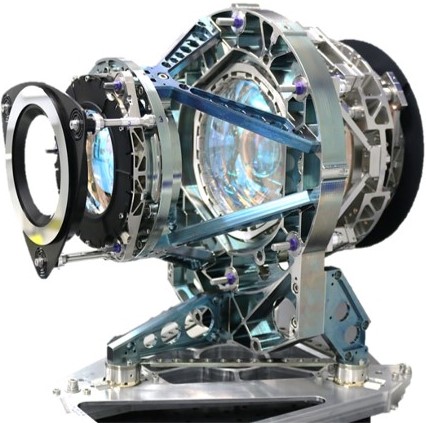}
    \caption{
    \ac{NI-OA}, consisting of \ac{CoLA} (left), \ac{CaLA} (right), and a filter dummy, mounted on an Invar structure in between, ready for the cryogenic test. The baffle in front of \ac{NI-OA}, an aperture stop, is mounted at the telescope's exit pupil.
    }
    \label{fig:NI-OA_Photo}
\end{figure} 
\begin{figure*} 
    \centering
    \includegraphics[scale=0.78]{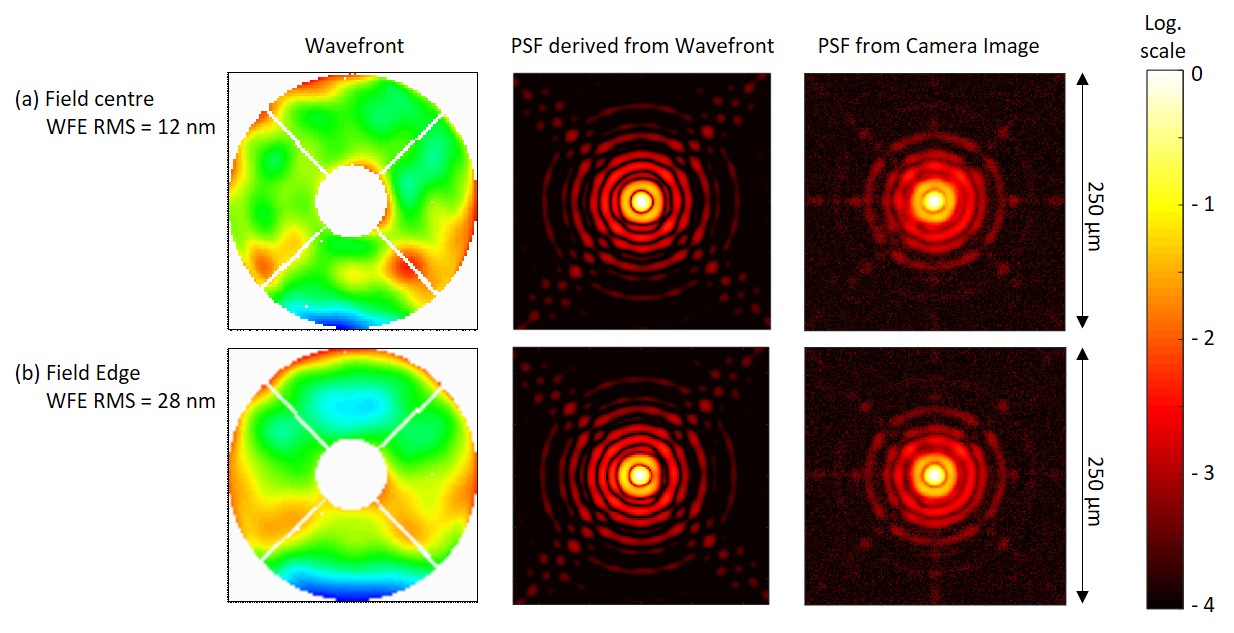}
    \caption{ 
    Optical performance of \ac{NI-OA} at $\lambda=960$\,nm, determined at two positions corresponding to the field centre (\textit{top row}) and a corner of the NISP field of view (\textit{bottom row}). The wavefront maps in the first column are based on a \acl{SHS} measurement, the \acp{PSF} in the second column are derived from these wavefronts, and the \acp{PSF} in the third column are directly observed with a CCD camera. Note the logarithmic scale over four orders of magnitude.
    }
    \label{fig:Measurement-WF&PSF}
\end{figure*} 

An estimate of \ac{NI-OA}'s in-flight performance is only possible in cryogenic conditions at operating temperatures of 133\,K, in a vacuum, and had to done before its integration into the NISP instrument.  For this purpose, a temporary mechanical Invar hexapod mount was developed to align \ac{CaLA} and \ac{CoLA} as shown in Fig.~\ref{fig:NI-OA_Photo}. The assembly also included a filter dummy without coating -- and thus without filter functionality -- as well as an aperture stop at the telescope's exit pupil, and thus formed an optically complete module for the test.

The imaging quality was evaluated at several \ac{FOV} positions with two complementary approaches \citep{10.1117/12.2235210}, namely a wavefront reconstruction based on a \acf{SHS} measurement and a direct observation of the \ac{PSF} with a cooled low-noise \ac{CCD} camera with additional optical magnification. All measurements were made with a superluminescent-diode at $\lambda=960$\,nm. Here we only present key results. For more details see \citet{2019SPIE11116E..0YB}, including the complete cryogenic experimental setup and discussions of further results, such as the flatness of the image plane. 

The first column in Fig.~\ref{fig:Measurement-WF&PSF} shows the wavefront in the centre and at the edge of the \ac{FOV}; central obscuration and spiders are caused by the measurement setup. They mimick \Euclid's telescope, which has however not four, but only three spider arms.
The typical WFE RMS over the entire \ac{FOV} range between $\lambda/60$ and $\lambda/30$, which exceeds the diffraction limit \citep[Eq.~30-53 in][]{gross2006} of $\lambda/14 = 69$\,nm by a factor of roughly 2 to 4.

The second column of Fig.~\ref{fig:Measurement-WF&PSF} shows the \ac{PSF} derived from the wavefront shown in the first column, and the third column shows the \ac{PSF} as measured by the \ac{CCD} camera. Due to the large intensity range of four orders of magnitude we averaged 40 individual images  to reduce the noise at low intensities present in individual camera measurement images. The slight visible blur is due to minor mechanical offsets between images -- both measurements methods are fully consistent, even in the fine structures of the aperture's complex diffraction pattern.  

\begin{figure*}[ht] 
    \centering
    \includegraphics[scale=0.72]{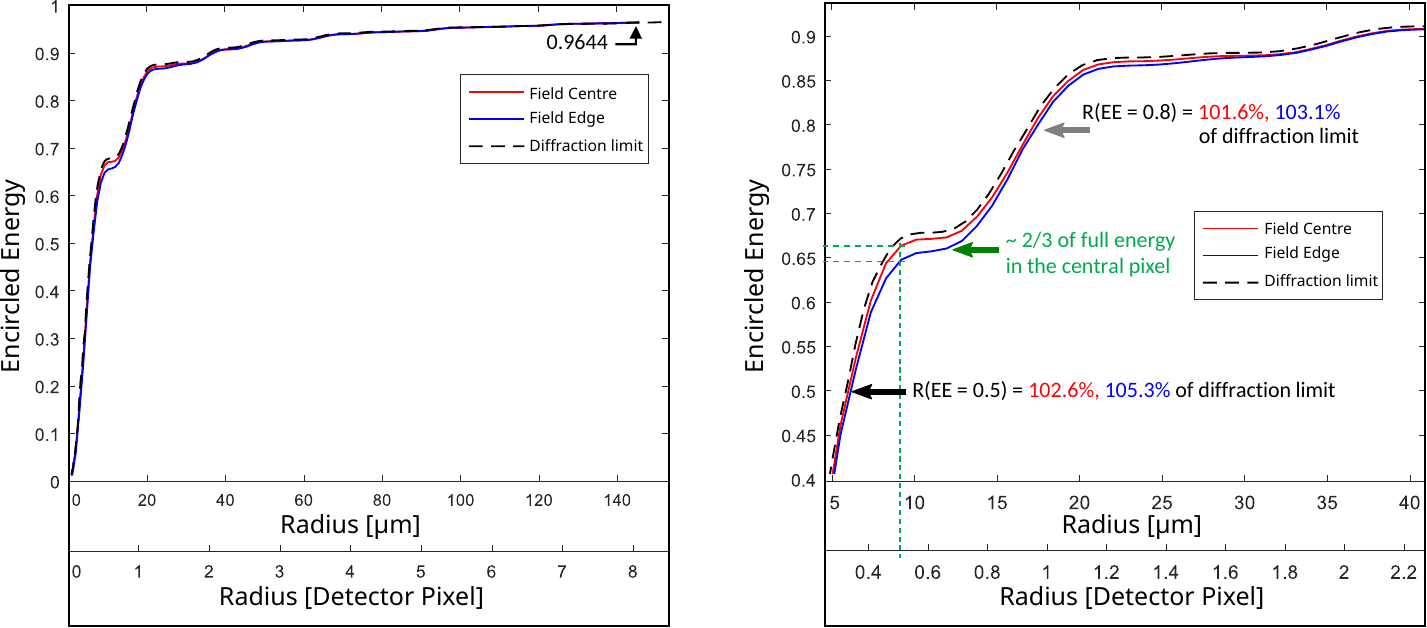}
    \caption{ 
    {\it Left:} \ac{EE} of the \ac{NI-OA} at $\lambda=960$\,nm based on \ac{SHS} measurements. 
    Shown in red is the \ac{EE} for the field centre, and in blue for the field edge.
    The ideal aberration-free or diffraction-limited case is given by the dashed black line. {\it Right:} An enlarged section of the full range of radii on the left. The radii for the 50\% and 80\% \ac{EE} values are expressed as a percentage of the diffraction-limited case. The $x$-axis is represented in physical units and in pixels.
    }
    \label{fig:Measurement-EE}
\end{figure*}

Lastly, in Fig.~\ref{fig:Measurement-EE}, we show the \ac{EE} derived from the \ac{SHS}-based \ac{PSF} in Fig.~\ref{fig:Measurement-WF&PSF}, for different field positions. The \ac{EE} is the proportion of the total energy that lies within a circle of radius $R$ around the centre of the \ac{PSF}. For comparison, we also display the ideal case for an aberration-free pupil function. The radius is given in physical units and in pixels. The right panel in the figure shows that at $\lambda=960$\,nm about $2/3$ of the total energy are contained within a radius of 0.5 pixel, greatly beneficial for the detection of faint compact sources. This is just a few percent, at most, below the ideal diffraction-limited case, showing that \ac{NI-OA} has almost perfect optical performance. The \ac{EE} determined from the direct CCD-camera measurements (not shown) is consistent with the \ac{SHS}-based result. We note that with a pixel pitch of 18\,\micron, the NISP detectors undersample the \ac{PSF} in the sense of the Nyquist-Shannon theorem.

After integration into NISP the optical quality and proper focus were tested in a complex cryo-setup at \ac{LAM}, confirming the previous predictions and measurements at \ac{NI-OA}-level \citep{costille2019b}.

To summarise, the \ac{NI-OA} meets the highest demands on optical imaging quality with only minimal deviations from the ideal diffraction-limited case  \citep{2019SPIE11116E..0YB}. This provides the basis for NISP's excellent image and data quality.
%

\subsection{\label{sc:nifs}Filters}

\begin{figure*}[ht]
   \includegraphics[width=1.0\hsize]{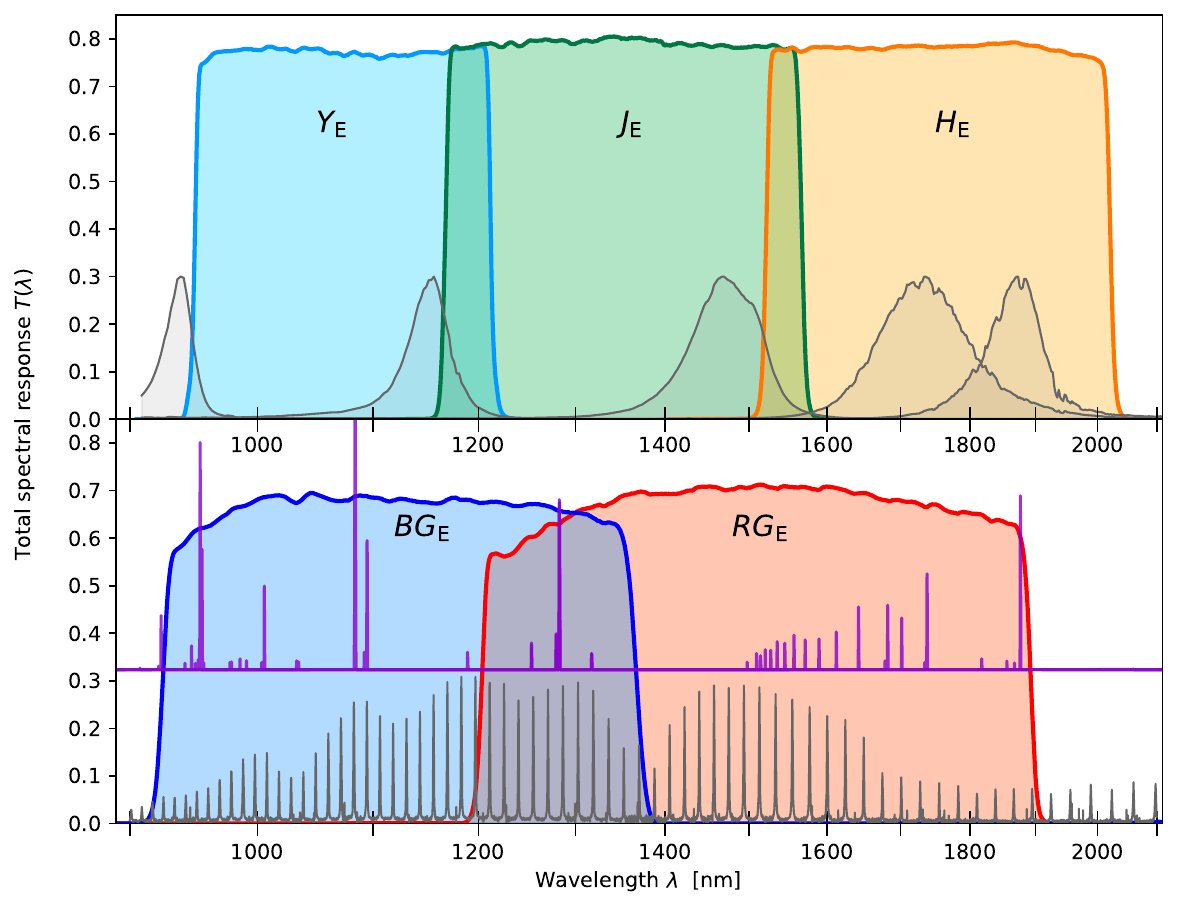}
    \caption{Total spectral response curves, accounting for mirrors, the dichroic element, all optical elements in NISP, as well as the detectors' mean \acl{QE}. \textit{Top panel}: The three filter passbands are shown together with the approximate emission curves of the five calibration lamps. \textit{Bottom panel}: Shown are the blue and red spectral passbands for the $1^\text{st}$-order. The etalon spectrum used for wavelength calibration on ground is overlaid as the grey curve at the bottom. The log-scaled spectrum of one of our compact planetary nebulae for in-flight wavelength calibration is shown in purple \citep{paterson2023}. All calibrator spectra are arbitrarily scaled in this figure.}
    \label{fig:trans}
\end{figure*}

\begin{figure*}[ht]
    \centering
    \centering
    \includegraphics[width=0.45\textwidth]{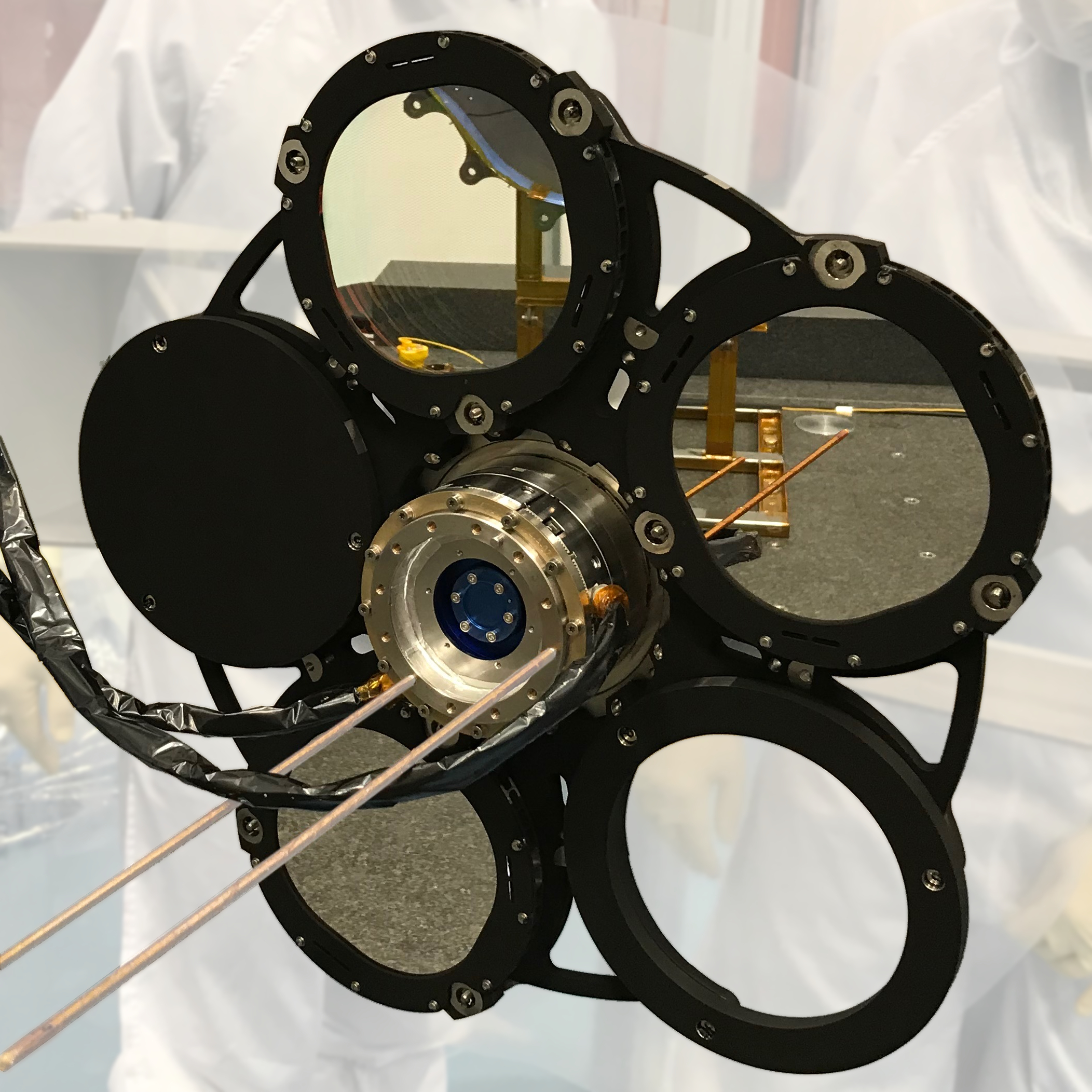}
    \includegraphics[width=0.45\textwidth]{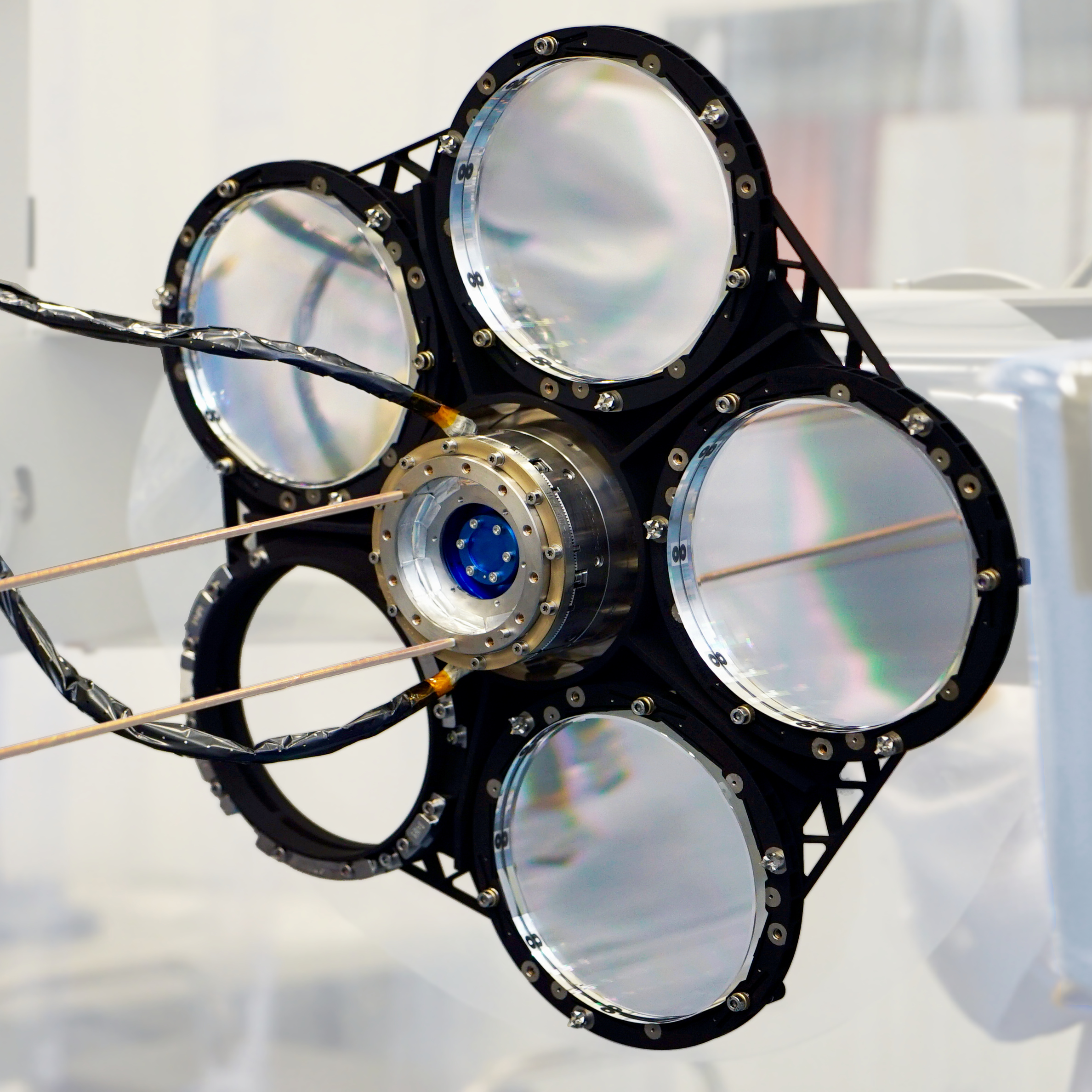}
    \caption{
    Flight models of the \acf{FWA} ({\em left}) and the \acf{GWA} ({\em right}) before integration into NISP. Both wheels contain the same cryo-motor and bearing system, but have their individual mechanical design. Each contains an open position to allow light from the respectively other mode to pass. The \ac{FWA} has one closed position, blocking the telescope beam into NISP for dark and flatfield exposures, and one filter each for the \YE, \JE, and \HE\ passbands. The \ac{GWA} has three grisms for the \RGE\ passband, at three orientations, and one for the \BGE\ passband.
    }
    \label{fig:fwa_gwa}
\end{figure*}

NISP provides three photometric passbands, \YE, \JE, and \HE, between 950 and $2021$\,nm. The properties of the filter substrates are described in Sect.~\ref{sec:filtergrismsubstrates}. The current best estimates of the passband edges are listed in Table~\ref{tab:photo_passbands} and plotted in Fig.~\ref{fig:trans}. A filter change is effected by the \ac{FWA} shown in the left panel of  Fig.~\ref{fig:fwa_gwa}. In addition, the \ac{FWA} has a light-tight closed position for various calibrations and to protect the detectors during field slews, and an open position to allow spectroscopic observations. 

The bandpass-forming dielectric interference layers were coated onto the substrates by formerly Optics Balzers Jena (Germany), now Materion Optics, using \ce{SiO2} and \ce{Nb2O5} as alternating low- and high-index materials, respectively. These coatings have very low sensitivity to radiation-induced ageing. The pessimistic upper limit is 2\% throughput loss until the end of the mission, with the actual change likely to be much smaller. The top layer of the layer stacks is always hard \ce{SiO2}, providing physical protection and allowing for cleaning. The \ac{PARMS} process was used to deposit between 72 and 188 layers per side, resulting in near-rectangular passband shapes. The stack height is about $20$\,\micron, with a good layer-thickness homogeneity of $\sim$\,0.25\% across the filter substrates. A pilot study based on \acl{IAD} yielded an insufficient homogeneity of $\sim1\%$. 

The stacks defining the high- and low-pass passband edges could in principle be separated on the substrates' entrance and exit surfaces. Owing to the wide wavelength-interval blocking requirements this would have resulted in stacks of considerably different thickness. This carried the risk of bending the substrate -- and hence degraded optical performance -- because of compressive stresses from the coating process, and due to different \ac{CTE} of the layer stack and the substrate. Hence the stacks were modified to have similar thickness on both substrate surfaces, between 19\,\micron\ and 22\,\micron\ depending on the filter. The maximum difference of the coating-stack heights on any filter is 1.2\,\micron.

While the NISP passband edges are defined by the filter coatings, the out-of-band blocking of photons down to 0.3\,\micron\ and up to 3.0\,\micron\ is the combined effort of the filter coatings, the dichroic, the \ac{NI-OA}, the telescope mirrors, and the detectors (see the chromatic selection function in Fig.~\ref{fig:chromatic_selection}). 
Within the NISP's main 0.9--2.0\,\micron\ range, the out-of-band blocking factor -- or total spectral response -- is typically on the order of a few $10^{-4}$ compared to the in-band transmission. Outside the NISP wavelength range the blocking factor is $10^{-5}$ to $10^{-9}$ or better. In practice this means that the out-of-band contamination of the NISP photometric measurements is at most 2\,mmag, and more typically $0.2$\,mmag. 

\begin{figure}[ht]
    \centering
    \includegraphics[width=\columnwidth]{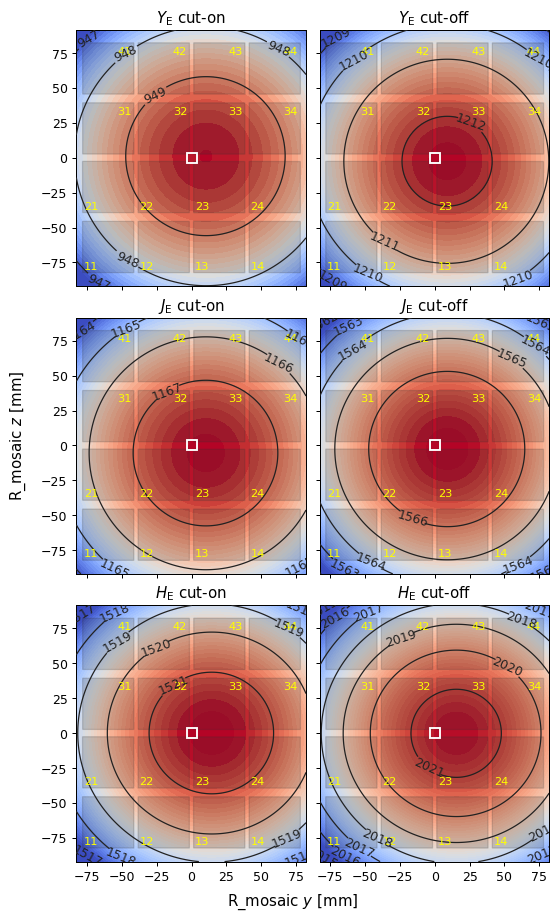}
    \caption{
    Variation of the passbands' cut-on ({\em left}) and cut-off ({\em right}) wavelengths with field position. The grey squares in the background mark the 4$\times$4 detector grid, with the \ac{SCA} positions shown in yellow. The white square shows the centre of the \ac{FOV}, with its applicable passband wavelengths listed in Table~\ref{tab:photo_passbands}. Note: This is the view from the sky towards the \ac{FPA}, see Fig.~\ref{fig:fpa_layout} for \ac{FPA} coordinate system details.
    }
    \label{fig:filter_variation}
\end{figure}

The as-built passband boundaries are given in Table~\ref{tab:photo_passbands}, and their spatial variations from coating inhomogeneities and angle-of-incidence variations, are shown in Fig.~\ref{fig:filter_variation}. More information about the passbands, including tabulated curves, as well as the underlying measurements can be found in our detailed study of the NISP photometric system and its knowledge uncertainties in \cite{schirmer2022}.

Please note that filter changes occur by rotation of the \ac{FWA}. The \ac{FWA} positioning is not forced by a clutch or a similar device, but NISP works without an extra mechanism and positions the \ac{FWA} -- and \ac{GWA} -- by commanding a certain number of motor steps, then keeping a position simply by bearing friction. Therefore this positioning can vary by a few 0\fdg1 of wheel angle between different instances of positioning a given filter. Since the NISP filters are having slight optical power, this impacts the exact distortion of the field -- correspondingly this can slightly vary between exposures.

\begin{table}[htb]
    \centering
       \caption{NISP photometry passband characteristics \citep[from][]{schirmer2022}. The passbands have similar width in $\Delta\lambda$/$\lambda$. The `50\% cut-on/off' wavelengths refer to the \ac{FWHM} passband edges of the total spectral response. $\sigma_\mathrm{50}$ is the uncertainty in the cut-on/off wavelengths, the first number refers to the field-of-view centre, the second to the corners of the field. $\sigma_\mathrm{cen}$ is the uncertainty in the passband central wavelength. A visualisation of the spatial cut-on/off wavelength shifts is shown in Fig.~\ref{fig:filter_variation}.} 
\addtolength{\tabcolsep}{-0.15em}    
\begin{tabular}{c| c c c c c}
    \hline\hline
      \noalign{\vskip 1pt}
         Filter   & 50\% cut-on & 50\% cut-off & centre & $\sigma_\mathrm{50}$ & $\sigma_\mathrm{cen}$\\
               & [nm] & [nm] & [nm] & [nm] & [nm]\\
               \hline  
               \noalign{\vskip 1pt}
           \YE &  949.6 & 1212.3 & 1080.9 & 0.8/0.8& 0.6\\
           \JE & 1167.6 & 1567.0 & 1367.3 & 0.8/0.9& 0.6\\
           \HE & 1521.5 & 2021.4 & 1771.4 & 0.8/0.9& 0.6\\
           \hline
    \end{tabular}
    \label{tab:photo_passbands}
\end{table}

\subsection{\label{sc:nigs}Grisms}

\begin{figure}[!b]
    \centering
    \includegraphics[width=\columnwidth]{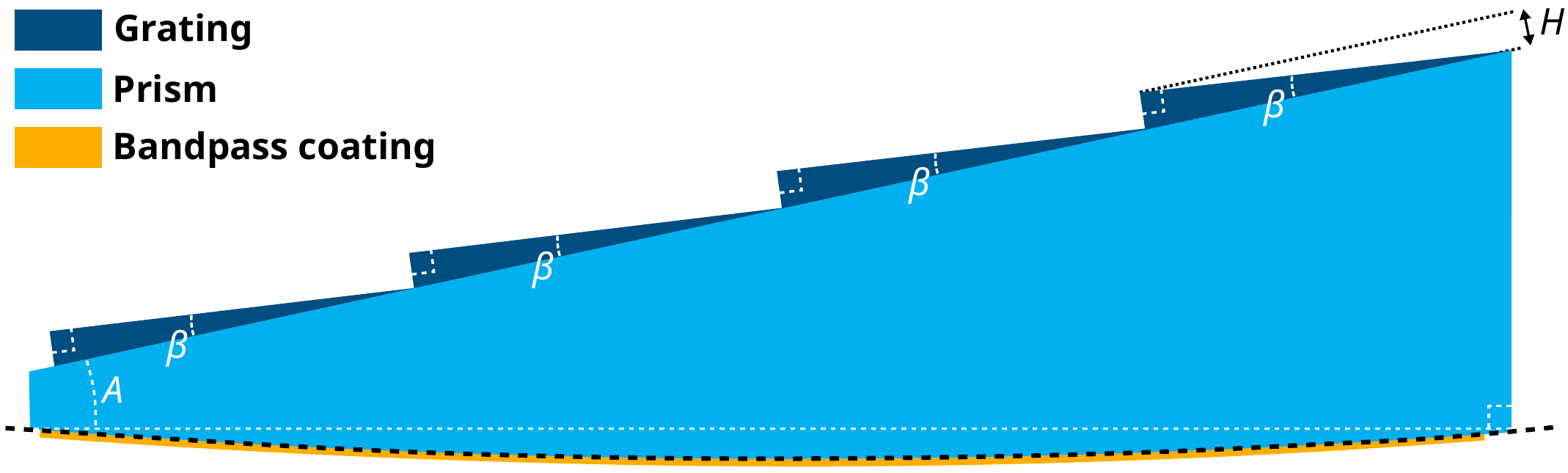}
    \caption{Schematic cross section of a NISP grism with the blazed dispersion grating (dark blue), the prism (light blue), and the passband filter (yellow). Prism angle $A$, groove height $H$, and blaze angle $\beta$ are listed along other fundamental grism characteristics in Table~\ref{tab:nisp_grism}.}
    \label{fig:grism-xSec}
\end{figure}
For its spectrometric measurements, NISP uses three `red' and one `blue' grism, referred to as RGS000/180/270 and BGS000, respectively. Their main characteristics are summarised in Table~\ref{tab:nisp_grism}. The grisms are mounted in the \ac{GWA}, shown in the right panel of Fig.~\ref{fig:fwa_gwa}, which is itself enclosed inside a cryogenic cavity made of \ac{SiC} together with the \ac{FWA}. The grisms, whose cross-section are shown in Fig.~\ref{fig:grism-xSec}, combine three different optical functions \citep{Costille_SPIE_2019a}. First, light dispersion: this is provided by the grism itself, a combination of a blazed dispersion grating and a prism, shown as dark blue and light blue in Fig.~\ref{fig:grism-xSec}, respectively. The grism places the 1st-order spectra of sources in the same detector region as the imaging filters, allowing the use of a common focal plane for both channels (Sect.~\ref{sc:nifs}). Second, focusing: a contribution to an optimal focus is provided by the convex surface of the base of the prism (see Sect.~\ref{sec:optics}). Third, bandpass filtering: a multi-layer coating deposited onto the surface of the base of the prism (yellow in Fig.~\ref{fig:grism-xSec}) defines the spectroscopic passbands, together with the transmission characteristics of the \ac{NI-OA} and the dichroic. This results in \ac{FWHM} wavelength ranges of \BGrangenew\,nm and \RGrangenew\,nm for the \BGE\ and \RGE\ passbands, respectively, as shown in Fig.~\ref{fig:trans}.

\begin{table*}[th]
    \caption{Optical properties of the flight model of the NISP grisms.}
    \centering
    \begin{tabular}{l|ccc}
        \hline\hline
          \noalign{\vskip 1pt}
          \multirow{2}{*}{Parameters} & \multicolumn{2}{c}{Red grisms} & Blue grism\\
                                      & RGS000/180        & RGS270       & BGS000\\
         \hline
           \noalign{\vskip 1pt}
         \ac{FWHM} bandpass [\si{\nano\meter}] & \multicolumn{2}{c}{\RGrangenew} &  \BGrangenew \\
         Prism angle $A$ & \multicolumn{2}{c}{$\ang{2.145;;}\pm\ang{;;30}$} &  $\ang{1.77;;}\pm\ang{;;30}$ \\
         Mean groove height $H$ [\si{\micro\meter}]& \multicolumn{2}{c}{$3.16\pm0.08$}  &  {$2.25\pm0.1$} \\
         Mean blaze angle $\beta$ & \ang{2.49} & \ang{2.48} &  \ang{1.94} \\
         Mean pitch $P$ of grooves [\si{\micro\meter}]& {$72.55$} & {$72.96$} &  $66.22$ \\
         Groove density [\si{\grv/m\meter}]& $13.75$ & $13.7$ &$15.1$ \\
         Averaged $0^\text{st}$-order transmission & \TRzero\% & \TRRzero\% &  \TBzero\% \\
         Averaged $1^\text{st}$-order transmission & \TRfirst\% & \TRRfirst\% &  \TBfirst\% \\
         Un-deflected wavelength [\si{\micro\meter}]& \multicolumn{2}{c}{$\simeq$\,1.2} &  $\simeq$\,0.9 \\
         Spectral resolution $\left|{d\lambda}/{dx}\right|$ [nm/pix]& 1.372&  -- & 1.239\\
         Resolving power $\mathcal{R}$& $>$\,480 &  -- & $>$\,400 \\
         \quad for \ang{;;0.5} diameter object & & & \\
         \hline
    \end{tabular}
    \label{tab:nisp_grism}
\end{table*}

The grisms' complex optomechanical design was derived through a research \& development programme funded by the \ac{CNES} from 2010 onward, including the first delivery of the grisms' qualification models and up to the flight models in 2017. The main complexities were the grism size of \SI{140}{m\m} diameter, the low groove frequency of the grating at $<$\,16\,grooves\,mm$^{-1}$, and a small blaze angle of $<$\,\ang{3}. Together with stringent requirements this led to highly specialised manufacturing specifications. Further challenges arose from the grisms' operating conditions in space, as the NISP detectors (Sect.~\ref{sc:NIDS}) require the optics to operate at a temperature of $\sim$\,135\,K to reduce thermal background.\footnote{A second reason is that 135\,K is the maximum temperature for the whole structure of NISP, which drives the minimal temperature the NISP detectors can reach. Higher detector temperatures would mean a worse performance due to for example higher dark current and read noise.} To minimise intrinsic thermo-mechanical stresses, the grisms were built each from a single fused-silica block with the grating directly engraved onto the hypotenuse surface of the prism, possible through the cumulative etching technology of SILIOS Technologies SA\footnote{\url{https://www.silios.com}} \citep{10.1117/12.2304211}. 

Another design driver was the required redshift precision of $\sigma(z)<0.001(1+z)$ for \Ha emission-line galaxies with a \ac{FWHM} of \ang{;;0.5} and a $3.5\sigma$ detection limit at $2\times10^{-16}$\,erg\,s$^{-1}$\,cm$^{-2}$ \citep{scaramella2022}. The red grisms were built with a prism angle of $A=\ang{2.145}$ and a groove frequency of $13.7$\,grooves\,mm$^{-1}$. The blue grism has $A=\ang{1.77}$ and $15.1$\,grooves\,mm$^{-1}$. The grooves were engraved onto the prism surface following complex curves. This allows provides a wavefront correction that compensates for the tilted NISP \ac{FPA} and the aberrations of the optical design, resulting in a nearly constant dispersion law and uniform image quality \citep{Costille_SPIE_2019a}. 

The grism raw material Suprasil 3001 \citep{heraus2023} is a water-free synthetic fused silica with \ce{OH} and metallic impurities lower than 1\,ppm, for maximum transmission in the NIR. The grating blaze angle maximises transmission of the $1^\text{st}$ order, with a peak transmission at the central wavelength of the grisms passband, which is at  $\sim1.5$\,\micron\ for the red grisms and at $\sim1.1$\,\micron\ for the blue grism. This minimises the flux losses near the passbands cut-on and cut-off wavelengths.
With $85\%$ in the $1^\text{st}$-order, the measured transmission of the finalised grisms is considerably larger than the requirement of $>65\%$. The out-of-band transmission is below 2\% \citep{2018SPIE10698E..2BC,Costille_SPIE_2019a}. Transmission in the $0^\text{th}$ order was measured to be $2\%$ on average; it is important for wavelength calibration to have enough flux in this order, as it is a reference for wavelength calibration. The mean total spectral response of the $1^\text{st}$-order as shown in the bottom of Fig.~\ref{fig:trans}, as well as for the $0^\text{th}$-order, is available online.\footnote{\url{https://euclid.esac.esa.int/msp/refdata/nisp/NISP-SPECTRO-PASSBANDS-V1}} The data include the contributions from the mirrors, the dichroic, the grisms, the \ac{NI-OA}, and the mean detector \acf{QE}. 

The NISP spectroscopic performance at instrument level was assessed during ground test campaigns in 2019 and 2020. A \ac{LED} light source combined with a Fabry-Perot etalon with $\simeq$\,34 transmission peaks in each grism passband -- its spectral energy distribution represented in grey in the bottom of Fig.~\ref{fig:trans} -- was used to determine the spectroscopic dispersion of each grism. 
The measured NISP image quality is nearly diffraction-limited. The spectral resolution of the grisms was measured to be $1.239$\,nm\,pixel$^{-1}$ for the blue grism BGS000, $1.372$\,nm\,pixel$^{-1}$ for the red grisms RGS000, and $-1.372$\,nm\,pixel$^{-1}$ for RGS180. Here, the negative sign shows that the the RGS180 disperses in the opposite direction as the RGS000.
The NISP resolving power has a minimal value along each spectrum of $\mathcal{R}_{\BGE}>400$ and $\mathcal{R}_{\RGE}>480$, enabling redshift measurements with an error $\sigma(z)<0.001(1+z)$. These numbers for $\mathcal{R}$ are based on ground measurements, exact values in flight will depend on the details of data reduction and spectral extraction. A detailed analysis of the dependency on wavelength, object size, etc.\ based on in-orbit data will be presented in a future paper, an initial assessment based on simulations is made in \citet{Gabarra-EP31}.

Further tests of NISP were repeated at payload-module level, with all of \Euclid's telescope mirrors and the dichroic in the light path. These tests demonstrated an excellent telescope and that the presence of these additional components -- specifically the dichroic beamsplitter -- do not alter the excellent spectroscopic performance. 

That said, please note that NISP carries three red grisms (Table~\ref{tab:nisp_grism}). Owing to a coordinate system misunderstanding during integration, the grisms were erroneously glued into their holding structures rotated by 180$^\circ$. While for the RGS000 and RGS180 grisms this only creates an inversion of the dispersion direction, the tilted focal plane along RGS270's dispersion direction means that observations with the RGS270 are out of focus outside its central wavelength. The RGS270 can therefore not be used for spectrum overlap decontamination as was its original main purpose. At the source this issue could only have been corrected by producing a new set of grisms and grism holders. This would have induced an estimated mission-delay of 1--2 years and posed extra risk from dismantling the instrument as well as degrading the initial integration that was outperforming expectations. Instead a procedure was developed without the RGS270, using the RGS000 and RGS180 grisms also with a 4$^\circ$ rotation of the \ac{GWA}, which was shown to be a functioning approach to decontaminate spectral overlaps in the \Euclid Wide Survey. Hence while along with the other grisms the RG270 has been launched to maintain spacecraft balance, it will not be used in the survey.

NISP-S blue grism is used to extend the lower redshift limit for H$\alpha$ to $z\,{=}\,0.41$. It will solely be used for observations of the \Euclid Deep and Auxiliary fields \citep{scaramella2022}, where multiple visits over time will observe these field at different orientation angles for overlap decontamination. The main purpose of the blue grism is to provide a large reference sample of galaxies with 99\% redshift completeness and 99\% purity required to characterise the typical \Euclid galaxy population, and to maximise the legacy value of these fields \citep{euclid2024}.

As for the filters above (Sect.~\ref{sc:nifs}) the NISP grisms are selected by rotating the \ac{GWA}, and the same wheel positioning uncertainties apply. This has the consequence of inducing slight angles of the dispersion direction with respect to the nominal orientation. This angle can vary by a few 0\fdg1 between observations with the same grism, if the \ac{GWA} has been moved in between.

\subsection{\label{sc:NIDS}Detector system and cold electronics}

\begin{figure*}[hbt]
    \centering    \includegraphics[height=60mm]{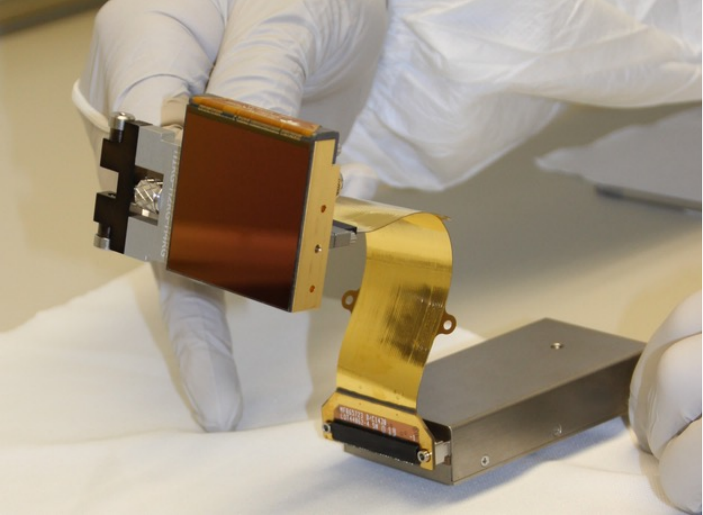}\quad  \includegraphics[height=60mm]{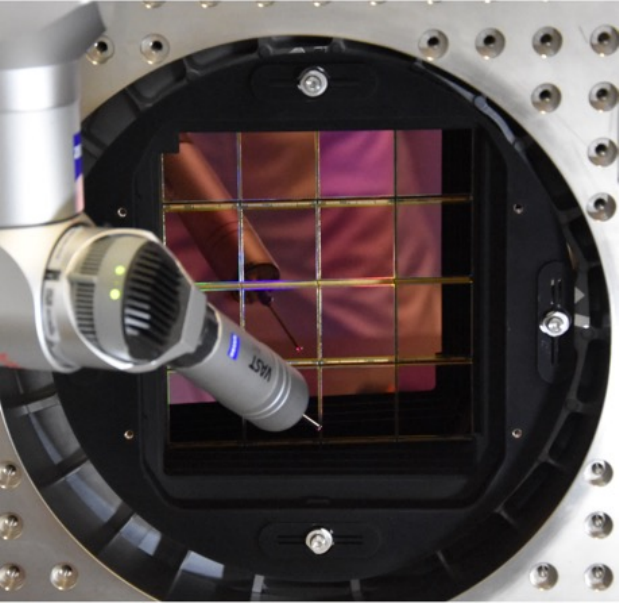}    
    \caption{{\it Left}: \ac{SCS} cold electronics triplet delivered by NASA, composed of the H2RG detector arrays (\ac{SCA}), flex cable (\ac{CFC}) and \ac{SCE} SIDECAR ASIC \citep{Holmes_SPIE_2022}. {\it Right}: Metrological measurements of the detectors mounted in their baffle.}
    \label{fig:NI-SCS}
\end{figure*}

The 16 H2RGs in the NISP \ac{DS} are shown to the right in Fig.~\ref{fig:NISP_FM}.
Each detector is part of a \ac{SCS}, delivered by NASA JPL, and is composed of three items (Fig.~\ref{fig:NI-SCS}, left panel). 

First, the \ac{SCA} is a \ac{MCT} 2048$\times$2048 photodiode array hybridised to a H2RG \acl{ROIC}. The \acp{SCA} are designed and manufactured by Teledyne Imaging Sensors. The long-wavelength cut-off was set to 2.3$\,$\micron\ by tuning the ratio of mercury to cadmium in the \ac{MCT}. This choice of cutoff reduces the total dark current to a negligible level $<$0.01\,e$^-$\,s$^{-1}$ at the operating temperature of 100\,K and maintains the highest \ac{QE} across the NISP wavelength range. A total of 34 working \acp{SCA} were made and through a rigorous test programme downselected to 16 flight models and 4 flight spares \citep{bai2018,waczynski2016}. 

Second, the \ac{SCE} is a cold electronics package used to provide timing, biases, communications, and data conversion for the operation of the \ac{SCA}, as well as serving as an interface to the NISP warm electronics.

The SCEs include an application-specific integrated circuit (ASIC) termed the  `SIDECAR'  \citep{loose2007} that was developed by Teledyne Imaging Systems.  The design, construction, acceptance testing, and performance testing of the SCEs were performed through a collaboration among NASA JPL, NASA GSFC, and the larger NISP team, and are described in \citet{Holmes_SPIE_2022}. The flight units and flight spares were tested in combination with non-flight SCAs. The units incorporated in the NISP flight detector system were selected from a larger collection of constructed units based on their noise performance and other characteristics.

Third, the \ac{CFC} connects the \ac{SCA} to the \ac{SCE} while keeping the thermal conductance between the 95\,K of the \ac{FPA} and the 135\,K of the cold electronics as low as possible. The measured thermal conductance is 0.85\,mW\,K$^{-1}$ over the range 135\,K to 95\,K, well below the specification of 1.5\,mW\,K$^{-1}$ \cite{Holmes_CEC2019}.

A detailed description of the detector chain as well as detector effects such as nonlinearity and persistence seen in the NISP flight \ac{SCS} triplets can be found in \citet{barbier2018}. The latest results on inter-pixel capacitance, which is of the order of 0.6\%, is reported for one flight SCS in \citet{legraet2022}. 

The H2RGs are arranged in a 4$\times$4 mosaic as described and displayed in Fig.~\ref{fig:fpa_layout} in Appendix~\ref{sec:fpa_layout}. The SCEs are placed at the back of the \ac{FPA}, in each two rows of four at the top and bottom, respectively. A picture of a single \ac{SCS} is shown in Fig.~\ref{fig:NI-SCS} before (left) and the 4$\times$4 array after integration into the \ac{DS} (right). The array position numbering is indicated in Fig.~\ref{fig:filter_variation}.

The \ac{SCE} flight firmware was delivered by Markury Scientific and validated by NASA \citep{MarkusLoose}. The firmware implements the science \ac{MACC} modes and calibration modes, such as the single-pixel reset along the pixel grid, while keeping the power consumption within specifications.       

To first order, the primary driver of the detector-chain sensitivity performance for faint objects is the \ac{SNR}. It is mainly driven by the noise in the science acquisition modes, that is the noise of the flux estimator -- the combination of read-out hardware and algorithm estimating the flux in each pixel from multiple reads -- in spectro- and photo-readout modes, and by the \ac{QE} across each passband. Table~\ref{tab:nisp_fpa} summarises the main features of the \ac{FPA}. Noise properties for both photometry and spectroscopy modes (Table~\ref{tab:macc}) were measured during both testing of each \ac{SCA} in a joint campaign of the \ac{CPPM} and the \ac{IP2I} as well as the first thermal vacuum test at instrument level at \ac{LAM}. These extensive campaigns -- each one month per detector, and 1.5 months for the instrument as a whole -- had the overall goal to create a general characterisation of all detector components, that would both allow to develop and test models, e.g.\ for non-linearity and persistence effects, as well as to establish ground reference maps for dark current, read noise, dead pixels, gain, and other parameters. These were being used for initial work by the \ac{SGS} as well as are providing a reference for comparison of in-flight behaviour. During these measurement campaigns also the implemented standard \ac{MACC} readout modes for photometry and spectroscopy were used, providing reference observations under various conditions for later comparison, as well as for preparation of final operation parameters, testing of the on-board data processing, and calibration activities.

\begin{table*}[th]
    \caption{Detector chain performance of the Flight Model NISP focal plane, for all 16 SCS, identified by their \ac{SCA} number. `Position' refers to position in the 4$\times$4 array as defined in Fig.~\ref{fig:fpa_layout}, `QE' is the quantum efficiency in the respective photometry passbands, with an absolute accuracy of 5\% \citep{waczynski2016}. `Noise' is the read noise in the standard survey \ac{MACC} modes for photometry and spectroscopy. `Disconnected pixels' gives the number of pixels per detector that do not produce a signal due to a missing or faulty indium bump. If not specified then values given are medians over the array.}
    \centering
    {
    \addtolength{\tabcolsep}{-0.25em}
    \begin{tabular}{ll|cccccccccccccccc}
\hline\hline
  \noalign{\vskip 1pt}
   \multirow{2}{*}{Mosaic}  & Position      & 11  & 12  &  13  & 14   & 21  & 22  &  23  & 24   & 31   &  32 &  33  & 34   & 41  & 42  &  43  & 44 \\ 
                             &  \ac{SCA} 18XXX  & 453 & 272 &  632 & 267  & 268 & 285 &  548 & 452  & 280  & 284 &  278 & 269  & 458 & 249 &  221 & 628 \\              
                             \hline
                               \noalign{\vskip 1pt}
 \multirow{3}{*}{QE\ [\%]}
  & \YE-band & 93 & 96 & 88 & 95 & 92 & 94 & 94 & 93 & 91 & 92 & 94 & 94 & 92 & 93 & 96 & 92 \\
 & \JE-band & 95 & 97 & 95 & 96 & 93 & 96 & 95 & 95 & 93 & 94 & 95 & 95 & 96 & 94 & 98 & 95 \\
 & \HE-band & 96 & 98 & 95 & 97 & 94 & 96 & 96 & 95 & 93 & 94 & 95 & 96 & 95 & 94 & 99 & 95  \\
\hline
  \noalign{\vskip 1pt}
 \multirow{2}{*}{Noise [e$^-$] }
 & Spectro  & 8.2 & 9.0 & 7.8 & 9.2 & 8.2 & 8.9 & 8.3 & 8.4 & 8.1 & 7.9 & 8.0 & 8.7 & 7.7 & 7.9 & 8.1 & 8.7 \\
 & Photo    & 6.0 & 6.2 & 5.9 & 6.4 & 6.3 & 6.7 & 6.4 & 6.4 & 5.9 & 5.8 & 5.6 & 6.4 & 5.7 & 5.6 & 6.1 & 6.1 \\
\hline
  \noalign{\vskip 1pt}
\multicolumn{2}{l|}{Disconnected pixels [\#]}&
 135&3223 &589 &252 &313  &2580 & 1512  & 146 & 1420& 269 & 262 & 970  & 198 & 334 & 222  & 255 \\
\hline
    \end{tabular}
    }
    \label{tab:nisp_fpa}
\end{table*}

\begin{figure}[!hb]
    \centering
    \includegraphics[width=\columnwidth, viewport=48 34 801 507, clip]{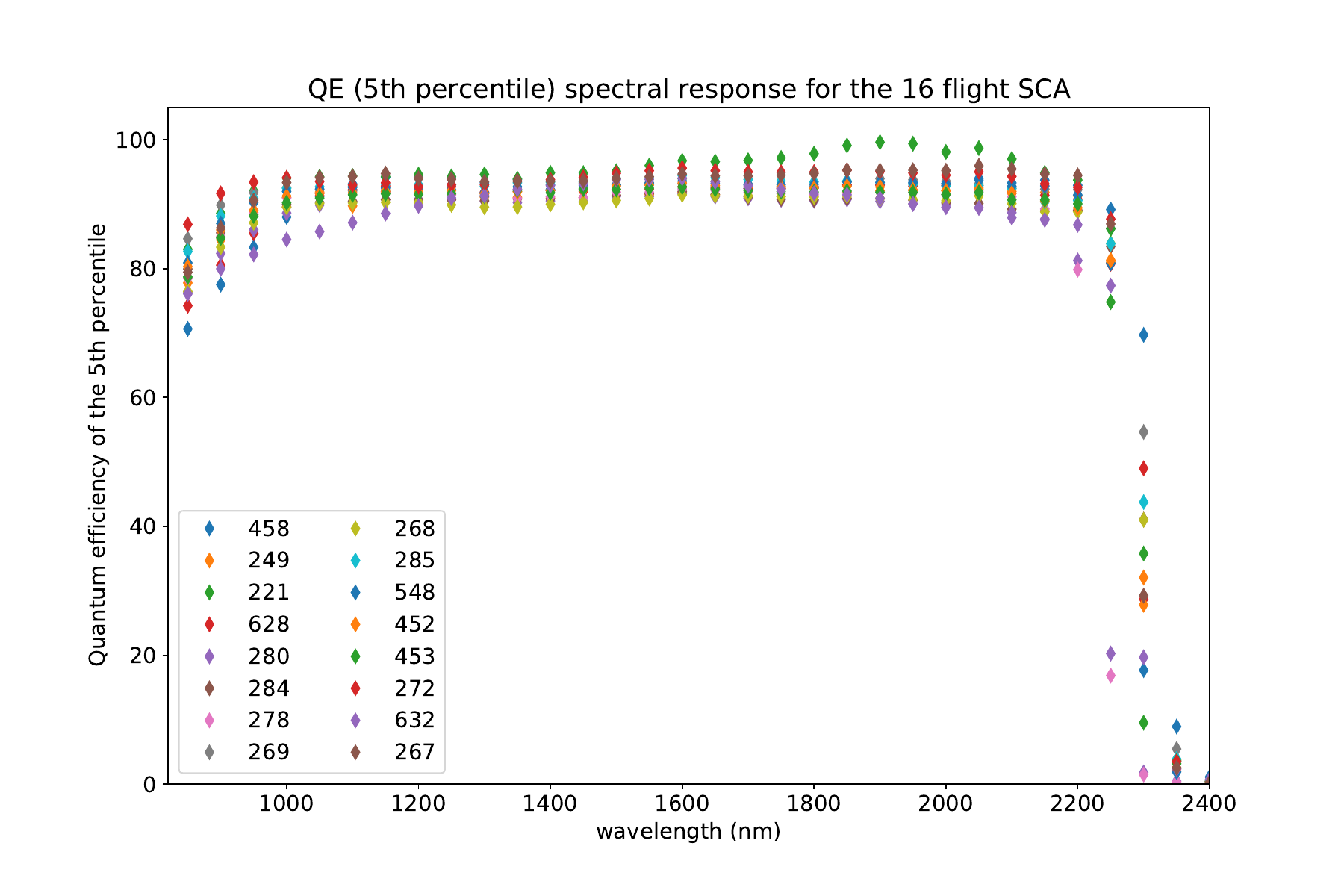}\\[3mm]
    \includegraphics[width=\columnwidth, viewport=-8 0 746 485, clip]{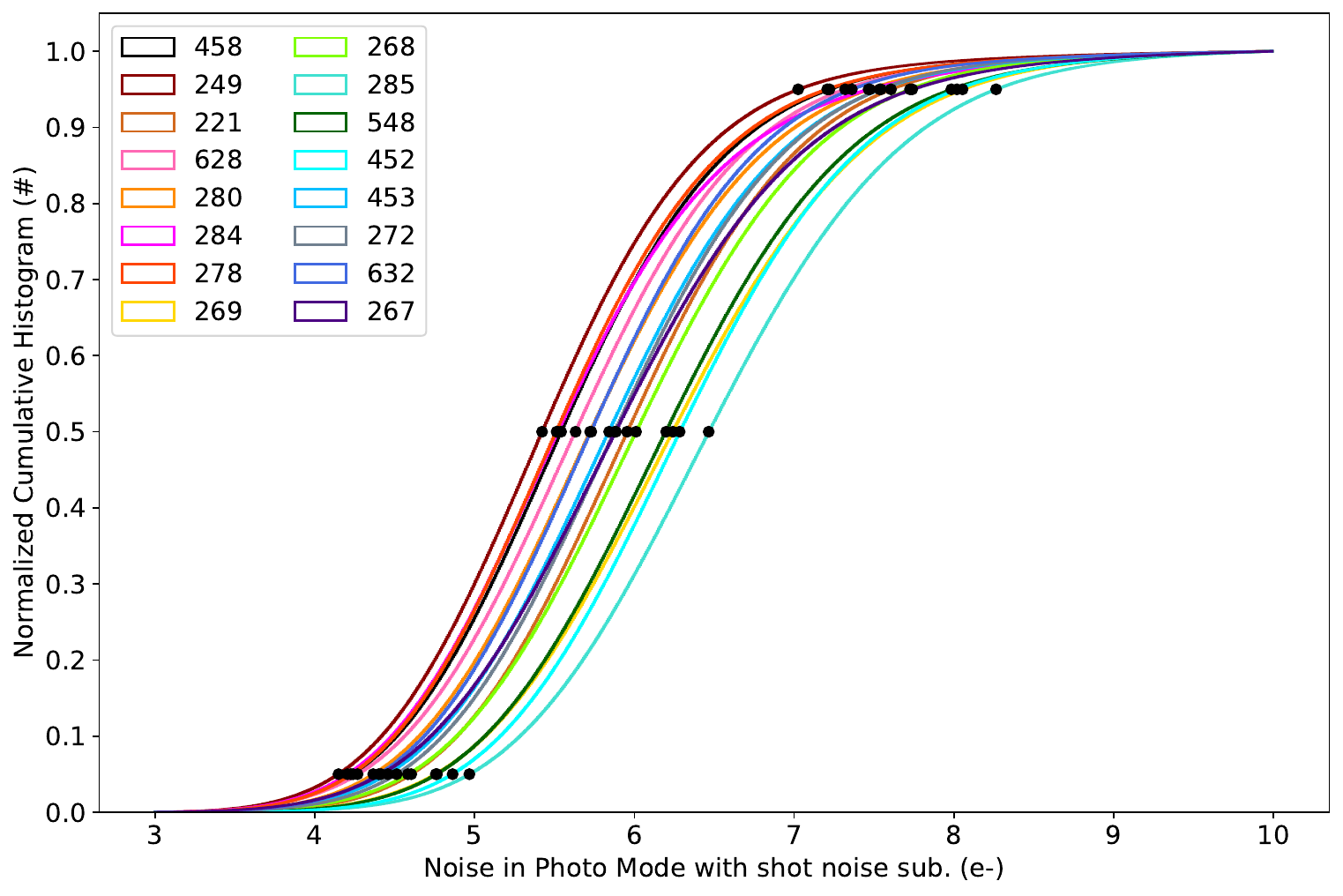}\\[3mm]  
    \includegraphics[width=\columnwidth, viewport=-8 0 746 485, clip]{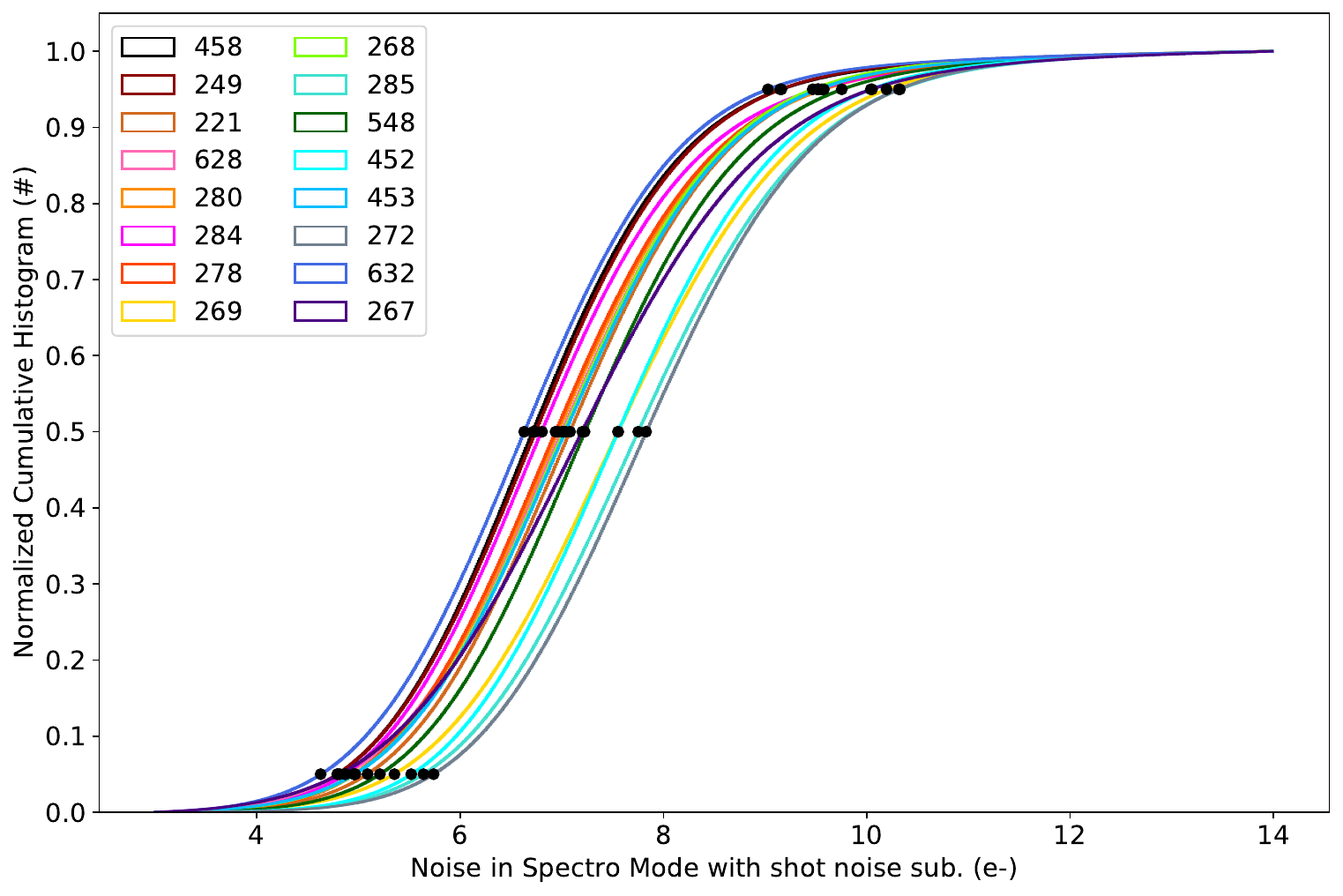}
    \caption{{\em Top:} Wavelength-dependent \ac{QE} of the 16 flight \ac{SCA}, 5-percentile values measured at NASA (95\% of pixels have at least this \ac{QE}). {\em Centre+bottom:} Noise cumulative distribution of the 16 flight triplet \ac{SCS}. Black points are the 5, 50, and 95 percentiles of the cumulative distribution. \ac{SCS} and \ac{SCS} are identified by their \ac{SCA} number 18XXX as in Table~\ref{tab:nisp_fpa}. }
    \label{fig:NI-SCS-QE-Noise}
\end{figure}

Figure~\ref{fig:NI-SCS-QE-Noise} shows the measured noise properties as well as the wavelength-dependent \ac{QE}, measured at NASA JPL as part of a special characterisation campaign. 
The top panel shows the excellent \ac{QE} spectral response against a cold calibrated photodiode, reported on the level reached by 95\% of pixels. 
The centre and bottom panel show the cumulative distribution of the spectro and photo noise across all detector pixels for the 16 selected flight \ac{SCS}, obtained during ground testing at \ac{LAM}. 

With respect to other space missions the 2.3$\,$\micron\ sensitivity cutoff reduces some of the thermal background that would be incurred with the more common 2.5$\,$\micron\ cutoff \ac{SCA} used for example in the \acs{JWST} NIRCam. Aside from the near-optimal \ac{QE} -- at 1$\,$\micron\ slightly better than NIRCam -- and low noise properties, what stands out about the NISP \ac{FPA} is the sheer number of detectors needed to populate the NISP focal plane. The 16 flight detectors, and the batch of 29 detector they were selected from, were drawn from a much larger set of H2RG that were actually manufactured, but where many did not pass the rigorous requirements set for \Euclid by ESA, NASA, and the NISP development team. 

The overall downselection and subsequent characterisation were only possible by a substantial and automated unattended testing effort with streamlined data analysis. The characterisation campaigns performed at the \ac{SCS} level are described in \citet{secroun2016} -- as mentioned above, they alone were a considerable effort with three months of testing in a row with two cryostats, each cryostat testing two \ac{SCA}s at a time. Nearly 1 PB of ground characterisation data were taken for the 16 flight and four spare \ac{SCS}, which facilitates a deep understanding of various detector effects.

\subsection{Calibration source}
\label{sc:NICU}
The NISP calibration concept (see Sect.~\ref{sec:cal}) requires the ability to measure the detector response and linearity pixel-by-pixel both in the integrated instrument on ground as well as in-flight.
For this purpose the instrument has the \ac{NI-CU}, a calibration lamp that directly illuminates the \ac{FPA} with one out of five custom-manufactured \acp{LED}. Their peak-emission wavelengths are distributed between 940 and 1870\,nm \citep[at an operating temperature of 135\,K; full description in][]{hormuth2024}, with their emission characteristics shown in the upper panel of Fig.~\ref{fig:trans}.

Each of the \acp{LED} has a comparatively narrow emission spectrum over a well-defined wavelength range. Their advantage over traditional incandescent tungsten lamps is that that their spectral temperature-dependence is extremely small: for example, NISP's nonlinearity calibration requires the capability to shine light at different fluences onto the detector. To use a single tungsten lamp for various brightness levels, one would vary the current flowing through the filament, which would  dissipate more heat, change its temperature, and therefore would have shifted the central black-body wavelength. This is not the case with \acp{LED}, where the wavelength is solely given by the semiconductor band gap, as opposed to blackbody emission. There is a negligibly small temperature dependence of the emitted spectrum, much smaller than for a tungsten element. At the same time operating the \acp{LED} leads to minimal energy dissipation facilitating a stable thermal equilibrium.

\begin{figure}[hb]
    \centering
    \includegraphics[height=0.5\columnwidth, angle=90]{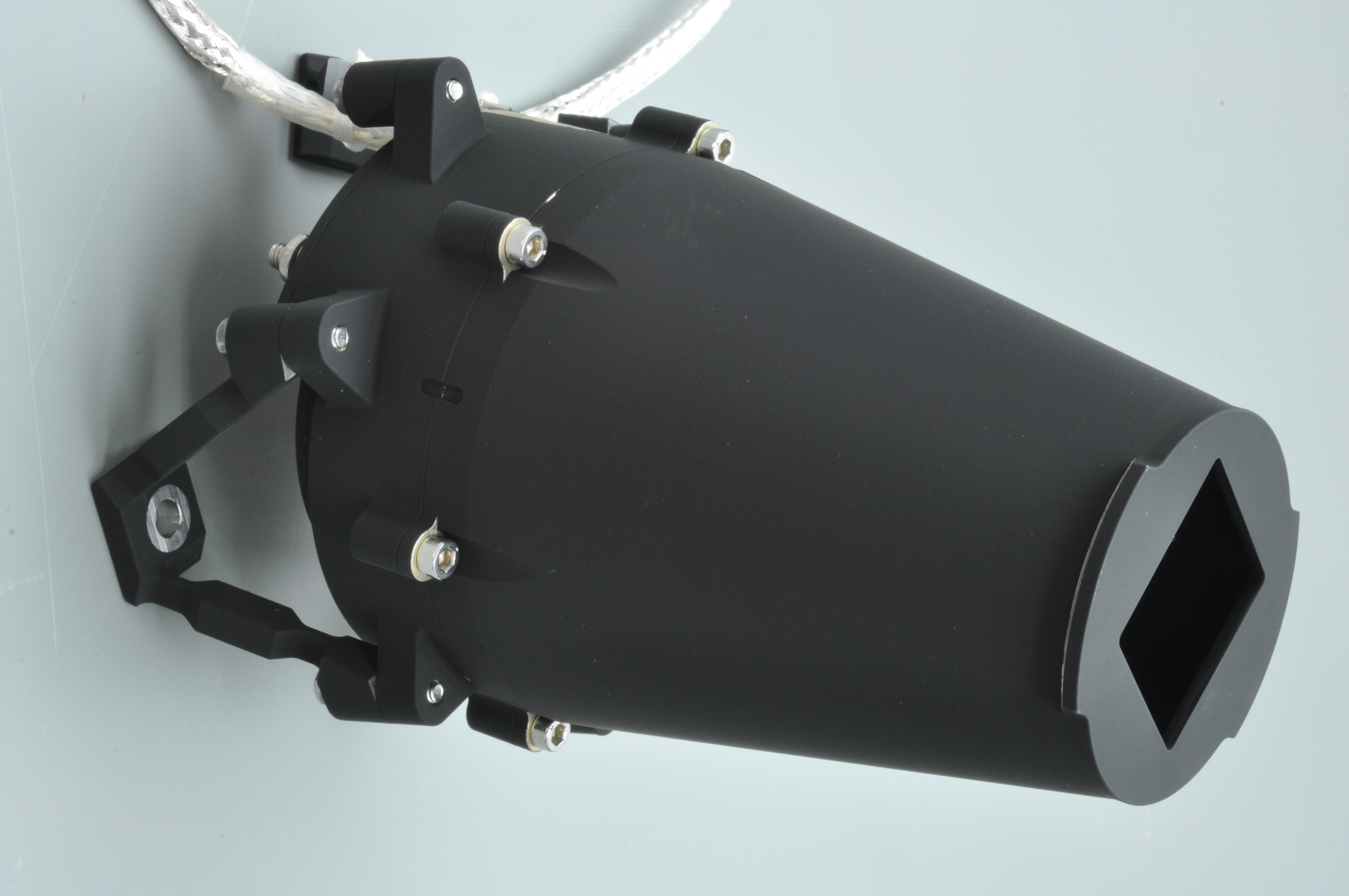}
    \caption{Flight model of the calibration source \ac{NI-CU} before integration into NISP. Two supply harnesses can be seen, feeding the nominal and redundant side, respectively, of \ac{NI-CU} with each 5  \acp{LED} emitting at different wavelengths. Inside the base any \ac{LED} illuminates the same \ac{PTFE} reflector patch that in turn directly illuminates the \ac{FPA}, with a field shaped by a set of five baffles, the last one visible at the top. The full \ac{NI-CU} height is approximately 150\,mm.}
    \label{fig:NICU}
\end{figure}

Inside \ac{NI-CU}, at each point in time any one of the five \ac{LED} can be operated, by the same power source in the warm electronics (Sect.~\ref{sec:warmelectronics}), but not several  \acp{LED} simultaneously. When operated, an  \ac{LED} illuminates a common central 3\,mm\,$\times$\,3.5\,mm oval reflective patch of space-grade Spectralon, a porous \ac{PTFE} plastic. Spectralon is a commonly used, near-perfect diffusive Lambertian reflector, owed to multiple internal scattering of light inside its several millimetre high body. The reflector patch in turn directly illuminates the \ac{FPA} with a field footprint shaped by five consecutive baffles along \ac{NI-CU} (Fig.~\ref{fig:NICU}).

\ac{NI-CU} is mounted off-axis, next to CoLA (see Fig.~\ref{fig:NISP_FM}), that is it illuminates the \ac{FPA} directly, bypassing all optical elements including the filters. Hence \ac{NI-CU}'s purpose is not to calibrate the NISP \YE, \JE, and \HE\ passbands, but solely the detector response. A mapping to detector properties within a specific passband takes place by combining calibration images from one or several \acp{LED} with the \ac{QE} maps obtained on ground (top panel of Fig.~\ref{fig:NI-SCS-QE-Noise}).

The combination of \Euclid's off-axis optical system and \ac{NI-CU}'s placement in the instrument results in an angle between the axis from NI-CU to the \ac{FPA} and the \ac{FPA} surface normal of {$\sim$\,\ang{12;;}}. To improve uniformity in the illumination of the \ac{FPA}, the Spectralon patch is tilted by $\sim$\,\ang{30;;} with respect to NI-CU's central axis, tilting the Lambertian emission cone. The resulting flux homogeneity across the \ac{FPA} is $\sim$\,12\%, providing similar flux and thus \ac{SNR} for all pixels. Any large-scale variations that might be imprinted on the data by \ac{NI-CU} or the telescope optics itself are captured by a photometric flat calculated from dithered on-sky data (Sect.~\ref{sec:cal}).

\ac{NI-CU} contains two sets of  \acp{LED}, one nominal, one redundant, connected to the corresponding sides of the NISP warm electronics power supply and control units (Sect.~\ref{sec:warmelectronics}). They are operated in a pulsed mode with a constant frequency but variable pulse width (\aclu{PWM}, \ac{PWM}), and variable current. The pulse width can be varied between 5\% and 50\% duty cycle, the current between typically 10\,mA and maximally 100\,mA, providing a dynamic range of $>$\,100 in fluence yet safely away from the upper operational limits that might reduce the  \acp{LED} lifetime, and away from very small currents and duty cycles that are difficult to drive while guaranteeing stability. As a result \ac{NI-CU} can provide very stable illumination levels over typical exposures of at least $\sim$\,100\,s, at fluence rates between $\sim$\,15\,ph$^{-1}$\,s$^{-1}$\,pixel$^{-1}$ and $>$\,1500\,ph$^{-1}$\,s$^{-1}$\,pixel$^{-1}$, for all  \acp{LED}. All but one \ac{LED} channel has a stability rating for drive currents $\ge$\,10\,mA for linearity calibration to less than 0.2\% linear drift and remaining $\sim$\,0.1\% RMS after removal of the linear drift component over 1200\,s. \ac{NI-CU} can and is being used with a current as low as 1\,mA, but at the price of lower stability, below a current of 10\,mA the temporal drift of the power supply and hence \ac{LED} fluence can increase proportionally.

Even though \ac{NI-CU}'s illumination levels and illumination pattern will be quite stable over time, due to the mechanical and optical design, a temporally perfectly fixed illumination pattern on the \ac{FPA} and a fixed absolute flux per given \ac{PWM} and current setting is not guaranteed, and illumination patterns can in principle show small differences between the five  \acp{LED}. NISP's calibration approach is designed correspondingly, assuming varying photon fluences and variable illumination patterns.\footnote{Absolute calibration lamps with perfectly known fluence levels at each time from internal information are a chicken-and-egg problem, requiring calibration of the fluence levels with an extra device like a photodiode, which itself needs a calibration mechanism. NISP has therefore decoupled small-scale spatial calibration (LED flatfields), large-scale calibration (on-sky data), and absolute calibration (celestial reference source).}

\subsection{Warm electronics}
\label{sec:warmelectronics}

\begin{figure}[htb]
    \centering
    \includegraphics[width=1\hsize, angle=0]{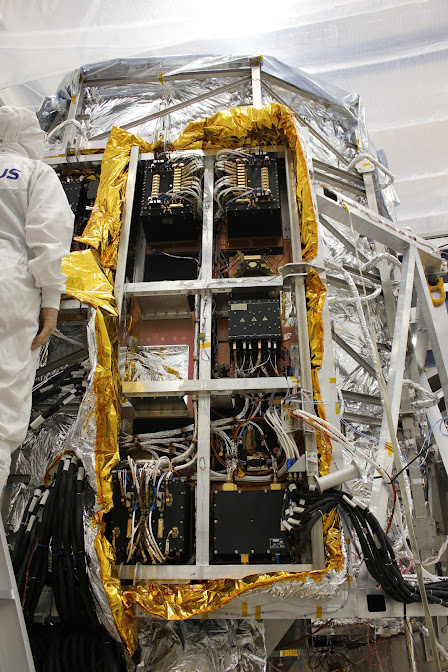}
    \caption{NISP Warm Electronics: the identical \acp{DPU} can be seen on the top of the picture, while the \ac{ICU} is located at the centre (right) of the so-called Spacecraft Y panel. Image courtesy Airbus Defence and Space}
    \label{fig:NIWE}
\end{figure}

The NISP warm electronics is a system composed of the \ac{ICU} and two identical \acp{DPU}, operating at $\sim$\,293\,K. The \acp{DPU} handle the raw-data acquisition of the NISP \ac{FPA} -- the exposure start times of the 16 detectors are synchronised to better than 10\,ns -- the onboard processing, the compression, and the transmission to the spacecraft's \ac{MMU}. The warm electronics units in their flight configuration are shown in Fig.~\ref{fig:NIWE}.

Each \ac{DPU} hosts eight \acp{DCU} that pre-process independently and in parallel the \acp{SCE}' live data stream; three radiation-tolerant class-S CPU boards (Maxwell$^{TM}$ 3$\times$SCS750F$\textsuperscript{\textregistered}$-PPC) with a 400 MHz clock error detection running the VxWorks 5.1 real-time operative system; memory banks that consist of error-logging SDRAM of 256\,MB (Reed-Solomon protected), non-volatile E2PROM of 8\,MB (Error Correction Code protected), a transit data buffer board of 6\,GB, and a 127\,MB Space Wire router 
board; and a power-supply board.

All boards inside each \ac{DPU} -- besides the \acp{DCU} -- have a `cold'-redundant counterpart, and three \acl{CPU} boards. The latter work jointly but in case of a failure the \ac{DPU} could also work with only two of three. Each \ac{DPU} is equipped with two communication interfaces: a 1553 MILBUS I/F for receiving telecommands with a maximum rate of 1\,Hz ($\sim$\,512\,bit) and transmitting telemetry with a maximum rate of 40\,Hz ($\sim$\,18\,kbit) with the \ac{ICU}, and a Space Wire I/F for the science data stream, internally connecting with the \acp{DCU}, and externally with \Euclid's \ac{MMU} \citep{10.1117/12.2561530}. 

\subsubsection{Command structure}
\label{sec:we_structure}

The NISP warm electronics comprises two different systems that are located in the warm environment of the \ac{SVM} -- the \ac{ICU} for main control, and the \acp{DPU} for data handling. The \ac{ICU} together with its on-board application software orchestrates the NISP operations, controls the NISP wheels and the calibration-unit electronics (Sect.~\ref{sc:NICU}), and regulates the heating elements within NISP based on the readings of various thermal sensors (Sect.~\ref{sec:thermal}). The \ac{ICU} interfaces with the two \acp{DPU} that each synchronously command half of the focal plane. It also provides the main commanding interface with the satellite, receiving telecommands from and sending telemetry data to the spacecraft. Data from the \acp{FPA} flows to the \acp{DCU}, then to the \acp{DPU} for processing and compression, and from there to the \ac{MMU}. The \ac{ICU} has two identical sections, one nominal and one redundant, connected with the corresponding versions of the wheel motors, \ac{NI-CU} \acp{LED}, and NISP thermal control. Either one can be operated at a time, following a full `cold' redundancy scheme where always one section is fully `off' and has to be explicitly activated when needed, by switching off one side and starting the other.

\subsubsection{Detector read-out and processing}
\label{sec:we_processing}

NISP detectors have a nominal frame rate of 1.45408\,s. While \Euclid already has a large data-downlink budget, it still does not permit to downlink every observed frame. NISP's standard \ac{ROS}, that is used in both the Wide and Deep Surveys, consists of the repetition of four dithers, each made of one spectroscopic exposure of about 549.6\,s of integration time followed by three consecutive photometric exposures of 87.2\,s integration time each \citep[Table~\ref{tab:macc}; full description in][]{scaramella2022}. For NISP transmitting each read frame would represent an accumulated data volume of $\simeq$\,20\,Tbit per day -- factors of many beyond NISP's available downlink budget of 290\,Gbit/day for NISP. 

To reduce the amount of data to an acceptable volume while minimising the exposure read-noise, the \acp{DPU} perform on-board data processing to estimate photons -- or rather charge -- accumulated in the pixels during each exposure \citep[see][]{2016SPIE.9904E..5RB}. For this purpose NISP detectors acquire data through a \ac{MACC} sampling scheme, using non-destructive readout. A number of groups $g$ each consists of several consecutively read-out frames $r$ that are averaged by the \ac{DCU}. Between groups, a number $d$ of frames is `dropped' -- that is not read out -- while integration of photons continues. The \ac{MACC} parameters used in the \ac{ROS} for the spectroscopy and photometry channels are shown in Table~\ref{tab:macc}. 

\begin{table}[th]
    \caption{MACC($g$,$r$,$d$) mode parameters for spectroscopy and photometry in the \acf{ROS}. Here $g$ is the number of groups in the slope fit, $r$ the number of reads per group, and $d$ the number of drops after each but the last group. Both grisms use the same \ac{MACC} settings, as do all three photometry bands. For both \ac{MACC} modes the integration time $t_\mathrm{int}$ (effective time accumulating light; the time relevant for science) and exposure time $t_\mathrm{exp}$ (total duration of frame including one reset frame before start of charge integration; the time relevant for exposure planning) are given, using the frame duration of 1.45408\,s.}
    \centering
    \begin{tabular}{l|ccc|cc}
        \hline\hline
          \noalign{\vskip 1pt}
 & $g$&$r$&$d$&$t_\mathrm{int}$\,[s]& $t_\mathrm{exp}$\,[s] \\
\hline\noalign{\vskip 1pt}
Spectroscopy &15 & 16& 11& 549.6 &574.4\\
Photometry & 4 & 16& 4& 87.2 & 112.0\\
         \hline
    \end{tabular}
    \label{tab:macc}
\end{table}

At the end of the exposure, the \ac{DPU} corrects each averaged group for the detector baseline offset (Sect.~\ref{sec:cal_baseline}) based on the group average of the un-illuminated reference pixels. It then estimates for each pixel a signal slope through a straight-line least-square fit to the averaged group differences, that is the signal between consecutive groups, which optimally reduces the read-out and shot noise \citep{kubik2016}. This value -- with a fixed digital offset of 1024\,ADU added -- is downlinked to ground.

In addition to estimating the accumulated signal, the \ac{DPU} evaluates a statistical quality factor of the fit \citep[see][]{kubik2016}. In case of the spectroscopic mode this quality factor is an 8-bit $\chi^2$ integer value, and for photometry mode it is a 1-bit flag. It provides feedback whether a pixel had a truly linear response during the exposure, enabling the on-ground processing pipeline to detect spurious events like cosmic-ray hits or random telegraph noise \citep{kohley2018}, as well as pixels often characterised by high  non-linearity. 

The processing times for photometric and spectroscopic exposures were evaluated from ground test campaigns to be of the order of 7\,s and 14\,s per detector, respectively \citep{2022SPIE12180E..1LM}, where each \ac{DPU} processes 8 detectors in sequence.

After onboard processing, both the science data and quality-factor data are compressed with a loss-less Rice compression algorithm\footnote{The {\tt RiceComp} routine from NASA's {\tt CFITSIO} library(5) is available at \url{https://heasarc.gsfc.nasa.gov/fitsio/fitsio.html}} \citep{Rice1971.1090789,2009PASP..121..414P} adapted to the \ac{DPU} application software by enforcing the required coding standards and test procedures. Data are then packaged with headers and detector telemetry to be sent to the \ac{MMU}. The compression factor for the science images varies from 2.8 to 3.4 depending on the exposure mode. The mean size of a nominal NISP dither with one NISP-S and three NISP-P exposures is about 378\,MB, or about 31\,$\pm$\,3\,GB/day for 20 \ac{ROS} pointings of four dithers each. These 248\,Gbit/day fit with some margin into the NISP's allocated downlink data volume of 290\,Gbit/day.

\subsubsection{NISP Thermal Control}
\label{sec:thermal}

The thermal design of NISP is based on a double passive radiator system to provide the main temperature references for the NISP opto-mechanical system and detector system.
The first radiator provides a $\sim$\,130\,K heat-sink for the structure, optics, cryo-motors, and the \ac{SCE} cold electronics. The second colder radiator sets the reference temperature for the infrared detectors at about 95\,K. 
The thermal interfaces have been designed to achieve a thermal stability of $\pm$2\,K throughout the mission, allowing to stabilise the optics to much better than that. Specifically for the detector radiator interface, an even stricter requirement is in place, mandating a temperature fluctuation of less than $\pm$0.1\,K over a period of 3 hours, and less than $\pm$4\,mK over 1210\,s, about the duration of a nominal dither. From a radiative perspective, NISP is situated within the Euclid instrument's cavity, where temperatures range from 130\,K to 135\,K. This environment is characterised by surfaces all having an emissivity greater than 0.9, a specification driven by optical requirements.

To achieve the designated thermal performance, the NISP thermal management system is made up of a set of thermal sensors, a heating system, and an \ac{MLI} blanket. The thermal sensors use 12 PT100 thermistors, six located on the mechanical structure, and each two on the cryo-motors, the SCE thermal straps, and on the \ac{FPA}. The thermistors are read by the ICU, and their value reported in the NISP housekeeping telemetry. In addition to thermal sensors the system contains two heaters, installed near the optics on the NISP structure, functioning in an open-loop configuration to supply the necessary power to maintain the target temperature. Finally, the instrument is wrapped in a six-layer, black-coated \ac{MLI} blanket to isolate the inner instrument from the external radiative environment.

Preliminary data, collected during the Euclid in-flight performance verification phase, indicate that NISP temperatures can be stable at the level of 2\,mK (RMS) for the optics and 1\,mK for the NISP focal plane during steady-state operations.

\subsubsection{Contingency and recovery}
\label{sc:contingency}

NISP has multiple levels of fault detection encoded in its warm electronics, to take safety measures autonomously when needed in order to protect the instrument at all times and in all situations. The \ac{ICU} continuously monitors instrument telemetry data including telemetry from the detectors sent by the \ac{DPU}. In case of deviations with respect to expectations it triggers the instrument-level \ac{FDIR}, changing NISP's operation state depending on the severity and location of the incident.

The \ac{DPU} application software again monitors any potential \ac{DPU} anomalies, including all \ac{DPU} subsystems and the NISP \ac{FPA}. In case of a detected anomaly it can change the instrument operational mode to {\em SAFE}, powering off the \ac{FPA}, or {\em PARKED}, disabling the \ac{FPA} commanding capability but leaving it powered on. These operations are done autonomously and independently by each \ac{DPU}, while signalling the anomalies to the \ac{ICU} through alarms and housekeeping parameters. The \ac{ICU} transmits the anomaly status to the spacecraft and sets NISP's resulting operation mode according to the criticality of the anomaly.

The \acp{DPU} have all of the hardware failure modes, effects, and the specification of the criticality-analysis for all its subsystem-hardware encoded in its \ac{FDIR} system, and are able to consider each detector chain independently. The event signalling and isolation is performed by the application software. The \ac{FDIR} system covers anomalies in temperature, currents and voltages exceeding thresholds of all hardware components, as well as memory upsets. Also the main processes managed by the \ac{DPU} software implement several \ac{FDIR} checks based on programmable watchdog systems, used for the monitoring of the correct operations and especially timing of the dynamical processing chain. This means it constantly tests if the relative timing of image exposure, science data collection, science data processing, and science data product transmission to the \ac{MMU} work out.  
Additionally, as mentioned above, both the \ac{ICU} and \ac{DPU} have cold-redundancy counterparts. Thus we can independently switch from the nominal to the redundant side of the \ac{ICU} and \ac{DPU} chain in case anomalies are detected, at full functionality.


\section{\label{sec:cal}Calibration approach}

The instrumental and data-processing capabilities of NISP flow down from the required \ac{FOV} and depth. Together, they facilitate an efficient coverage of the targeted 14\,000\,$\deg^2$ down to 24.0\,AB\,mag  -- which led to the instrument design as just described -- but at the same time set requirements on accuracy. 
At a higher level these requirements differ for the two NISP channels, but also affect the same properties, for example in the common NISP detector system.

For NISP-P the main requirement is a relative photometric accuracy of 1.5\% in the fully processed and calibrated data. NISP-S must achieve a relative spectrophotometric accuracy of 0.7\%, and a wavelength accuracy of 5\,\AA~or 0.3 NISP pixel anywhere in the \ac{FPA}. Both channels need accurate distortion correction on the order of \ang{;;0.1}: first, for the unambiguous matching with the VIS astrometric reference system, which in turn is based on Gaia \citep{gaia2016}; second, for the decontamination of the overlapping spectra orders in the slitless spectroscopy exposures. 

A total of 31 calibration products are defined for NISP that are maintained during flight, in addition to further calibration and characterisation files -- such as \ac{QE} -- that could only be generated on ground. Calibrating the VIS and NISP instruments and the telescope is a complex task with a substantial number of dependencies and limitations set by (i) the spacecraft, (ii) the instruments' hardware, (iii) the instruments' on-board processing capabilities, (iv) the data-downlink volume, and (v) timescales of spacecraft degradation by space weathering and molecular outgassing \citep{schirmer2023}.

Within the scope of this paper we provide an overview of the principal effects and the calibration strategies. The associated calibration products can be divided into three categories: those that are common to both channels, and those that are individual to NISP-P and NISP-S. Calibration products were updated during the initial \ac{PV} phase that followed commissioning and extended over three months. Those calibration products that we consider to likely evolve over time are monitored with appropriate cadences. The stability of the other calibration products can be assessed with regular in-flight science and calibration data, and corrective actions taken if necessary. Concrete results from the \ac{PV} phase will be reported elsewhere.

\subsection{Calibration products common to NISP-P and NISP-S\label{sec:cal_common}}
Common calibration products apply mostly to the detector level and the read-out electronics. They are independent of the illumination provided by the NISP optics. The following calibration products are either used to monitor evolution of detector properties, or are applied to data in the \Euclid \ac{SGS} data pipeline.

\subsubsection{Detector baseline\label{sec:cal_baseline}}
The baseline is a bias value set for each pixel after reset. Reference and science pixels have respective pedestals of about 5.5 and 10\,kADU, to ensure that the pixels operate in the linear range of the analogue-digital chain and to encompass the large variation across pixels. The available dynamic range is reduced accordingly. The baseline has a temperature dependence of about $-120$\,ADU\,K$^{-1}$, was redetermined for each pixel during commissioning once the detectors stabilised to their actual in-flight temperatures, and once more during \ac{PV}. The baseline is estimated from 512 reference pixels per read-out channel for every \ac{MACC} exposure and automatically removed by the onboard processing before the ramp-fitting process. Baseline instabilities are monitored using regular dark and flat-field calibration frames.

\subsubsection{Dark current\label{sec:cal_dark}}
The dark current for the NISP detectors is very low with $\sim$\,0.02\,e$^-$\,s$^{-1}$\,pixel$^{-1}$, requiring hundreds of exposures for a high-\ac{SNR} measurement. The goal of the dark maps is to characterise dark current levels and to identify warm and hot pixels that need to be masked. Dark maps were obtained during \ac{PV} at a time when persistence charges  (Sect.~\ref{sec:cal_persistence}) had decayed, and are subtracted from the science data in the pipeline.

Monitoring of hot pixels is provided by so-called `slew darks', one of which is taken every 1.5\,h during the slews from one survey field to the next. Slew darks are affected by persistence charge from the previous exposure history.

\subsubsection{Detector cross-talk\label{sec:cal_crosstalk}}
Cross-talk is caused within the detector as well as the detector electronics. The amplitudes inferred on ground are on the order of 0.001\%, meaning that cross-talk is detectable only for saturated field stars. Intra-chip cross-talk between readout-channels, as well as cross-talk between detectors, are determined from in-flight observations of bright stars. Victim pixels are masked if the cross-talk flux violates the relative photometric accuracy requirement.

\subsubsection{Interpixel capacitance (IPC)\label{sec:cal_ipc}}
\acsu{IPC} is a localised form of electronic cross-talk that is independent of wavelength. Prior to read-out, it spreads charge from the pixel in which it was collected to the adjacent pixels because of the capacitive coupling between pixels. \ac{IPC} broadens the \ac{PSF} and smooths the Poisson noise, and hence must be corrected before determining for example the gain. The \ac{IPC} is expected to change very little over time. It is temperature-dependent, though, and was therefore  determined once the detectors reached their stable in-flight temperature \citep[for more details see][]{legraet2022}. It can be evaluated by measuring the cross-talk when charge is artificially injected in a pre-defined pattern of pixels, enabled by a special calibration feature implemented in the NISP detectors.

\subsubsection{Persistence\label{sec:cal_persistence}}
Persistence is an achromatic effect caused by charge traps in the detector. The traps remove photo-electrons from the readout of the current exposure and release them over time in an exponentially decaying manner, leading to ghost images in subsequent exposures \citep{smith2008}. NISP ground and in-flight data show visible persistence signals from previous observations. This is applies most visibly to spectroscopy traces in subsequent photometry images, but also vice versa. The amount of persistence depends on the number density and types of charge traps, its illumination history, and it is also considerably different below and above full-well saturation. We adopt a simple persistence model where each pixel is characterised by two time constants representative for short- and long-duration traps, and a persistence amplitude that scales with the local trap density. Based on the fluence of a pixel in an exposure the model then determines for how long -- that is for how many subsequent exposures -- that pixel needs to be masked; a time that can easily exceed several hours.

A more complex persistence model would account for the illumination history, that is bright objects falling on the same pixel in several exposures. Whether that pixel would be masked in a future exposure could then also depend on whether it falls on a bright or a faint source, the latter being relatively more affected. This will be considered at a later time when more experience is available with the in-flight system. We might consider actual pixel-level correction rather than masking, depending on whether high-quality persistence models can be built.

We assume that the time constants and amplitudes are stable over time as the detectors are radiation-hard and receive a total dose of only 2.5\,krad during the mission. Therefore, persistence is characterised during the PV phase by flat-field illumination followed by several hours of short- and long dark exposures, capturing both time constants as well as amplitudes. The persistence model can be validated with every single survey exposure.

\subsubsection{LED flats\label{sec:cal_led}}
The NISP detectors can be illuminated directly with the five \acp{LED} in the NI-CU calibration source, bypassing all optical elements. These \ac{LED} images -- or lamp flats -- provide a nearly monochromatic illumination that is uniform to better than 0.2\% on spatial scales of $\le$\,100\,pixel. They are taken on a monthly basis to compute effective flat fields for the NISP-P and NISP-S modes (Sects.~\ref{sec:cal_flat_nispp} and \ref{sec:cal_flat_nisps}).

\subsubsection{Brighter-fatter effect\label{sec:cal_bfe}}
The \ac{BFE} is caused by the increasing electrostatic field in a pixel as it accumulates photo-electrons. The changing field strength shifts the effective pixel boundaries in the sense that newly created photo-electrons are pushed into neighbouring pixels. The \ac{FWHM} of stars, which would be independent of source flux for a perfect detector and only determined by the optics, therefore increases with brightness. The \ac{BFE} is measurable over large parts of the dynamic range, and it is not fully flux-conserving. The \ac{BFE} also affects the pixel variance in flat fields, and is characterised from \ac{LED} flats with different fluences. A more in-depth description can be found in \cite{plazas2018} and \cite{hirata2020}.

\subsubsection{Nonlinearity\label{sec:cal_nonlinearity}}
The measured signal in a H2RG pixel is increasingly nonlinear with increasing charge-filling level of that pixel, independent of wavelength. To calibrate this nonlinearity, we illuminate the detectors with the calibration lamp while obtaining \ac{MACC} ramps; the filter wheel is closed during these observations, so that only flux from the lamps is observed. Contrary to nominal science observations, where the slope-fit to the individual \ac{MACC} group images is obtained on-board and downlinked (Sect.~\ref{sec:warmelectronics}), we downlink the individual group images. Nonlinearity manifests as a deviation from the expected linear trend in the increasing fluence of the group images as they approach saturation. The data volume exceeds the nominal telemetry budget, so that this calibration has to be obtained for two detectors at a time instead of all 16 detectors simultaneously. 

The MACC(4,16,4) and MACC(15,16,11) modes for standard photometry and spectroscopy observations (Table~\ref{tab:macc}) provide four and 15 group images, respectively, meaning that the dynamic range would be sparsely sampled with just a single \ac{LED} flux level. Therefore we observe five different \ac{LED} flux levels, and obtain independent nonlinearity calibrations for both photometry and spectroscopy readout modes for every pixel. Nonlinearity calibrations are obtained on a monthly basis, with a full set achieved every four and six months for MACC(4,16,4) and MACC(15,16,11), respectively. We note that \ac{BFE} (Sect.~\ref{sec:cal_bfe}), IPC (Sect.~\ref{sec:cal_ipc}), persistence (Sect.~\ref{sec:cal_persistence}), and reciprocity failure (Sect.~\ref{sec:cal_reciprocity}) are also forms of nonlinearity, calibrated separately. 

\subsubsection{Reciprocity failure\label{sec:cal_reciprocity}}
Reciprocity failure is a flux-dependent nonlinearity caused by charge trapping; as such it is related to persistence (Sect.~\ref{sec:cal_persistence}). A brighter source will fill -- and thus passivate -- charge traps in a pixel more quickly than a fainter source, therefore resulting in earlier linear pixel response. We expect this flux nonlinearity to be independent of wavelength, an assumption to be verified with the \ac{PV} data.  There, we observed the same stellar field in different filters first without and then with increasingly higher additional flux from the calibration lamps. The measured fluxes of the same sources without and with additional \ac{LED} background then carry the reciprocity-failure signal. This means we can only sparsely sample the detectors, and only an average correction can be achieved instead of a pixel-based correction such as in Sect.~\ref{sec:cal_nonlinearity}. During the nominal mission, reciprocity failure will be calibrated on a yearly basis as we expect only a slow evolution of the charge-trap density due to radiation damage (Sect.~\ref{sec:cal_persistence}).

\subsection{Calibration products specific to NISP-P}
\subsubsection{Absolute flux calibration\label{sec:cal_absphot_nispp}}
For absolute flux calibration we observed the \ac{WD} standard star J175318+644502 ($\JE=18.56$\,AB\,mag) in the self-calibration field. This star was identified by us using available multi-colour photometry and long-term photometric monitoring, using among others Pan-STARRS, Gaia, and the Zwicky Transient Factory (ZTF). The star was spectroscopically confirmed as a \ac{WD} by us using the Gemini Multi-Object Spectrograph (GMOS). We excluded considerable short-term variability on time-scales of about 100\,s--1000\,s using fast photometry with GMOS. Accordingly, this star is stable to better than 1\% from time scales of 45\,s to several years. The data we have collected for various \Euclid spectrophotometric WD standard stars will be discussed in a different paper.

The WD was placed on 16 positions per detector. To establish the absolute flux scale we obtained accurate \HST (HST) imaging and spectroscopic flux measurements of this star -- and several others which will be observed periodically in the \Euclid Deep Fields as validation -- with the Wide-Field Camera 3 (WFC3) and Space Telescope Imaging Spectrograph (STIS) at optical and near-infrared wavelengths (Proposal 16702, cycle 29: PI.\ P.~Appleton, and Proposal 17442, cycle 31, PI: S.~Deustua). Hence the NISP-P absolute photometry is based on a very well-understood photometric reference system.\footnote{The VIS instrument in \Euclid uses the same standard star.}

\subsubsection{Astrometry and optical distortions\label{sec:cal_astrometry_nispp}}
The astrometric solution is based on a VIS astrometric reference catalogue for every survey field, which in turn is based on highly reliable Gaia \citep{gaia2016} data. The absolute positions of NISP-P sources must be known to better than \ang{;;0.1}, which is easily achievable using the VIS/Gaia reference catalogue for every single exposure. The latter is important, as the filters carry optical power and the filter wheel has an intrinsic positioning RMS of about \ang{0.15;;}, leading to a nontrivial change in optical field distortion between exposures.

\subsubsection{Small-scale flats\label{sec:cal_flat_nispp}}
To correct for pixel-response non-uniformity (PRNU) in the \YE, \JE\, and \HE\ images we use a weighted combination of the various \ac{LED} flats (Sect.~\ref{sec:cal_led}). The weighting factors are determined by the \ac{SED} of the zodiacal background and the detectors' \ac{QE} maps. These flats will correct any PRNU effects to better than 0.2\%. However, since the \ac{LED} flats have a Lambertian profile, the \ac{FPA} corners receive about 10\% less flux than the centre of the \ac{FPA}. The \ac{ZP} in the flat-fielded images varies accordingly, which is corrected by additional photometric flats (Sect.~\ref{sec:cal_photflat_nispp}).

\subsubsection{Photometric flat\label{sec:cal_photflat_nispp}}
While \Euclid's optics are essentially free of vignetting, apodisation may cause additional spatial variations of several percent in the illumination. These will result in a change of the \ac{ZP} across the \ac{FPA}. By observing the self-calibration field with a special pattern of 60 dither positions we can measure how the fluxes of individual sources vary as they are imaged onto different parts of the \ac{FPA}. From this information we then compute a photometric flat \citep[see e.g.][]{holmes2012} to correct any large-scale illumination non-uniformities to better than 0.5\%.

\subsubsection{Straylight and ghosting\label{sec:cal_straylight}}
During PV phase we validated the straylight models by placing very bright stars (\JE\,$\sim$\,0--2\,AB\,mag) at increasingly smaller distances outside the \ac{FPA} corners and edges. We also measured the positions of ghost images caused by internal double reflections in the dichroic element and the filters, and built suitable models for masking. Ghost images are masked once they exceed a certain surface brightness.

\subsection{Calibration products specific to NISP-S}
\subsubsection{Absolute flux calibration\label{sec:cal_absphot_nisps}}
For absolute flux calibration during the PV phase, we observed the WD standard star GRW+70 (\JE=14.33\,AB\,mag) from the HST CALSPEC database on 5 positions per detector and per grism. Mean sensitivity functions -- one for the blue and four for each red grism/grism tilt combination -- are calculated based on the observations. The density of field sources around this star is fairly low so that clean slitless spectra that are not contaminated by other sources could be observed. Once the absolute flux scale is established, no more observations are required as any throughput variations are monitored by the monthly self-calibration observations. The absolute flux scale will be further validated by observations of other spectrophotometric standards over the course of the mission, as well as monitoring relative flux changes using the self-calibration field. 

\subsubsection{Dichroic ghost model\label{sec:cal_dichroic_nisps}}
As for NISP-P, the dichroic-ghost model must be validated for NISP-S, so that affected pixels can be masked. The mapping from bright source positions to ghost positions is given by a ray-tracing model and can be validated with every on-sky exposure.

\subsubsection{0-th order SED dependence\label{sec:cal_0th_SED}} 
The 0th order in the spectra is an important reference point for the wavelength solution, and it is double-peaked for two reasons. First, the prism in the grism has some dispersive power even in the 0th order, resulting in a 10 pixel long spectrum of the 0th order. Second, the grating is blazed for a specific wavelength, maximising the number of photons for that wavelength in the 1st order and minimising it in the 0th order. The double-peaked shape also has some field dependence. 

The reference 0th-order position for wavelength calibration is given by the mean of the two peak positions, which depend on the \ac{SED} of the source. The main impact of the source \ac{SED} is to change the ratio of the 2 peaks, and therefore the inferred mean value. The effect is expected to be at the level of one pixel and below. The calibration will be built up as the \acl{EWS} grows and sources with very different \acp{SED} are mapped at different positions in the focal plane.

\begin{figure*}[!ht]
    \centering
    \includegraphics[width=0.9\hsize]
{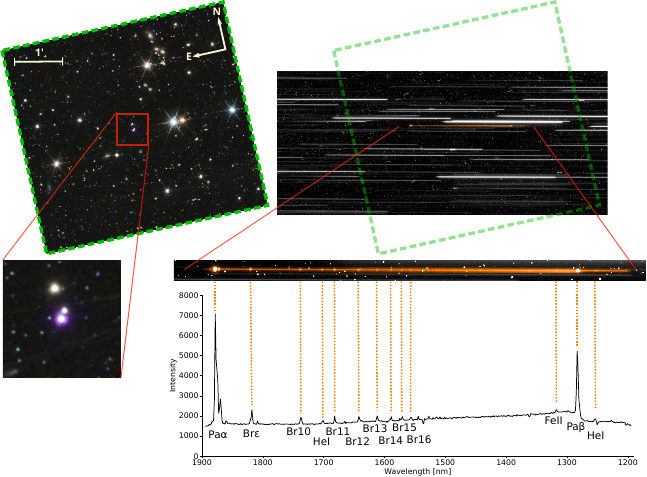}
    \caption{NISP-S grism slitless spectroscopy mode. {\em Left:} NISP-P colour-composite image of the planetary nebula (PN) SMC-SMP-20 \citep{paterson2023}, observed during the PV phase for wavelength calibration. The PN is located at the image centre; with a diameter of \ang{;;0.40} it appears unresolved. {\em Right:} Raw NISP-S 2D-spectrogram of the same area, the photometry field is sketched again with a green box. Each horizontal stripes is a spectrogram of a bright object. Small dots between these spectrograms are mostly cosmic rays hitting the detector. The PN can easily be identified in the 2D-spectrogram by the distinct emission lines above a weak continuum; it has been colourised for emphasis. The zoom below repeats the 2D-spectrum together with the extracted 1d-spectrum. Identified emission lines are marked. We note that the direction of the wavelength axis will differ on the sky depending on the specific grism used. The visible strong Pa$\beta$ emission in \YE-band and the Pa$\alpha$ line in the \HE-band lend a distinct purple colour to the PN and its prominent diffraction ring in the image on the left.}
    \label{fig:SMC-SMP-20}
\end{figure*}

\subsubsection{Wavelength calibration\label{sec:cal_wavelength}}
Determining the dispersion law of NISP-S involves a complex calibration during the PV phase consisting of several steps: (i) a mapping from the astrometric sky as defined by VIS to the 0th order position in the NISP-S dispersed images; (ii) an offset from the 0th order position to a reference wavelength in the 1st order; (iii) the spatial curvature of the spectral trace; and (iv) the nonlinear wavelength dispersion within the 1st order. The calibrated spectral distortion laws consist of fourth-degree Chebyshev polynomials to each of these mappings as a function of field position. The wavelength dispersion itself is measured using the emission lines of a point-like planetary nebula \citep[PN;][see also Fig.~\ref{fig:SMC-SMP-20}]{paterson2023}. We expect the wavelength calibration to be stable and will check it after one year into the survey by observing the PN once more. Furthermore, we transfer the PN-based wavelength solution to stellar absorption-line systems in the self-calibration field, so monthly consistency checks are possible.

\subsubsection{Flat-fielding\label{sec:cal_flat_nisps}}
Like NISP-P, the slitless spectroscopy channel must also be corrected for PRNU (Sect.~\ref{sec:cal_flat_nispp}) and large-scale illumination effects (Sect.~\ref{sec:cal_photflat_nispp}). The corresponding pixel-correction maps will be built in an analogous fashion. For each grism's transmission band a flatfield map will be built for three different wavelengths, to be subsequently combined.

\section{\label{sc:performance}NISP performance}

\begin{figure*}[ht]
    \centering
    \includegraphics[width=0.9\textwidth]{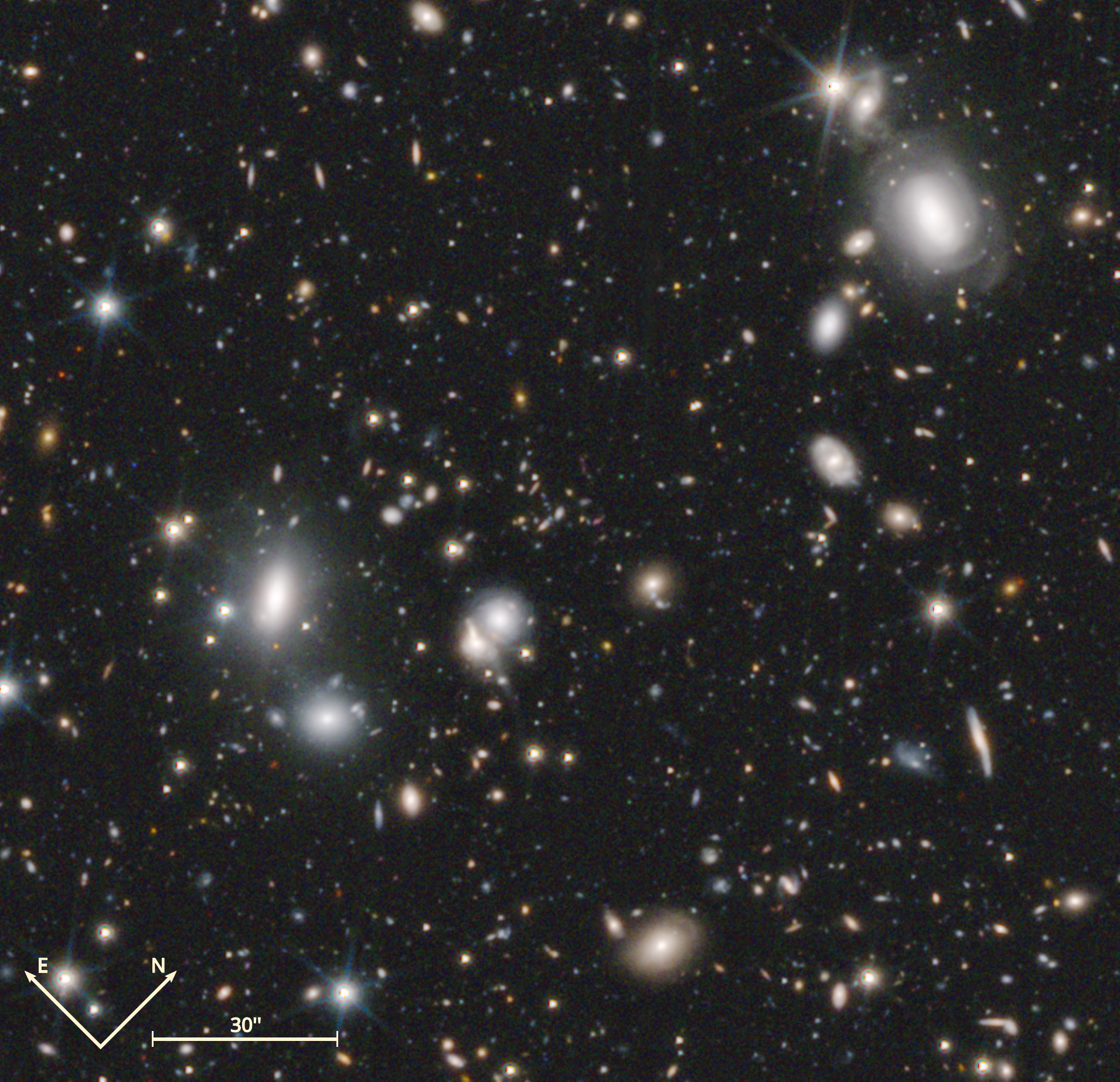}
   \caption{ 
Quality of the NISP imaging channel. Shown is a $\ang{;3.0;}\times\ang{;3.0;}$ wide area in the self-calibration field. The depth corresponds to that of the Euclid Deep Survey, that is about 26.3\,AB\,mag for $5\sigma$ point sources, achieved in 3.7\,h integration time per filter. For this composite we mapped the \HE, \JE, and \YE-passbands to RGB colour channels. (Near) saturated stars appear with characteristic spikes and diffraction rings differing in size as a function of wavelength. The Euclid Deep Survey will in total cover about 53\,deg$^2$, more than 20\,000$\times$ the area shown here.
    }
    \label{fig:P_examples}
\end{figure*}

\Euclid was launched on 1 July 2023 and during the initial commissioning and performance verification phases a first assessment of the instrument's performance was carried out. Overall the pre-launch predictions and expectations are met or exceeded. 
NISP is producing excellent data in both channels, as can be seen for NISP-S spectroscopy in Fig.~\ref{fig:SMC-SMP-20} and for NISP-P imaging in Fig.~\ref{fig:P_examples}.

\subsection{PSF}
\begin{figure}[ht]
    \centering
    \includegraphics[width=\columnwidth]{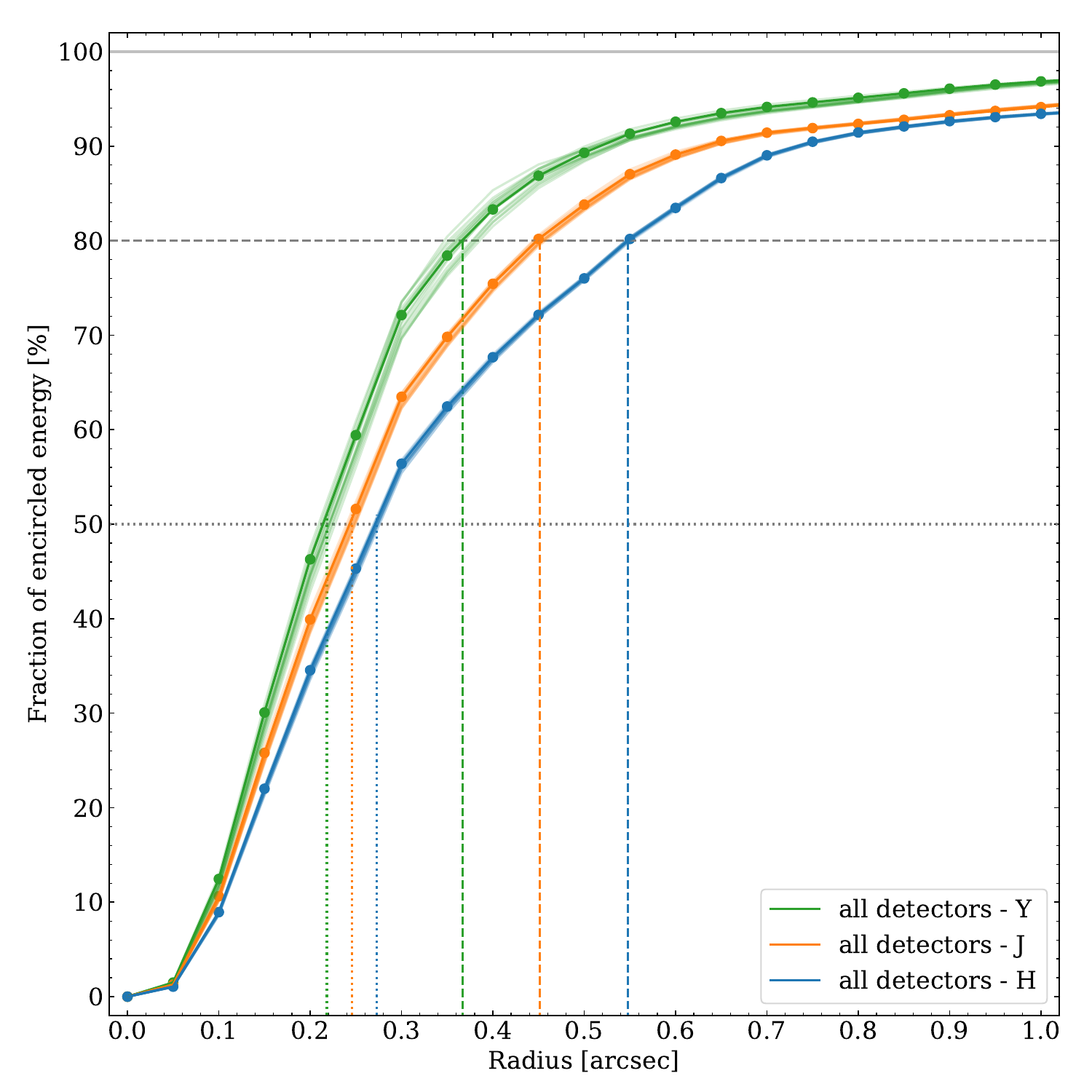}
    \caption{ 
Modelled \acf{EE} of the \Euclid NISP PSF in the three photometry channels, from data observed in the self-calibration field during the PV phase, and based on simplified PSF models. Shown is the cumulative \ac{EE} of point sources as a function of radius from the objects' centre. The three colours are curves for the \YE-band (green), \JE-band (orange), and \HE-band (blue). For each passband there is one line for each of the 16 detectors in the \ac{FPA}, in each case describing the mean detector EE curve -- which is very similar for all detectors -- and an overall average. Vertical lines mark the radius encircling 50\% (rEE50) and 80\% (rEE80) of the the total flux.
    }
    \label{fig:inflight-EE}
\end{figure}

The NISP \ac{PSF} as measured on-sky is very compact, with initial estimates of the 50\% and 80\% encircled-energy radii rEE50 and rEE80 of 0\farcs21--0\farcs27 and 0\farcs37--0\farcs55, respectively, as shown in Fig.~\ref{fig:inflight-EE} (for photometry) and tabulated for both channels in Table~\ref{tab:zp}. As expected, both rEE50 and rEE80 are larger after projection onto the detector than the pure optical component at 960\,nm measured in the lab (Fig.~\ref{fig:Measurement-EE}): (i) The \YE-band has a central wavelength of 1081\,nm \citep[see Table~\ref{tab:photo_passbands} and][]{schirmer2022}, $\sim$\,13\% longer than the lab reference, and hence produces a PSF also 13\% wider, given that the optical PSF is practically diffraction limited (Sect.~\ref{sec:opt_perf_test}). (ii) In flight the pointing jitter by the spacecraft amounts to $\sim$\,35\,mas, which directly broadens every image. (iii) On top come the effects of \ac{IPC} (Sect.~\ref{sec:cal_ipc}), which additionally broadens the \ac{PSF} upon readout, most pronounced for the steep centre, and (iv) \ac{BFE} (Sect.~\ref{sec:cal_bfe}), which broadens the \ac{PSF} more for brighter stars. Overall the \ac{PSF} measured on the \ac{FPA} is consistent with a diffraction-limited image delivered by the NISP optics.

\begin{table}[htb]
    \centering
       \caption{Initial in-flight estimates of PSF 50\% and 80\% energy radii, PSF FWHM, AB-mag zeropoints, and limiting sensitivities for the three photometric and two spectroscopic passbands of NISP, where available. Photometry sensitivity is given in AB-mag for 5$\sigma$-detections of point-sources using a model fit. For red grism spectroscopy this is a 3.5$\sigma$-detection of a point-source line with a putative width of two resolution elements of each 13.4\,\AA, given in erg\,s$^{-1}$\,cm$^{-1}$. For the blue grism a limiting sensitivity was not yet available. Both throughputs and limiting sensitivities are snapshots at the time of writing and might change over time as both the instruments and image reduction pipeline evolve. } 
    \begin{tabular}{llllll}
    \hline \hline\noalign{\vskip 1pt}
Filter & rEE50 & rEE80 & FWHM & \ac{ZP} & Lim.\ sens.\\ 
\hline\noalign{\vskip 1pt}
\YE & 0\farcs22& 0\farcs37& 0\farcs35&24.95 & 24.6\\
\JE & 0\farcs24& 0\farcs45& 0\farcs34&25.19 & 24.6\\
\HE & 0\farcs27& 0\farcs55& 0\farcs35&25.11 & 24.5\\
\BGE & 0\farcs21 & -- & -- & -- & N.A.\ \\
\RGE & 0\farcs24 & -- & -- & -- & $\sim$\,2$\times$10$^{-16}$\\ 
\hline
    \end{tabular}
    \label{tab:zp}
\end{table}

\begin{figure}[ht]
    \centering
    \includegraphics[width=\columnwidth]{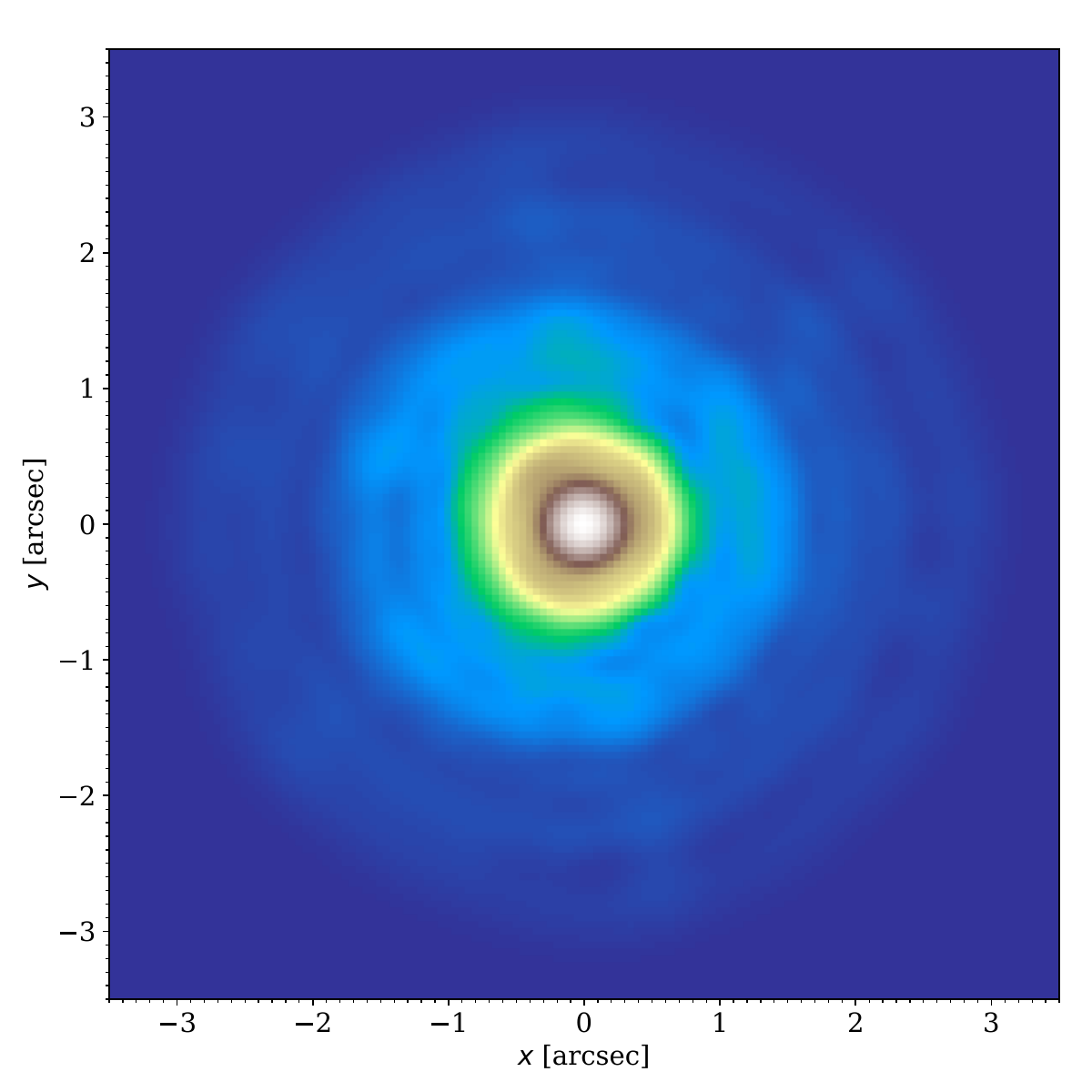}
    \caption{
    NISP-P example \ac{PSF} in the \HE-passband. This shows an initial shapelet reconstruction basen $\sim$\,1400 unsaturated point sources, on a 6$\times$ oversampled grid (0\farcs05) compared to the native NISP pixel sampling. Clearly visible is the trefoil component that is present in all NISP images as well as VIS. A 6-fold spike pattern becomes visible for brighter point sources towards or beyond saturation.
    }
    \label{fig:inflight-PSF}
\end{figure}

We note that the NISP pixels (\ang{;;0.3}) strongly undersample the \ac{PSF}. However, a reference NISP-P PSF (Fig.~\ref{fig:inflight-PSF}) and encircled-energy curves in full resolution have been derived by modelling $\sim$\,1400 unsaturated stars from 76 different exposures with Gauss-Laguerre shapelet vectors \citep{massey2005}, using 6-fold subsampling per pixel. The \ac{PSF} has a small trefoil component stemming from the \Euclid primary mirror, resulting in the visible slightly triangular shape. 

\subsection{Zero point and sensitivity}
A first assessment of the in-flight \acp{ZP} is given in Table~\ref{tab:zp}. For photometry they are within 0.1\,mag from the pre-launch zeropoints derived in \citet{schirmer2022}.
Preliminary ${\rm \ac{SNR}}=5$ limiting sensitivities for point sources in the three passbands were calculated using \ac{PV} data and are listed in Table~\ref{tab:zp}. They are averages over a \ac{ROS} pointing footprint, where the majority of area has a depth of either 3 ($\sim$\,40\%), 4 ($\sim$\,42\%), or even $>$\,4 ($\sim$\,8\%) individual dither exposures, using the standard S-dither pattern \citep[see Sect.~4 of][]{scaramella2022}. These are 0.1\,mag to 0.3\,mag fainter than the earlier predictions for limiting sensitivities listed in \citep{scaramella2022}, and these initial in-flight \ac{ZP} and depth values demonstrate that NISP's throughput and sensitivity are as expected. 

For spectroscopy, the initial estimates for the red grism also conform with expectations and requirements. The 3.5\,$\sigma$ point-source sensitivity for emission lines lies around 2$\times$10$^{-16}$\,erg\,s$^{-1}$\,cm$^{-1}$, for integration of a line width of two resolution elements of each 13.4\,\AA.

We note that both throughputs and limiting sensitivities are snapshots of the system at the time of writing based on data taken during the \ac{PV} phase and will change with both an evolving spacecraft and \ac{SGS} data reduction pipeline.

\subsection{Optical ghosting\label{sec:ghosting}}

As for all optical systems with refractive elements, NISP shows `ghost' images from multiple reflections in different parts, as well as reflections from outside the nominal light path towards the \ac{FPA}. These arise in different situations, and depending on their nature require different treatment in data reduction.

\subsubsection{Dichroic ghost}
The dichroic ghost images are caused by a double reflection inside the dichroic beamsplitter. They are doughnut-shaped, showing the central obstruction by the secondary mirror (Fig.~\ref{fig:ghosts}). The images are very weak, with a ghost ratio -- that is the ratio $f_{\rm g}$ between the total flux of a bright star and the ghost -- of $3\times10^{-8}$. This has the consequence that dichroic ghosts are only apparent for very bright saturated stars. The strength of the ghosts decreases with wavelength, that is they are brightest in \HE~and faintest in \YE.

We model the position of the ghosts with respect to their stars with a third-degree 2D-polynomial. The angular size of the ghosts is constant as a function of position, but increases with the brightness of the star. The diameter is $\sim$5\arcsec. The surface brightness of the ghosts is highly variable across the \ac{FOV} and thus they cannot be subtracted; we mask them when they violate the relative photometry requirements of faint galaxies.

\subsubsection{Filter ghosts}
Filter ghosts are considerably brighter than the dichroic ghost images. They are caused by a double-reflection between the planar and curved side of the NISP filters. Their images are approximately circular, with a characteristic cusp caustic in their centre (Fig.~\ref{fig:ghosts}) that could be mistaken for an extragalactic source. 
The size of the filter ghosts is $\sim$\,7--12\arcsec and depends on the position in the \ac{FOV}. As for the dichroic, their relative position with respect to the bright stars is modelled with a third-degree 2D-polynomial as a function of star position. Separate models are required for each NISP filter. The ghosts must be masked as for dichroic ghost images. 

Any other reflections between optical surfaces inside NISP become very large in area once they reach the focal plane. Their surface brightness is therefore so low that they are not recognisable even for the brightest stars (4\,mag\,AB) allowed inside the \ac{FOV}.

\subsubsection{Reflections on mechanical parts}
\begin{figure}[t]
    \centering
    \includegraphics[width=1.0\hsize]{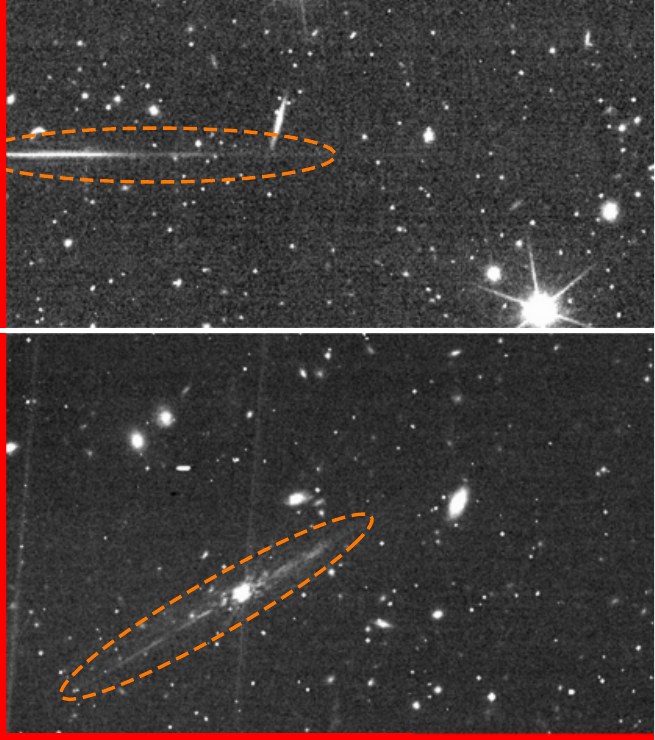}
    \caption{Glints in NISP images, when a bright star falls onto the edge (\textit{top panel}) or near the corner (\textit{bottom}) of the baffle just outside the \ac{FPA}. The red line marks the edges of the detectors.}
    \label{fig:glints}
\end{figure}

Ghost images also occur when short streaks or glints are created by bright stars falling onto reflective spots just outside the detectors. In case of baffle edges, bright streaks of about 1\,arcmin appear in the corners or along the edges of the \ac{FOV} (Fig.~\ref{fig:glints}). They are considerably brighter than the dichroic and filter ghosts and must be masked in all cases.

\subsubsection{Reflections on NI-OA}
\begin{figure*}[t]
    \centering
    \includegraphics[width=1.0\hsize]{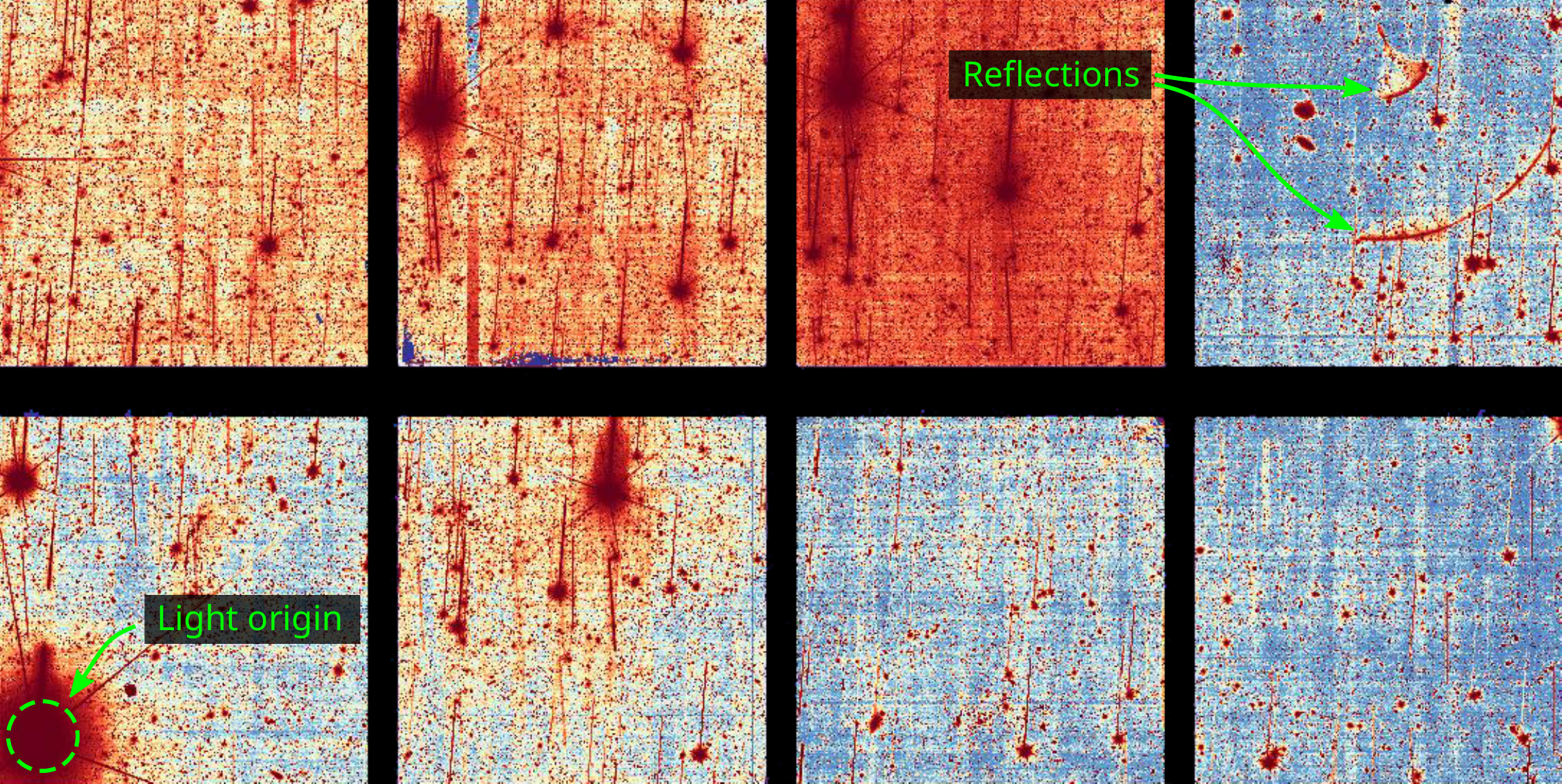}
    \caption{The centre two detector rows, showing examples of large point-source caustics. The bright star at the bottom-left edge causes a pair of concentric arcs in the upper-right detector, diametrically opposite the centre of the \ac{FOV}. Characteristically, the caustics' symmetry axes do not point to the star, but at a \ang{90;;} angle to it.}
    \label{fig:ghost_arclike}
\end{figure*}

Large and complex caustic-shaped reflections occur for very bright stars due to multiple reflections within \ac{NI-OA} (Fig.~\ref{fig:ghost_arclike}). The optical surfaces involved have not yet been identified. The star and its ghosts are diametrically opposite the centre of the FPA. The ghost's morphology and strength are highly dependent on the star's position, necessitating ray-tracing for the masking models. At this point we can likely rule out an involvement of the detector surfaces, as these ghosts also appear when the star moves into the gap between detectors.

\subsubsection{Persistence ghosts from wheel actuations}
\begin{figure}[t]
    \centering
    \includegraphics[width=1.0\hsize]{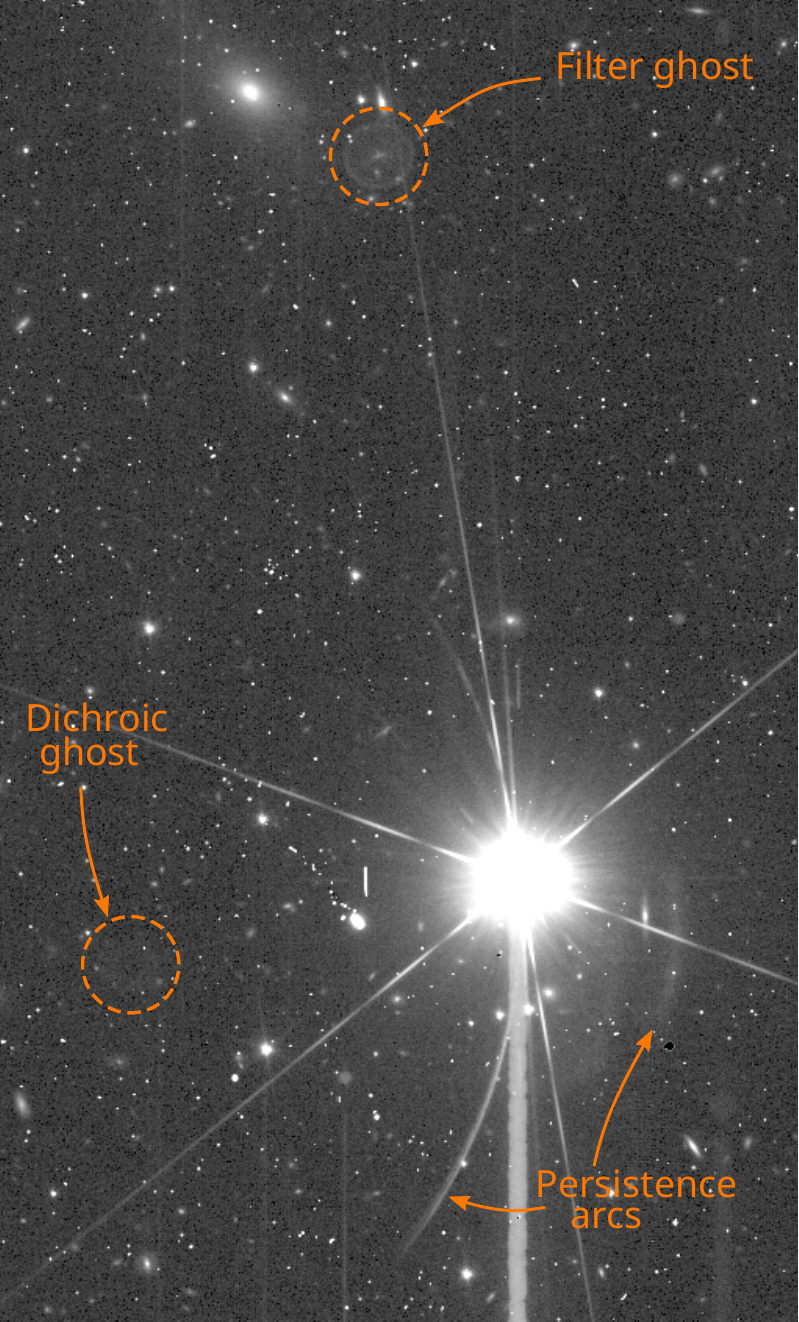}
    \caption{Appearance of the common optical ghosts in a NISP \YE-band image. The bright star has about 8.7\,AB\,mag in the 2MASS $J$-band. Most other linear features -- vertical or near-vertical -- are persistence effects from preceding spectroscopic exposures.}
    \label{fig:ghosts}
\end{figure}

The last characteristic features in NISP images are bright arcs with a curvature radius of approximately 1.7\,arcmin, running through the centres of very bright stars (Fig.~\ref{fig:ghosts}). These are caused when NISP switches between filter and grism observations. Since NISP does not have a separate shutter, bright-star images fall on the detector while the wheels are moving. Since both filters and grisms carry optical power, the images move on the detector plane while the wheels rotate. The ghosts are either focused and of high surface brightness, or defocused and  diffuse and of low surface brightness. They also need to be masked. A model for them still is still in development.

\section{\label{sec:operations}Operational flexibility and limits}
\Euclid is not a classical observatory such as HST or JWST, but by design of its operation closer to an experiment with a very constrained but often-repeated set of observations. The \Euclid Wide and Deep Surveys will consist of basically 50\,000 identical observations, the \ac{ROS}, just at different positions on the sky. Executed once per field on the sky for the Wide Survey, many times for the Deep Fields.

This repetitive nature of survey activities allowed the design of two more robust and accurate instruments, but led to a limited flexibility in commanding NISP. In the following we discuss the commanding and hardware limits -- mainly to manage expectations and options for what kind of other programmes could be possible with NISP in the future, in case that time-slots were to become available during idle periods of survey operations, or after the surveys are completed. Neither is guaranteed, and specific calls would provide much more detailed options and constraints in due time.

{\bf Lead times:} Overall, planning cycles are long and observations have a long lead time, of at least 4 weeks, and are limited by the need to design, approve, and upload commanding sequences. Any non-standard observations would in addition require substantial individual vetting and testing in order to not run into any instrumental limits of onboard processing, storage capacity, data rate, or interference with the VIS instrument's operations. \Euclid is best for surveys and not suited for example for rapid-response target-of-opportunity observations.

{\bf Standard and non-standard operation modes:}
\Euclid science operation and calibration has been optimised for the standard \ac{ROS} observing modes. For NISP this is the \ac{MACC} readout mode with several groups of read images that are averaged, and then a slope is fit to these values to derive a count-rate of signal in each pixel. Effectively only the slope value and a value for the quality of the slope fit is linked down.

As mentioned already in Sect.~\ref{sec:cal} there are also other modes that are not used for science, but regularly for calibration or were used
for diagnostics during initial commissioning. These include a mode downlinking not the fitted slope but the value of each group. For a standard spectroscopy observation using MACC(15,16,11) this would provide 15 datapoints instead of one.

Even more information can be obtained with a `engineering raw mode', which could read and downlink all group-average frames of a MACC observation, not just the fitted slope. For the \ac{ROS} there is even a routine downlink of up to five individual rows for each read of each detector, as a diagnostic if non-nominal behaviour is observed for an exposure, and for monitoring of the \ac{FPA} properties over time.

Using non-ROS or non-standard modes for science is in principle feasible, but requires substantial work beforehand: For each non-ROS configuration the timing of commanding, timing of data processing and internal transfer, data storage buffer sizes, data downlink rates, and many other dependencies have to be simulated on ground, and be planned in conjunction with VIS and overall \Euclid operations and commanding. For example engineering raw-mode observations can usually only be done for single detectors at a time instead of 16 in the \ac{FPA} due to sheer data volume. MACC observations with fewer than 16 reads per group could bring on-board data processing to its limits, for example by violating timing constraints.

Aside from this, corresponding calibration data are routinely only created for ROS data. This means any non-ROS MACC or other non-standard observing modes need their own plan to create required calibration data and concepts. It is unlikely that a simple scaling from ROS calibration to other observing modes can be achieved beyond a rather low level of accuracy.

So while other than ROS MACC-mode options exist, they can be quite complex to implement. They also depend on resources on ground that are not always guaranteed.

{\bf Bright and faint limits:}
Further core constraints interesting beyond \Euclid's surveys are NISP's observational bright and faint limits. An imaging ROS MACC(4,16,14) will start saturating for $\sim$16.0\,AB\,mag point sources, differing by $\sim$\,0.3\,mag between passbands due to different levels of undersampling, and with some variation across the \ac{FOV}. Something that is ultimately fixed for NISP is the frame time: each read will take a fixed 1.45408\,s, set by the read-out clocks. For shorter exposures one will therefore have to consider other MACC parameters. From the architecture of the warm electronics, the shortest integration possible with NISP is a MACC(2,1,1) exposure, that is two groups of one frame each, with one dropped frame in the middle. This corresponds to an integration time of $t_\mathrm{int}=2.91$\,s. For photometry the point-source saturation limit would then be $\sim$3.6\,mag brighter than for a standard MACC(4,16,4), around 12.0\,mag--12.5\,mag. This is a hard bright limit for NISP photometry for the full \ac{FOV} and using standard observing modes. Corresponding limits apply when using a grism.

Long exposures are constrained by the maximal allowed number of 15 groups and 16 reads per group. Even though there is no limit on the number of dropped frames, there is a maximal duration of a spacecraft pointing of 50\,min. At some point it will be more efficient to simply obtain several subsequent exposures at the same location, especially given non-zero backgrounds that are adding up over time. If the telescope is slightly dithered in between then this will also aid cosmic ray rejection. This is the approach of the \Euclid Deep Survey with up to $\sim$\,50 \ac{ROS} observations per sky location.

{\bf Time-resolution:}
The second interesting parameter is the potential time-resolution of NISP data. As just stated, the shortest regular observation is $t_\mathrm{int}=2.91$\,s. At this exposure time up to seven exposures can be recorded in direct succession -- with an extra reset frame before each of these exposures -- and then after a buffer time of 3\,s the next block of seven. With the engineering raw mode a MACC(15,1,1) can be programmed, that downlinks each of the 15 single `group' frames, each followed by a dropped frame. Here every other frame is read and downlinked -- but, as noted, owing to the data volume only for a single detector instead of 16. Hence this in principle also provides an option for shorter exposures of 1.45408\,s and cadence of 2.91\,s, for 15 frames, but for a much smaller area on the sky. At this point it seems unlikely that \Euclid and NISP with their explicit survey focus will provide substantially improved capabilities over other facilities.

\section{\label{sec:outlook}Summary and outlook}
\Euclid has launched and NISP is working at L2 as envisioned. NISP's performance is excellent, in both spectroscopy and imaging channels, and the instrument will create two unprecedented datasets that will be the new standard for NIR data for decades to come. Post-launch expectations for \Euclid are an operational lifetime of substantially longer than the nominal 6-year survey, possibly up to ten years in total.

As \Euclid was launched not too long from the Solar maximum, NISP will experience most of its exposure to Solar cosmic rays rather early in its operational lifetime, but the impact will be small. All of the optics are resistant to radiation, as are the detectors, which are expected to degrade very little. A formal 5\% loss in \ac{QE} is budgeted for NISP's performance, but the actual impact is expected to be substantially less.

NISP operations cover both nominal survey operations as well as regular calibration observations to update and improve calibration products, and to monitor any evolutionary effects. In addition there will be options for `decontamination' periods, in case too much water outgassing from the various \Euclid components deposits onto mirrors, or NISP's lenses or \ac{FPA}. The latter is the coldest part of the instrument and specifically prone to ice deposition. Plans for a brief partial warmup of \Euclid are in place and have already been executed, which can be repeated when needed. 

All this provides the prospect of excellent data during \Euclid's wide and deep surveys, and potentially beyond. Despite its undersampling of the PSF, NISP is an enormous advance from any previous near-infrared instrument capability currently available. Only the combination of field area and very high throughput of telescope and instrument make it possible to carry out the planned surveys, for which HST or JWST would take more than 100 years. All of NISP's data will be released to the public in regular intervals along the mission timeline \citep[see][and \url{https://www.euclid-ec.org}]{euclid2024}, calibrated and ready to use for any astrophysical project -- and for anyone in the world.


\begin{acknowledgements}

\AckEC
\end{acknowledgements}

%
%

\bibliography{euclid_nisp.bib}
%

\begin{appendix}
  \onecolumn 
%



\newpage

\section{NISP focal plane layout}
\label{sec:fpa_layout}

\begin{figure*}[hb!]
    \centering
    \includegraphics[width=0.9\textwidth]{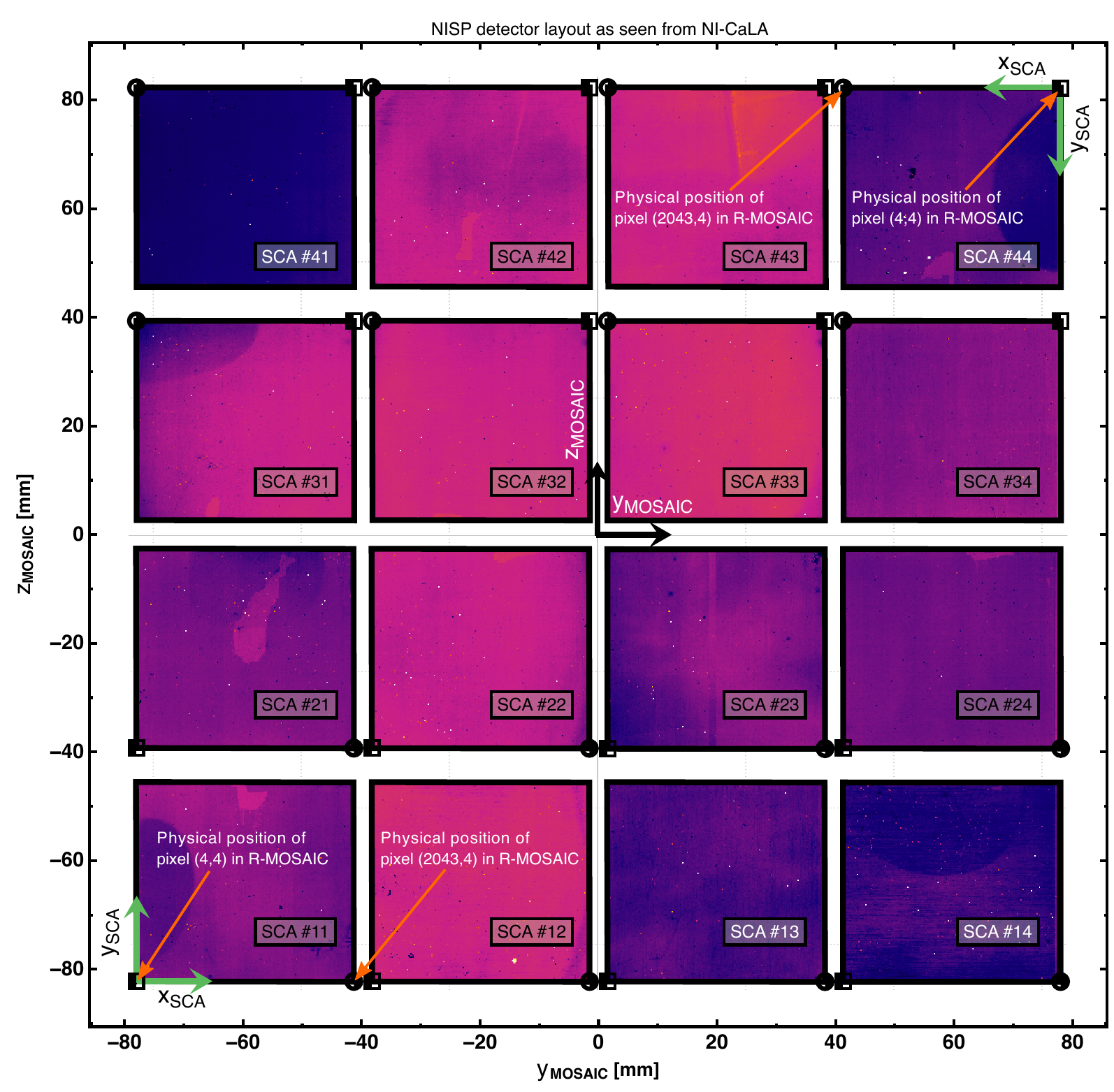}
    \caption{Layout of the NISP \ac{FPA}, orientated as looking towards the \ac{FPA} from the sky and NISP optics. Shown are the positions 11 to 44 of the SCS in the focal plane mosaic (see Table~\ref{tab:nisp_fpa}), the origin point and orientation of the physical R-MOSAIC coordinate system, as well as the pixel coordinate system for each of the \ac{SCA} detectors. For the latter please note the 180$^\circ$ rotation between the upper and lower half of detectors. For a view towards the sky see Fig.~\ref{fig:fpa_layout_sky}.}
    \label{fig:fpa_layout}
\end{figure*}

\begin{figure*}[hb!]
    \centering
    \includegraphics[width=0.9\textwidth]{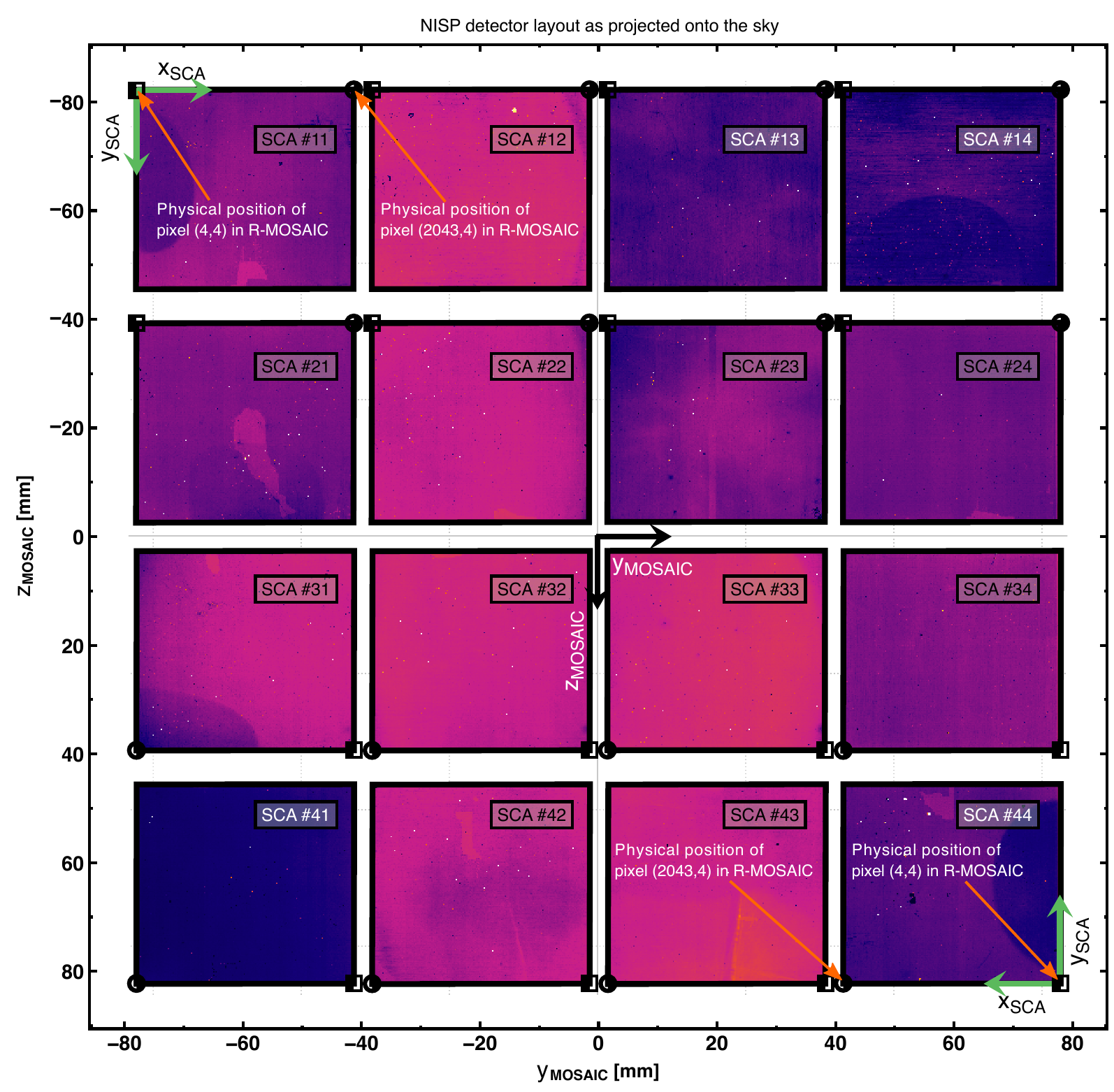}
    \caption{Layout of the NISP \ac{FPA}, orientated as looking towards the sky. Shown are the positions 11 to 44 of the SCS in the focal plane mosaic (see Table~\ref{tab:nisp_fpa}), the origin point and orientation of the physical R-MOSAIC coordinate system, as well as the pixel coordinate system for each of the \ac{SCA} detectors. For the latter please note the 180$^\circ$ rotation between the upper and lower half of detectors. For a view from the sky towards the \ac{FPA} see Fig.~\ref{fig:fpa_layout}. An internal definition of a position angle value of zero would rotate this view on the sky so that North is upwards and East is to the left. }
    \label{fig:fpa_layout_sky}
\end{figure*}

NISP's focal plane array (\ac{FPA}) and detector system (\ac{DS}) comprises the 16 sensor chip systems (\ac{SCS}) in a 4$\times$4 array, each being a triplet of sensor chip array (\ac{SCA}), sensor chip electronics (\ac{SCE}), connected by a cryo flex cable (\ac{CFC}). In order to associate the positions of these \ac{SCS}, with their properties listed in Table~\ref{tab:nisp_fpa} and shown in Fig.~\ref{fig:NI-SCS-QE-Noise}, and with data products created by NISP, we show two versions of the physical layout of the \ac{FPA}: Fig.~\ref{fig:fpa_layout} is the layout as looking from the sky or NISP optics towards the \ac{FPA}, Fig.~\ref{fig:fpa_layout_sky} is the reverse view, the \ac{FPA} as it would appear projected onto the sky.

Both versions show the same information, the \ac{SCA} positions from 11 to 44 in the \ac{FPA} detector mosaic, which can also be found in headers of all FITS files associated with detector-based data products created by NISP. Also indicated are the frame orientations for each \ac{SCA} in the physical `R-MOSAIC' coordinate system, with units in mm. Each \ac{SCA} has its own pixel coordinate system: light-sensitive pixels range from 4 to 2043 in each of the two dimensions ($x_\mathrm{SCA}$,$y_\mathrm{SCA}$), with count starting at 0. Rows/columns 0--3 and 2044--2047 contain reference pixels that are not sensitive to light. Note that the \ac{SCA}-coordinate system orientation is rotated by 180$^\circ$ between the upper and lower 8 detectors. The black square for each detector marks the physical location of pixel (4,4), the black circle of pixel (2043,4) in the `R-MOSAIC' coordinate system.

On the side the R-MOSAIC physical coordinate system ($y_\mathrm{MOSAIC}$,$z_\mathrm{MOSAIC})$, using units of mm, is shown for reference.

\end{appendix}

\end{document}